\documentclass[10pt,onecolumn]{IEEEtran}
\usepackage{cite}
\ifCLASSINFOpdf
  \usepackage{graphicx}
\else
\fi

\usepackage{amsfonts}
\usepackage{amsmath}
\usepackage{amssymb}
\usepackage{amscd}
\usepackage{mathrsfs}
\usepackage{color}
\usepackage{amsthm}
\usepackage{mathtools}
\allowdisplaybreaks
\usepackage{algorithmic}
\usepackage{array}
\usepackage{stfloats}
\usepackage{url}
\usepackage{eqparbox}
\usepackage{fancyhdr}
\usepackage[bookmarks=false]{hyperref}

\newtheorem{thrm}{Theorem}
\newtheorem{prop}{Proposition}
\newtheorem{Lemma}{Lemma}

\usepackage{etoolbox}
\makeatletter
\patchcmd{\@makecaption}
  {\scshape}
  {}
  {}
  {}
\makeatletter
\patchcmd{\@makecaption}
  {\\}
  {.\ }
  {}
  {}
\makeatother

\begin{document}

\title{Detecting an Odd Restless Markov Arm with a Trembling Hand}
%
%
%

\author{P. N. Karthik and
        Rajesh Sundaresan

        \thanks{P. N. Karthik is with the Department of Electrical Communication Engineering at the Indian Institute of Science, Bangalore 560012, Karnataka, India. Rajesh Sundaresan is with the Department of Electrical Communication Engineering and the Robert Bosch Centre for Cyber Physical Systems at the Indian Institute of Science, Bangalore 560012, Karnataka, India. Email: (periyapatna, rajeshs)@iisc.ac.in.}

        \thanks{This work was supported by the Science and Engineering Research Board, Department of Science and Technology (grant no. EMR/2016/002503), by the Robert Bosch Centre for Cyber Physical Systems and the Centre for Networked Intelligence at the Indian Institute of Science.}
        
        \thanks{A shorter version of this paper was presented at the 2020 IEEE International Symposium on Information Theory (ISIT).}
        }
\maketitle

\begin{abstract}
In this paper, we consider a multi-armed bandit in which each arm is a Markov process evolving on a finite state space. The state space is common across the arms, and the arms are independent of each other. The transition probability matrix of one of the arms (the odd arm) is different from the common transition probability matrix of all the other arms. A decision maker, who knows these transition probability matrices, wishes to identify the odd arm as quickly as possible, while keeping the probability of decision error small. To do so, the decision maker collects observations from the arms by pulling the arms in a sequential manner, one at each discrete time instant. However, the decision maker has a trembling hand, and the arm that is actually pulled at any given time differs, with a small probability, from the one he intended to pull. The observation at any given time is the arm that is actually pulled and its current state. The Markov processes of the unobserved arms continue to evolve. This makes the arms restless.

 For the above setting, we derive the first known asymptotic lower bound on the expected time required to identify the odd arm, where the asymptotics is of vanishing error probability. The continued evolution of each arm adds a new dimension to the problem, leading to a family of Markov decision problems (MDPs) on a countable state space. We then stitch together certain parameterised solutions to these MDPs and obtain a sequence of strategies whose expected times to identify the odd arm come arbitrarily close to the lower bound in the regime of vanishing error probability. Prior works dealt with independent and identically distributed (across time) arms and rested Markov arms, whereas our work deals with restless Markov arms. 
\end{abstract}


\section{Introduction}\label{sec:introduction}
%
The problem of odd arm identification deals with identifying an anomalous (or \emph{odd}) arm in a multi-armed bandit as quickly as possible, while keeping the probability of decision error small. Here, the term \emph{anomaly} simply means that the law, say $\psi_1$, of one of the arms is different from the common law, say $\psi_2$, of each of the other arms. We assume that the arms are independent of each other. A decision maker, who may or may not have prior knowledge of $\psi_1$ and $\psi_2$, and whose goal it is to identify the index of the odd arm, samples the arms in a sequential manner, one at a time.  The process of sampling the arms continues until the decision maker is sufficiently confident of which arm is odd, at which time he stops further sampling and declares the index of the odd arm. In forming his  decision about the odd arm, it is important for the decision maker to ensure that his error probability is low (below a pre-specified threshold). It is natural to expect that smaller the pre-specified error probability threshold, longer the decision maker will have to wait before declaring the odd arm location. The main objective of this paper is to identify the asymptotic growth rate of the decision time as a function of the error probability, where the asymptotics is as the error probability goes to zero.  

 Prior works on odd arm identification consider the cases when either each arm yields independent and identically distributed (iid) observations \cite{Vaidhiyan2017, prabhu2017optimal, vaidhiyan2017learning}, or when each arm yields Markov observations from a common finite state space \cite{pnkarthik2019learning}. When each arm yields iid observations, $\psi_1$ refers to the law of a random observation coming from the odd arm, while $\psi_2$ refers to the law of a random observation coming from any of the non-odd arms. When each arm yields Markov observations, $\psi_1$ refers to the transition law of the Markov process of the odd arm, while $\psi_2$ refers to the transition law of the Markov process of each of the non-odd arms. When the state space is discrete, the transition laws $\psi_1$ and $\psi_2$ may be specified equivalently by the respective transition probability matrices, say $P_1$ and $P_2$, where $P_1\neq P_2$. We use the term `observation' in place of the commonly used term `reward' because our focus is on early identification of the odd arm in contrast to reward maximisation or regret minimisation.
   

An important feature of the setting in \cite{pnkarthik2019learning} is that the Markov process of any given arm evolves by one time step only when the arm is selected, and does not evolve otherwise; this is known as the setting of \emph{rested} arms. In this paper, we partially extend the results of \cite{pnkarthik2019learning} to the more difficult {\em restless} arms setting in which the Markov process of each arm continues to evolve whether or not the arm is selected. The continued evolution of the Markov process of each arm makes it necessary for the decision maker to keep a record of (a) the time elapsed since each arm was previously selected (called the arm's \emph{delay}), and (b) the state of each arm as observed at its previous selection time (called the \emph{last observed state} of the arm). Notice that the notion of arm delays is superfluous when the arms are {rested} as in \cite{pnkarthik2019learning} since the unobserved arms remain frozen at their previously observed states. It is also superfluous in the special case of the restless setting when each arm yields iid observations (as in \cite{Vaidhiyan2017, prabhu2017optimal, vaidhiyan2017learning}) because the last observed state of each arm is independent of the arm's current state. Therefore, the notions of arm delays and last observed states are strikingly new features of the setting of general restless Markov arms. 

For the rest of this paper, we assume that the transition matrices $P_1$ and $P_2$ of the odd arm and the non-odd arm Markov processes are known to the decision maker. Further, we assume that the common state space of the Markov process of each arm is finite as in \cite{pnkarthik2019learning}. All the essential conceptual difficulties related to restless arms remain despite these simplifications.  New tools are needed to overcome the difficulties, and these are highlighted in Section \ref{subsec:key_contributions}. The case when $P_1$ and $P_2$ are unknown is beyond the scope of this paper and is currently under study.

\subsection{Motivation and the Notion of a Trembling Hand}
Our motivation to study the restless odd Markov arm problem comes from the desire to extend, to more general settings, the decision theoretic formulation of a certain visual search experiment conducted by Sripati and Olson \cite{sripati2010global} and analysed in Vaidhiyan et al. \cite{Vaidhiyan2017}. In this  experiment, human subjects were shown a number of images at once, with one {\em oddball} image in a sea of {\em distracter} images. The goal of the experiment was to understand the relationship between (a) the average time taken by the human subject to identify the oddball image, and (b) the dissimilarity between the oddball and distracter images as perceived by the human subject. The images used in the above experiment were static images. Vaidhiyan et al. also conducted experiments with dynamic drifting-dots images (movies), similar to the ones conducted by Krueger et al. \cite{krueger2017evidence}, in which the dots in each movie location executed Brownian motions with fixed drifts. Further, the drifts were identical in all the distracter movie locations, and were different from the drift in the oddball movie location. In this context, what are optimal strategies to identify the oddball movie? A systematic analysis of this question, along the lines of \cite{Vaidhiyan2017}, requires an understanding of the {restless} odd Markov arm problem which forms the main subject of this paper.

It is often the case in such visual search experiments that though the subject (or decision maker) intends to focus his attention at a certain location, the actual focus location differs from the intended focus location with a small probability. We model this in our multi-armed bandit setting as a {\em trembling hand} for the decision maker: with probability $1-\eta$, the decision maker pulls the intended arm, but with probability $\eta$, the decision maker pulls a uniformly randomly chosen arm (we use the phrases `arm pulls' and `arm selections' interchangeably). Up to Section \ref{sec:main_result}, we assume that $\eta>0$, as is often the case in visual search experiments such as that described above. The case when $\eta=0$ is dealt with separately in Section \ref{sec:case_eta_=_0}.

{\color{black} Our assumption about the uniform sampling of the arms under the trembling hand model is merely for convenience, and any probability distribution on the arms that puts a strictly positive mass on each of the arms may be used in place of uniform distribution. The values of all the expectations and probabilities (in particular, the lower bound of Section \ref{sec:lower_bound}), which rely on the uniform sampling assumption, will accordingly differ.

For a related example in the cognitive radio setting (no trembling) in which the number of anomalous arms may be more than one, see \cite{zhao2008myopic}.}

\subsection{Prior Works on Restless Markov Arms}
The topic of restless Markov arms has been studied extensively in the literature in the context of reward maximisation (or equivalently, regret minimisation). In such works, each arm is assumed to yield, upon being sampled, an immediate `reward' based on the arm's current state. Regret is then defined as the difference between the expected sum of rewards obtained under a particular arm selection scheme and that obtained by a scheme that knows which arm yields the highest expected reward. Whittle \cite{whittle1988restless} 
refined and extended the results of Gittins \cite{gittins1979bandit} on the optimality, in the setting of rested arms, of a certain index-based policy. Whittle \cite{whittle1988restless} demonstrated that Gittins's policy in \cite{gittins1979bandit} is not necessarily optimal in the context of restless arms, introduced a new index (now called \emph{Whittle's index}) which could be computed if each arm satisfied an \emph{indexability} condition, and demonstrated that the new index coincides with Gittins's index in the rested setting. 
Yet, as Whittle showed, the new index-based policy is not necessarily optimal for the general setting of restless arms.

Whittle's results require the Markov transition laws of each of the arms to be known beforehand. Extensions of Whittle's results to the case when the laws are not known beforehand appear in Liu et al. \cite{liu2012learning}. Ortner et al. \cite{ortner2012regret} provide a policy that, when the transition laws of the arms are unknown, gives a regret of the order $O(\sqrt{T})$ after $T$ time steps in relation to a policy that knows the Markov transition laws of all the arms. As Ortner et al. show in \cite{ortner2012regret}, an optimal policy for the restless bandit problem does not necessarily pick the arm with the largest stationary mean at each time instant\footnote{This is indeed the case in a multi-armed bandit problem with iid observations from each arm, as was shown in \cite{auer2002finite}.}, but instead switches between the arms in an optimal fashion. Working on this key idea, Gr\"unelwalder et al.  \cite{grunewalder2019approximations} provide conditions under which the problem of finding the arm with the largest stationary mean serves as a ``good'' approximation to the original problem of finding the optimal arm switching strategy when each arm is a stationary $\phi$-mixing process and the arms are restless. The works \cite{ortner2012regret} and \cite{grunewalder2019approximations} deal with general state spaces (i.e., not necessarily finite or countable) and address the associated technical challenges.
 
In contrast to all the works mentioned above, this paper focuses on the \emph{stopping} problem of identifying the index of the odd arm as quickly as possible. 
 It is worth noting here, as also noted in \cite{Bubeck2011}, that policies which are optimal for the problem of minimising regret may not necessarily be optimal in the stopping problem context.  

For applications of the restless odd Markov arm problem, see \cite{pnkarthik2019learning}. For a related problem of best arm identification instead of odd arm identification, see \cite{Kaufmann2016, moulos2019optimal}. {\color{black} The recent papers  \cite{deshmukh2018controlled}, \cite{deshmukh2019sequential} and \cite{prabhu2020sequential} deal with more general problems of sequential hypothesis testing in multi-armed bandits, special cases of which are the problems of best arm identification and odd arm identification, in the context of iid observations from each arm. In contrast to these papers, our work deals specifically with the problem of odd arm identification in the context of restless Markov arms.}

\subsection{An Overview of the Results and Our Contributions}\label{subsec:key_contributions}

We now provide an overview of our results and highlight our contributions.
\begin{enumerate}
	\item We show that given a pre-specified error probability threshold $\epsilon>0$, the expected time taken by the decision maker to identify the index of the odd arm with probability of error at most $\epsilon$ grows as $\Theta(\log (1/\epsilon))$. We give a precise characterisation of the best (smallest) constant multiplying $\log (1/\epsilon)$, which we call $R^*(P_1,P_2)$, in terms of the Markov transition probability matrices $P_1$ and $P_2$. This is the first known characterisation of this constant for the setting of restless Markov arms. 
       See Section \ref{sec:lower_bound} for an exact mathematical expression. 
       We prove this by first showing a lower bound in Section \ref{sec:lower_bound} and then a matching asymptotic upper bound in Section \ref{sec:achievability}.
      
     \item An examination of the lower bounds in the prior works \cite{Vaidhiyan2017, prabhu2017optimal, vaidhiyan2017learning, pnkarthik2019learning} reveals that the best constant multiplier in these works is the solution to an optimisation problem having an outer supremum over all (unconditional) probability distributions on the arms, followed by an inner minimum over all alternative odd arm locations (i.e., a sup-min optimisation problem). A further examination reveals that when arm $h$ is the odd arm, there exists a probability distribution $\lambda_h^*$ on the arms, possibly depending on the odd arm location $h$,  that (a) attains the outer supremum, and (b) puts equal mass on each of the non-odd arm locations.
     
      Along lines similar to those of the prior works, we show that the best constant multiplier $R^*(P_1,P_2)$ is the solution to a sup-min optimisation problem in which the supremum is over all \emph{conditional} probability distributions on the arms, conditioned on arm delays and last observed states, and the minimum is over all alternative odd arm locations. We also show that the constant $R^*(P_1,P_2)$ is not a function of the actual odd arm location; this is due to symmetry in the structure of the arms. The constant $R^*(P_1,P_2)$ represents the amount of effort required to identify the true odd arm location by guarding against identifying the nearest, incorrect alternative odd arm location. 
     
     However, given an odd arm location $h$, the question of whether there exists a conditional probability distribution that attains the supremum in the expression for $R^*(P_1,P_2)$ is still under study.
	
	\item In order to derive the constant $R^*(P_1,P_2)$, we use the fact that the arm delays and the last observed states form a \emph{controlled Markov process}, with the arm selections playing the role of \emph{controls}. This approach of ours takes into account the delays and the last observed states of \emph{all} the arms jointly. In contrast, the approaches of \cite{Vaidhiyan2017, prabhu2017optimal, vaidhiyan2017learning, pnkarthik2019learning} suggest dealing with the delays and the last observed states of each of the arms separately, which we view as a `local' perspective of the arm delays and the last observed states. In Section \ref{subsec:infinite_LPP}, we show that this local perspective of arm delays and last observed states leads to an infinite dimensional, constrained, linear programming problem (LPP). The drawback of this approach is that it is not easy to find the tightest set of constraints for the LPP. As a consequence, the constant multiplier obtained as the solution to the LPP may not necessarily be the best (smallest). 
	
	On the other hand, our `lift' approach, which considers the delays and the last observed states of all the arms jointly, leads us naturally to a family of Markov decision problems (MDPs) and, in turn, provides the necessary perspective to arrive at the best constant multiplier $R^*(P_1,P_2)$.
	
	\item We show that under a \emph{stationary} arm selection policy (in which at each time, the arms are selected according to a certain conditional probability distribution on the arms, conditioned on the delays and last observed states at that time), the aforementioned controlled Markov process is, in fact, a Markov process. Additionally, we show that under every stationary arm selection policy, this Markov process is \emph{ergodic} when the trembling hand parameter $\eta>0$ (Lemma \ref{lem:pi_delta^lambda_is_an_SSRS}). It is this ergodicity property, together with the strict positivity of the trembling hand parameter $\eta$, that plays a crucial role in our analysis of the lower and the upper bounds. The case $\eta=0$ demands a careful examination since, in this case, such an ergodicity property is not readily available for every stationary arm selection policy. 	
	\item We show that for every arm selection policy of the decision maker, stationary or otherwise, what enters into the analyses of the lower and the upper bounds is the following statistic: for each possible value of arm delays $\underline{d}$, last observed states $\underline{i}$ and arm $a$, the long-term fraction of times the aforementioned controlled Markov process visits the state $(\underline{d},\underline{i})$ and arm $a$ is selected.  
	This fact, together with Theorem \ref{thrm:restriction_to_SRS_policies}, enables us to restrict attention only to stationary arm selection policies in arriving at the best constant multiplier $R^*(P_1,P_2)$.  
	
	In spite of the above simplification, the computability of $R^*(P_1,P_2)$ remains an issue since it involves a search over the space of all stationary arm selection policies. One must resort to \emph{Q}-learning in the context of restless Markov arms (see, for instance, \cite{avrachenkov2020whittle}) to compute $R^*(P_1,P_2)$. {\color{black} Under some circumstances, good approximations to $R^*(P_1, P_2)$ may be possible; see Section \ref{sec:conclusion}.}
	
	\item The question of whether the supremum in the expression for $R^*(P_1,P_2)$ is attainable is still under study, as mentioned in point 2 above. The arm delays, being positive and integer-valued, introduce a countably infinite dimension to the problem. As a consequence, it is not clear if the space of all conditional distributions on the arms, conditioned on the arm delays and the last observed states, is compact. In the iid and the rested Markov settings of the prior works, only unconditional distributions on the arms appear in the analysis of the lower and the upper bounds, and because of the finite nature of the number of arms, it follows immediately that the space of all unconditional distributions on the arms is compact. Such a compactness property plays a key role in showing that the supremum is attained.
	
	Notwithstanding the additional technical difficulty encountered in the setting of restless arms due to the presence of the countably infinite-valued arm delays, we show that the supremum in the expression for $R^*(P_1, P_2)$ may be approached arbitrarily closely by stitching together certain parameterised solutions to the MDPs mentioned in point 3 above. We present the details in Section \ref{sec:achievability}.

    \item The trembling hand model (with $\eta>0$) may be viewed as a regularisation that ensures stability of the aforementioned controlled Markov process (of arm delays and last observed states) for free. If $\eta=0$, one could deliberately add some regularisation parameterised by $\eta$, re-label the constant $R^*(P_1,P_2)$ in this case as $R_\eta^*(P_1,P_2)$ for each $\eta>0$, and analyse the limiting value of $R_\eta^*(P_1,P_2)$ as $\eta\downarrow 0$. We show that in this case, (a) the limit of $R_\eta^*(P_1,P_2)$ as $\eta\downarrow 0$ exists, and (b) the upper bound is governed by $\lim\limits_{\eta\downarrow 0}R_\eta^*(P_1,P_2)$, while the lower bound is governed by $R_0^*(P_1,P_2)$ (which is obtained by plugging $\eta=0$ in the expression for $R_\eta^*(P_1,P_2)$). So, the question then is, do these lower and the upper bounds match? In Section \ref{sec:case_eta_=_0}, we are only able to establish that $\lim\limits_{\eta\downarrow 0}R_\eta^*(P_1,P_2)\leq R_0^*(P_1,P_2)$. A key tool needed to establish equality in this inequality is the ``envelope theorem'' \cite[Theorem 2]{milgrom2002envelope}. A verification of the hypotheses of the envelope theorem for the setting of restless arms still remains open.
    
	\item We verify that the envelope theorem holds in the iid and rested Markov settings of the prior works \cite{Vaidhiyan2017, prabhu2017optimal, vaidhiyan2017learning, pnkarthik2019learning}, thus leading to matching upper and lower bounds in these works. Thus, sufficient conditions for the upper and the lower bounds to match are either (a) $\eta>0$, or (b) $\eta=0$ and the observations come from either iid or rested Markov arms.

\end{enumerate}

\subsection{Organisation of the Paper}
The rest of this paper is organised as follows. In Section \ref{sec:preliminaries}, we set up the notations that we use throughout the paper. In Section \ref{sec:MDP_preliminaries}, we provide some preliminaries on MDPs. In Section \ref{sec:lower_bound}, we present the lower bound on the expected time to identify the odd arm as a function of the error probability for the setting of restless Markov arms. In the same section, we also show that by following the conventional approaches available in the prior works, we arrive at an infinite-dimensional linear programming problem (LPP) with countably infinitely many constraints. In Section \ref{sec:achievability}, we present a sequence of strategies whose expected times to identify the odd arm approach that of the lower bound in the limit of vanishing error probabilities, following which we state the main result of this paper in Section \ref{sec:main_result}. We discuss the no trembling hand case in Section \ref{sec:case_eta_=_0}. 
We conclude the paper in Section \ref{sec:conclusion}.
{\color{black} The proofs of all the results are contained in Appendices \ref{appndx:proof_of_Lemma_pi_delta^lambda_is_an_SSRS}-\ref{appndx:exp_upper_bound_for_a_certain_term}. Appendix \ref{appndx:infinite_dimensional_LP} contains the description of an infinite dimensional linear programming problem that may be arrived at following the approaches of the prior works for deriving the lower bound, and Appendix \ref{appndx:an_important_theorem} contains the statement of an important theorem that is used in several places in the main body of the paper.}


\begin{table*}[t]
  \centering
  \caption{Table of important notations.}
	\begin{tabular}{|c|l|}
  \hline
  $K$ & The number of arms; we consider $K\geq 3$ for the problem of ``odd'' arm identification to be well-defined.\\
  \hline
  $\mathcal{S}$ & The common, finite state space on which the Markov process of each arm evolves.\\
  \hline
  $\mathcal{A}$ & The set of arms.\\
  \hline
  $\underline{d}(t)$ & The vector $(d_1(t), \ldots, d_K(t))$ representing the delays of the arms at time $t$; defined only for $t\geq K$.\\
    \hline
    $\underline{i}(t)$ & The vector $(i_1(t), \ldots, i_K(t))$ representing the last observed states of the arms at time $t$; defined only for $t\geq K$.\\
    \hline
    $\underline{d}$, $\underline{d}'$ & A generic vector of arm delays.\\
    \hline
    $\underline{i}$, $\underline{i}'$ & A generic vector of last observed states.\\
    \hline 
    $\mathbb{S}$ & The countably infinite set of all possible values of the pair $(\underline{d}, \underline{i})$.\\
    \hline
    $B_t$ & Arm intended to be selected at time $t$; defined for all $t\geq 0$.\\
    \hline
    $A_t$ & Actual arm selected at time $t$ for all $t \geq 0$; $P(A_t=a \mid B_t=b)=\frac{\eta}{K}+(1-\eta)~\mathbb{I}_{\{a=b\}}$, \quad $a, b \in \mathcal{A}$.\\
    \hline
    $\lambda(\cdot \mid \cdot)$ & A conditional probability distribution on the arms, conditioned on the arm delays and the last observed states\\
    & $=\{\lambda(a\mid \underline{d}, \underline{i}):(\underline{d}, \underline{i})\in \mathbb{S}, ~ a\in \mathcal{A}\}$.\\
    \hline
    $\pi^\lambda$ & An SRS policy which selects arm $B_t$ at time $t$ according to $\lambda(\cdot \mid \underline{d}(t), \underline{i}(t))$.  \\
    \hline
    $\mu^\lambda$ & The stationary distribution of the process $\{(\underline{d}(t), \underline{i}(t)): t\geq K\}$ under the SRS policy $\pi^\lambda$.\\
    \hline
    $\nu^\lambda$ & The ergodic state-action occupancy measure under the SRS policy $\pi^\lambda$.\\
    & $\nu^\lambda(\underline{d}, \underline{i}, a)=\mu^\lambda(\underline{d},\underline{i})\cdot \left(\frac{\eta}{K}+(1-\eta)~\lambda(a\mid \underline{d}, \underline{i})\right)$,\quad $(\underline{d}, \underline{i})\in \mathbb{S}$, $a\in \mathcal{A}$.\\
    \hline
    $N(n, \underline{d}, \underline{i}, a)$ & The number of times up to time $n$ that the state $(\underline{d}, \underline{i})$ is observed and arm $a$ is selected subsequently.  \\
    \hline
    $N(n, \underline{d}, \underline{i}, a, j)$ & The number of times up to time $n$ when the state $(\underline{d}, \underline{i})$ is observed, arm $a$ is selected subsequently, and state $j$ is observed.\\
    \hline
    $Z_{hh'}(n)$ & The log-likelihood ratio of all the intended arm selections, actual arm selections and observations obtained up to time $n$ \\
    & under the hypothesis $\mathcal{H}_h$, with respect to that obtained under the hypothesis $\mathcal{H}_{h'}$.\\
    \hline
    $P_h^a$ & Transition probability matrix of arm $a$ when arm $h$ is the odd arm.\\
    \hline
    $(P_h^a)^d(j|i)$ & $(i, j)$th entry of the transition probability matrix $\underbrace{P_h^a\times \cdots \times P_h^a}_{d\text{-fold product}}$.\\
    \hline
  \end{tabular}	
\label{table:table_of_notations}
\end{table*}

\section{Notations and Problem Formulation}\label{sec:preliminaries}
We consider a multi-armed bandit with $K\geq 3$ arms, and define $\mathcal{A}\coloneqq \{1,\ldots,K\}$ to be the set of arms. We associate with each arm an ergodic and discrete-time Markov process on a finite state space $\mathcal{S}$. Further, we assume that the Markov process of any given arm is independent of those of the other arms. The Markovian evolution of states on one of the arms (known as the \emph{odd} arm) is governed by a transition probability matrix $P_1$, and the evolution of states on each of the non-odd arms is governed by $P_2$, where $P_2\neq P_1$. We denote by $\mu_i$ the unique stationary distribution of $P_i$, $i=1,2$.

For any integer $d\geq 1$ and a transition probability matrix $P$ on $\mathcal{S}$, let $P^{d}$ denote the transition probability matrix obtained by multiplying $P$ with itself $d$ times. For $i,j\in\mathcal{S}$ and $d\geq 1$, we write $P_1^{d}(j|i)$ and $P_{2}^{d}(j|i)$ to denote the $(i,j)$th element of the matrices $P_1^{d}$ and $P_{2}^{d}$ respectively (the case $d=1$ corresponds to $P_{1}$ and $P_{2}$ respectively). We assume that for all $i,j\in\mathcal{S}$, (a) $P_1(j|i)>0$ if and only if $P_2(j|i)>0$. This assumption ensures that the decision maker cannot infer whether or not a given arm is the odd arm merely by observing certain specific state(s) or state-transition(s) on the arm. For $h\in\mathcal{A}$, we denote by $\mathcal{H}_h$ the hypothesis that $h$ is the odd arm location.

We assume that $P_1$ and $P_2$ are known to a decision maker, whose goal it is to identify the index of the odd arm as quickly as possible, subject to an upper bound on the probability of error. In order to do so, the decision maker devises a sequential arm selection strategy in which, at each discrete-time instant $t\in\{0,1,\ldots\}$, the decision maker first identifies an arm to pull; call this $B_{t}$. The decision maker however has a trembling hand and, as a consequence, the intended arm $B_{t}$ gets pulled with probability $1-\eta$ and a uniformly random arm gets pulled with probability $\eta$. The parameter $\eta$, which is fixed and strictly positive, governs the error in translating the decision maker's intention into an action. Write $A_{t}$ for the arm that is actually pulled. The decision maker observes $A_{t}$, therefore knows whether or not his hand made an error in pulling the intended arm. Further, the decision maker observes the state of the arm $A_{t}$, denoted by $\bar{X}_{t}$. The unobserved arms continue to undergo state evolution, making the arms \emph{restless}. Thus, for each $t\geq 0$, $B_{t}, A_{t}$ and $\bar{X}_{t}$ denote respectively the intended arm, the selected arm, and the observed state of the selected arm at time $t$. We use the shorthand notation $(B^{t}, A^{t},\bar{X}^{t})$ to denote the collection $(B_0,A_0,\bar{X}_0,\ldots,B_{t},A_{t}\,\bar{X}_{t})$.

{\color{black} We note here that the observations $\{\bar{X}_t: t\geq 0\}$ are noiseless. The case of noisy observations, e.g., hidden Markov models, is important and is left for future work.}

\subsection{Policy}
A policy prescribes one of the following two actions at each time $t$: Based on the history $(B^{t-1},A^{t-1},\bar{X}^{t-1})$,
\begin{itemize}
	\item choose to pull arm $B_t$ according to a deterministic or a randomised rule, or
	\item stop and declare the index of the odd arm.
\end{itemize}
We use $\pi$ to denote a generic policy, and let $\tau(\pi)$ denote the stopping time of policy $\pi$. Throughout this paper, all stopping times are defined with respect to the filtration $\mathcal{F}_t\coloneqq\sigma(B^{t-1},A^{t-1},\bar{X}^{t-1})$, $t\geq 1$ and $\mathcal{F}_{0}\coloneqq \{\Omega,\emptyset\}$. Let $\theta(\tau(\pi))$ denote the index of the odd arm declared by the policy $\pi$ at its stopping time $\tau(\pi)$.

Let $P_h^\pi(\cdot)$ and $E_h^\pi[\cdot]$ denote probabilities and expectations computed under policy $\pi$. For ease of notation, we drop the superscript $\pi$, and {\color{black} request the  reader} to bear the dependence on $\pi$ in mind. Given a target probability of error $\epsilon>0$, we define $\Pi(\epsilon)$ as the set
\begin{equation}
	\Pi(\epsilon)\coloneqq \{\pi:P_h(\theta(\pi)\neq h)\leq \epsilon\text{ for all }h\in\mathcal{A}\}\label{eq:Pi(epsilon)}
\end{equation}
of all policies whose probability of error at stoppage is below $\epsilon$ {for all possible odd arm locations}. We emphasise that policies in $\Pi(\epsilon)$ work for all possible odd arm locations. We anticipate from similar results in the prior works that $$\inf\limits_{\pi\in\Pi(\epsilon)}E_h[\tau(\pi)] = \Theta(\log(1/\epsilon)).$$
Our interest is in characterising the constant factor multiplying $\log(1/\epsilon)$ {\color{black} in the limit as $\epsilon\downarrow 0$. For simplicity, we assume that every policy starts with the observation that arm $1$ is observed at time $t=0$, arm $2$ is observed at time $t=1$, etc., and arm $K$ is observed at time $t=K-1$. This can be effected by sampling the arms uniformly until this event occurs. Clearly, for $\eta > 0$, this requirement will result in a finite delay almost surely which does not affect the asymptotic analysis as $\epsilon \downarrow 0$.}

\subsection{Delays and Last Observed States}
Recall that at each time $t\in \{0,1,\ldots\}$, the decision maker observes only one of the arms, while the unobserved arms continue to undergo state evolution. Therefore, the probability of the observation $\bar{X}_t$ on the {selected} arm $A_t$ is a function of (a) the time elapsed since the previous time instant of selection of arm $A_t$ (called the \emph{delay} of arm $A_t$), and (b) the state of arm $A_t$ at its previous selection time instant (called the \emph{last observed state} of arm $A_t$). Notice that when the arms are \emph{rested}, the notion of arm delays is superfluous since each arm remains frozen at its previously observed state until its next selection time instant. Also, the notion of arm delays is redundant in the setting of iid observations since, in this special case, the current state of the arm selected is independent of the state at its previous selection. Thus, the notion of arm delays is a key distinguishing feature of the setting of restless arms.

We now define a new and more convenient notion of a state, based on the delays and the last observed states of the arms. As we demonstrate below, this new notion of state results in a  Markov decision problem that is amenable to analysis.

For $t\geq K$, we denote by $d_a(t)$ and $i_a(t)$ respectively the delay and the last observed state of arm $a$ at time $t$. Write $\underline{d}(t) \coloneqq (d_1(t),\ldots,d_K(t))$ and $\underline{i}(t) \coloneqq (i_1(t),\ldots,i_K(t))$ for the delays and the last observed states, respectively, of the arms at time $t$. Note that arm delays and last observed states are defined only for $t\geq K$ since these quantities are well-defined only when at least one observation is available from each arm. We set $\underline{d}(K)=(K,K-1,\ldots,1)$. {\color{black} Thus, we observe that $d_a(t)\geq 1$ for all $t\geq K$, and that $d_a(t)=1$ if and only if arm $a$ is selected at time $t-1$}.

We follow the rule below for updating the arm delays and last observed states: if $A_{t}=a'$, then
\begingroup \allowdisplaybreaks\begin{align}
	{d}_a(t+1)=\begin{cases}
		d_a(t)+1, &a\neq a',\\
		1,& a=a',
	\end{cases} \qquad \qquad 
	i_a(t+1)=\begin{cases}
		i_a(t),& a\neq a',\\
		\bar{X}_{t},& a=a',
	\end{cases}
	\label{eq:specific_transition_pattern}
\end{align}\endgroup
where $\bar{X}_{t}$ is the state of the arm $A_t=a'$ at time $t$.

One thus has the sequence of intended arm pulls, actual arm pulls, observations, and states as follows: at each $t \geq K$, based on $(\underline{d}(t), \underline{i}(t))$, choose to pull $B_{t}$; due to the trembling hand, observe that $A_{t}$ is pulled; see the state $\bar{X}_{t}$ of arm $A_t$; then form $(\underline{d}(t+1), \underline{i}(t+1)) $. This repeats until stoppage, at which time we have the declaration $\theta(\tau(\pi))$ (under policy $\pi$) as the candidate odd arm.

\subsection{Controlled Markov Process and the Resulting Markov Decision Problem}
From the update rule in \eqref{eq:specific_transition_pattern}, it is clear that the process $\{(\underline{d}(t),\underline{i}(t)):t\geq K\}$ takes values in a subset $\mathbb{S}$ of the {countable} set $\mathbb{N}^K\times\mathcal{S}^K$, where $\mathbb{N}=\{1,2,\ldots\}$ denotes the set of natural numbers. The subset $\mathbb{S}$ is formed based on the constraint that at any time $t\geq K$, exactly one of the components of $\underline{d}(t)$ is equal to $1$, and all the other components are strictly greater than $1$. Note that for all $(\underline{d},\underline{i}) \in \mathbb{S}$ and $t\geq K$,
\begin{align}
	P(\underline{d}(t+1)=\underline{d},\underline{i}(t+1)=\underline{i}\mid (\underline{d}(s),\underline{i}(s)),B_s,~K\leq s\leq t)=P(\underline{d}(t+1)=\underline{d},\underline{i}(t+1)=\underline{i}\mid (\underline{d}(t),\underline{i}(t)),B_t).\label{eq:controlled_markov_chain}
\end{align}
On account of \eqref{eq:controlled_markov_chain} being satisfied, we say that under any policy $\pi$, the evolution of the process $\{(\underline{d}(t),\underline{i}(t)):t\geq K\}$ is \emph{controlled} by the sequence $\{B_t\}_{t\geq 0}$ of intended arm selections under policy $\pi$. Alternatively, we say that $\{(\underline{d}(t),\underline{i}(t)):t\geq K\}$ is a controlled Markov process, with $\{B_t\}_{t\geq 0}$ as the sequence of controls; the terminology used here follows that of Borkar \cite{borkar1988control}. Thus, we are in a Markov decision problem (MDP) setting. We now make precise the state space, the action space, the transition probabilities and our objective.

The state space of the MDP is $\mathbb{S}$, with the state at time $t$ denoted $(\underline{d}(t), \underline{i}(t))$. The action space of the MDP is $\mathcal{A}$, with action $B_t$ at time $t$ possibly depending on the previous actions $B^{t-1}$ and the previous states $\{(\underline{d}(s), \underline{i}(s)), K \leq s \leq t\}$. (It is easy to see that this is equivalent to taking an action based on $(B^{t-1}, A^{t-1}, \bar{X}^{t-1})$.) The transition probabilities for the MDP are given by
\begin{enumerate}
	\item the trembling hand rule
\begin{equation}
	P(A_t = a | B_t) = \frac{\eta}{K}+ (1-\eta) \,\,\mathbb{I}_{\{B_t = a\}}, \quad \forall a \in \mathcal{A},\label{eq:trembling_hand_rule}
\end{equation}
	\item the law associated with arm $A_t$, and
	\item the update rule \eqref{eq:specific_transition_pattern}.
\end{enumerate}
In \eqref{eq:trembling_hand_rule}, $\mathbb{I}$ denotes the indicator function. In order to write the transition probabilities of the MDP precisely, let us introduce some notations. Given $h,a\in\mathcal{A}$, let $P_h^a$ denote the transition probability matrix of the Markov process of arm $a$ under the hypothesis $\mathcal{H}_h$. That is,
\begin{equation}
	P_h^a=\begin{cases}
		P_1,&a=h,\\
		P_2,&a\neq h.
	\end{cases}\label{eq:P_h^a}
\end{equation}
Furthermore, for any integer $d\geq 1$, let $(P_h^a)^d$ denote the transition probability matrix obtained by multiplying $P_h^a$ with itself $d$ times. Then, given any $(\underline{d},\underline{i}), (\underline{d}',\underline{i}')\in\mathbb{S}$ and $b\in\mathcal{A}$, the transition probabilities for the MDP are given by 
\begin{align}
	&P(\underline{d}(t+1)=\underline{d}',\underline{i}(t+1)=\underline{i}'\mid \underline{d}(t)=\underline{d},\underline{i}(t)=\underline{i}, B_t=b)\nonumber\\
	&\hspace{4cm}=\begin{cases}
		\left(\frac{\eta}{K}+(1-\eta)\,\mathbb{I}_{\{b\}}(a)\right)\,(P_h^a)^{d_a}(i_a'|i_a),&\text{if }d_a'=1\text{ and }d'_{\tilde{a}}=d_{\tilde{a}}+1\text{ for all }\tilde{a}\neq a,\\
		&i_{\tilde{a}}'=i_{\tilde{a}}\text{ for all }\tilde{a}\neq a,\\
		0,&\text{otherwise},
	\end{cases}\label{eq:MDP_transition_probabilities}
\end{align}
where $d_a'$ and $i_a'$ in \eqref{eq:MDP_transition_probabilities} denote the component corresponding to arm $a$ in $\underline{d}'$ and $\underline{i}'$ respectively. Note that the transition probabilities defined in \eqref{eq:MDP_transition_probabilities} are stationary and independent of time. Also, for $a\in \mathcal{A}$, we have
\begin{align}
	&P(\underline{d}(t+1)=\underline{d}',\underline{i}(t+1)=\underline{i}'\mid \underline{d}(t)=\underline{d},\underline{i}(t)=\underline{i}, A_t=a)\nonumber\\
	&\hspace{4cm}=\begin{cases}
		(P_h^a)^{d_a}(i_a'|i_a),&\text{if }d_a'=1\text{ and }d_{\tilde{a}}'=d_{\tilde{a}}+1\text{ for all }\tilde{a}\neq a,\\
		&i_{\tilde{a}}'=i_{\tilde{a}}\text{ for all }\tilde{a}\neq a,\\
		0,&\text{otherwise}.
	\end{cases}\label{eq:MDP_transition_probabilities_A_t}
\end{align}
The left-hand sides of \eqref{eq:MDP_transition_probabilities} and \eqref{eq:MDP_transition_probabilities_A_t} differ in that $B_t$ in \eqref{eq:MDP_transition_probabilities} is replaced by $A_t$ in \eqref{eq:MDP_transition_probabilities_A_t}. We shall write $Q(\underline{d}',\underline{i}'|\underline{d},\underline{i},a)$ to denote the quantity in \eqref{eq:MDP_transition_probabilities_A_t}.

Our objective, however, is nonstandard in the context of MDPs, and more in line with what information theorists study. We are interested in determining, for each hypothesis $\mathcal{H}_h$, the following:
\begin{equation}
  \label{eqn:objective}
  \lim_{\epsilon \downarrow 0} ~ \inf_{\pi \in \Pi(\epsilon)} ~ \frac{E_h [\tau(\pi)]}{\log (1/\epsilon)}.
\end{equation}

In the next section, we provide some preliminaries on MDPs. The terminologies used follow Borkar \cite{borkar1988control}.

{\color{black} Table \ref{table:table_of_notations} provides a quick summary of the important notations used throughout the paper.}

\section{Preliminaries on MDPs}\label{sec:MDP_preliminaries}
Let $\pi$ be an arbitrary policy. Consider the controlled Markov process $\{(\underline{d}(t),\underline{i}(t)):t\geq K\}$, with the corresponding sequence of controls $\{B_t\}$, under the policy $\pi$. Note that for all $t\geq K$,
\begin{align}
	&P(\underline{d}(t+1)=\underline{d},\underline{i}(t+1)=\underline{i}\mid B^{t-1}, \{(\underline{d}(s),\underline{i}(s)),~K\leq s\leq t\})\nonumber\\
	&=\sum\limits_{b=1}^{K}P(B_t=b\mid B^{t-1}, \{(\underline{d}(s),\underline{i}(s)),~K\leq s\leq t\})\, P(\underline{d}(t+1)=\underline{d},\underline{i}(t+1)=\underline{i}\mid B_t=b, ~B^{t-1}, \{(\underline{d}(s),\underline{i}(s)),~K\leq s\leq t\})\nonumber\\
	&=\sum\limits_{b=1}^{K}P(B_t=b\mid B^{t-1},\{(\underline{d}(s),\underline{i}(s)),~K\leq s\leq t\})\cdot P(\underline{d}(t+1)=\underline{d},\underline{i}(t+1)=\underline{i}\mid (\underline{d}(t),\underline{i}(t)), B_t=b),\label{eq:completely_specify_transition}
\end{align}
where the last line above follows from \eqref{eq:controlled_markov_chain}. From \eqref{eq:completely_specify_transition}, it is evident that the policy $\pi$ may be described completely by specifying  $P(B_{t} | B^{t-1}, \{ (\underline{d}(s),\underline{i}(s)),K\leq s\leq t \})$ for all $t\geq K$. 
We say that a policy $\pi$ is a \emph{stationary randomised strategy} (SRS) if there exists a Cartesian product $\lambda$ of the form
\begin{equation}
	\lambda=\bigotimes \limits_{(\underline{d},\underline{i})\in\mathbb{S}} \lambda_{(\underline{d},\underline{i})},\label{eq:SRS_Phi_defn}
\end{equation}
with the component $\lambda_{(\underline{d},\underline{i})}(\cdot)$ being a probability measure on $\mathcal{A}$, such that for all $t\geq K$ and $b\in\mathcal{A}$, under the policy $\pi$,
\begin{equation*}
	P(B_{t} = b\mid B^{t-1}, \{ (\underline{d}(s),\underline{i}(s)),K\leq s\leq t \})=\lambda_{(\underline{d}(t),\underline{i}(t))}(b).\label{eq:prob_of_arm_selection_under_SRS_Phi}
\end{equation*}
Such an SRS $\pi$ will be denoted $\pi^{\lambda}$. Note that $\{(\underline{d}(t),\underline{i}(t)):t\geq K\}$ is indeed a \emph{Markov process} under the SRS $\pi^{\lambda}$. This follows from the relation \eqref{eq:completely_specify_transition} where the first probability term inside the summation in \eqref{eq:completely_specify_transition} is now a function only of $(\underline{d}(t),\underline{i}(t))$. Let $\Pi_{\textsf{SRS}}$ denote the set of all SRS policies. 


For convenience, we write $\lambda_{(\underline{d},\underline{i})}(\cdot)$ as $\lambda(\cdot | \underline{d},\underline{i})$ so that we may write $\lambda$ itself in the more familiar form $\lambda(\cdot | \cdot)$.

An immediate and important property of any $\pi^\lambda\in\Pi_{\textsf{SRS}}$ is the following.

\begin{Lemma}\label{lem:pi_delta^lambda_is_an_SSRS}
Let $\eta\in(0,1]$. For every $\pi^{\lambda}\in\Pi_{\textsf{SRS}}$, the controlled Markov process $\{\underline{d}(t),\underline{i}(t):t\geq K\}$ under the policy $\pi^{\lambda}$ is irreducible, aperiodic, positive recurrent, and hence ergodic.
\end{Lemma}
\begin{IEEEproof}
See Appendix \ref{appndx:proof_of_Lemma_pi_delta^lambda_is_an_SSRS}.
\end{IEEEproof}
The proof of Lemma \ref{lem:pi_delta^lambda_is_an_SSRS}  relies on the hypothesis that the trembling hand parameter $\eta>0$. 

As a consequence of Lemma \ref{lem:pi_delta^lambda_is_an_SSRS}, it follows that under every SRS policy, a unique stationary distribution exists for the Markov process $\{(\underline{d}(t),\underline{i}(t)):t\geq K\}$. Let us call this stationary distribution $\mu^\lambda$ corresponding to the SRS policy $\pi^\lambda$.

With the above ingredients in place, we state in the next section the first main result of this paper -- an asymptotic lower bound on the expected time to identify the odd arm. 

\section{Lower Bound}\label{sec:lower_bound}
We now present a lower	 bound for \eqref{eqn:objective}. Given two probability distributions $\mu$ and $\nu$ on the finite state space $\mathcal{S}$, the Kullback-Leibler (KL) divergence (also called the relative entropy) between $\mu$ and $\nu$ is defined as
\begin{equation}
	D(\mu\|\nu)\coloneqq \sum\limits_{i\in\mathcal{S}}\mu(i)\log \frac{\mu(i)}{\nu(i)},\label{eq:D(mu||nu)}
\end{equation}
where, by convention, $0\log \frac{0}{0}=0$.

\begin{prop}
    \label{prop:lower_bound}
	Let $\eta\in(0,1]$ and $h\in\mathcal{A}$ be fixed. Assume that $\mathcal{H}_h$ is the true hypothesis. Let $P_1$ be the transition probability matrix of the Markov process of arm $h$, and for each $a\neq h$, let $P_2$ be the transition probability matrix of the Markov process arm $a$. Then,
\begin{equation}
	\liminf\limits_{\epsilon\downarrow 0}\inf\limits_{\pi\in\Pi(\epsilon)}\frac{E_h[\tau(\pi)]}{\log(1/\epsilon)}\geq \frac{1}{R^*(P_1,P_2)},\label{eq:lower_bound}
\end{equation}
where $R^*(P_1,P_2)$ is given by
\begingroup \allowdisplaybreaks\begin{align}
R^*(P_1,P_2)
\coloneqq \sup\limits_{\pi^\lambda\in\Pi_{\textsf{SRS}}}\, \min\limits_{h'\neq h} ~ \sum\limits_{(\underline{d},\underline{i})\in\mathbb{S}}~\sum\limits_{a=1}^{K}  \nu^\lambda(\underline{d},\underline{i},a) \,\textcolor{black}{k_{hh'}(\underline{d}, \underline{i}, a)},\label{eq:R_delta^*(h,P_1,P_2)}
\end{align}\endgroup
with
\begin{equation}
	\textcolor{black}{k_{hh'}(\underline{d}, \underline{i}, a)}\coloneqq
	\begin{cases}
		D(P_1^{d_a}(\cdot|i_a)\|P_2^{d_a}(\cdot|i_a)),&a=h,\\
		{D(P_2^{d_a}(\cdot|i_a)\|P_1^{d_a}(\cdot|i_a))},&a=h',\\
		0,& a\neq h,h',
	\end{cases}\label{eq:k(a,d_a,i_a)}
\end{equation}
and
\begin{eqnarray}
 \nu^\lambda(\underline{d},\underline{i},a)\coloneqq \mu^\lambda(\underline{d},\underline{i})\left(\frac{\eta}{K}+(1-\eta)\,\lambda(a|\underline{d},\underline{i})\right), \quad
\forall (\underline{d},\underline{i},a)\in\mathbb{S}\times\mathcal{A}.
\label{eq:ergodic_state_action_occupancy_measure}
\end{eqnarray}
\end{prop}
\begin{IEEEproof}
See Appendix \ref{appndx:proof_of_prop_lower_bound}.
\end{IEEEproof}
The proof of the lower bound follows the outline in \cite{pnkarthik2019learning}, with necessary modifications for the setting of restless arms. The key ingredients are the data processing inequality for relative entropies, a Wald-type Lemma for Markov processes, and a recognition that, for any $(\underline{d},\underline{i})$, the long-term fraction of exits from the state $(\underline{d},\underline{i})$ matches the long-term fraction of entries into the state $(\underline{d},\underline{i})$. This forces the long-term probability of seeing the controlled Markov process in the state $(\underline{d},\underline{i})$ to be that under its unique stationary distribution, by ergodicity (Lemma \ref{lem:pi_delta^lambda_is_an_SSRS}). These observations lead to  \eqref{eq:lower_bound}.

Observe that the left-hand side of \eqref{eq:lower_bound} is evaluated by taking into consideration \emph{all} policies, including those that are not necessarily SRS policies, whereas the supremum in \eqref{eq:R_delta^*(h,P_1,P_2)} is only over SRS policies. This is a consequence of \cite[Theorem 8.8.2]{puterman2014markov}, a formal statement of which appears in Theorem \ref{thrm:restriction_to_SRS_policies} of Appendix \ref{appndx:an_important_theorem} as applicable to the context of this paper. For details on how Theorem \ref{thrm:restriction_to_SRS_policies} is used in the proof, see Appendix \ref{appndx:proof_of_prop_lower_bound}. 

Finally, note that the constant $R^*(P_1,P_2)$ in \eqref{eq:R_delta^*(h,P_1,P_2)} does not depend on the odd arm location $h$. This is due to symmetry in the structure of the arms.

\subsection{Our `Lift' Approach}
It may be a little surprising to the reader as to why the summation on the right-hand side of $R^*(P_1,P_2)$ in \eqref{eq:R_delta^*(h,P_1,P_2)} is over the delays and the last observed states of \emph{all} the arms when the function $\textcolor{black}{k_{hh'}(\underline{d}, \underline{i}, a)}$, as given in \eqref{eq:k(a,d_a,i_a)}, is a function only of $d_a$ and $i_a$, the delay and the last observed state of arm $a$. In fact, the prior works  \cite{Vaidhiyan2017, prabhu2017optimal, vaidhiyan2017learning, pnkarthik2019learning} suggest that it suffices to use $(d_a, i_a)$ in place of $(\underline{d}, \underline{i})$ for deriving the lower bound. Relabelling $k_{hh'}(\underline{d}, \underline{i}, a)$ as $k_{hh'}(d_a, i_a, a)$ and proceeding to derive the lower bound as suggested by the prior works leads to a linear programming problem (LPP) with countably infinitely many linear constraints; see Appendix \ref{appndx:infinite_dimensional_LP} for the details. However, it is not clear if the constraints of the above LPP constitute the tightest set of constraints. This is important because the optimal value of the LPP, say $R_1^*(P_1, P_2)$, may not necessarily be the smallest (best) constant for the problem at hand if the constraints are not tight, in which case we can only assert that $R_1^*(P_1, P_2) \geq R^*(P_1, P_2)$. In this case, it is not clear if this inequality is indeed an equality. 

In contrast to the approach of using only $(d_a, i_a)$ as suggested by the prior works, our `lift' approach of using $(\underline{d}, \underline{i})$ automatically captures all the constraints of the LPP and makes the problem amenable to analysis, thereby enabling us to assert that $R^*(P_1, P_2)$ is the best (smallest) constant for the problem at hand. For more details on the LPP, see Appendix \ref{appndx:infinite_dimensional_LP}.

\section{Achievability}\label{sec:achievability}

{\color{black} The question of whether the supremum in (13) is a maximum, i.e., whether there exists an SRS policy that obtains the supremum value, is under study.}
Recall that this supremum is over all $\pi^\lambda\in \Pi_{\textsf{SRS}}$ for $\lambda(\cdot|\cdot)$ which are conditional probability distributions on the arms, conditioned on the arm delays and the last observed states. This is in contrast to the works \cite{Vaidhiyan2017,prabhu2017optimal, vaidhiyan2017learning, pnkarthik2019learning}, where the corresponding supremum is over all \emph{unconditional} probability distributions on the arms. This is because, in those works, the arm delays are superfluous. The unconditional probability measures are elements of the probability simplex on $\mathcal{A}$, whereas the conditional probability measures are more complex due to the countably many possible values for the arm delays. In spite of this added complexity, we can come arbitrarily close to the supremum in \eqref{eq:R_delta^*(h,P_1,P_2)}. We shall use this fact in our achievability result, which is the topic of this section.

We begin with some notations. Given $h,h'\in\mathcal{A}$, with $h\neq h'$, and a policy $\pi$, let $Z_{hh'}(n)$ denote the log-likelihood ratio (LLR), under the policy $\pi$, of all intended arm pulls, actual arm pulls, and observations up to time $n$ under the hypothesis $\mathcal{H}_h$ with respect to that under the hypothesis $\mathcal{H}_{h'}$. Then, $Z_{hh'}(n)$ may be expressed as
\begin{align}
	Z_{hh'}(n)&=\log\frac{P_h(B^n,A^n,\bar{X}^n)}{P_{h'}(B^n,A^n,\bar{X}^n)}\nonumber\\
    &=\log\frac{P_h(B_0)}{P_{h'}(B_0)}+\log\frac{P_h(A_0|B_0)}{P_{h'}(A_0|B_0)}+\log\frac{P_h(\bar{X}_0|B_0,A_0)}{P_{h'}(\bar{X}_0|B_0,A_0)}\label{eq:Z_{hh'}(n)_0}\\
    &\quad \quad+\sum\limits_{t=1}^{n}\log\left(\frac{P_h(B_t|B^{t-1},A^{t-1},\bar{X}^{t-1})}{P_{h'}(B_t|B^{t-1},A^{t-1},\bar{X}^{t-1})}\right)\label{eq:Z_{hh'}(n)_1}\\
    &\quad \quad+\sum\limits_{t=1}^{n}\log\left(\frac{P_h(A_t|B^{t},A^{t-1},\bar{X}^{t-1})}{P_{h'}(A_t|B^{t},A^{t-1},\bar{X}^{t-1})}\right)\label{eq:Z_{hh'}(n)_2}\\ &\quad \quad+\sum\limits_{t=1}^{n}\log\left(\frac{P_h(\bar{X}_t|A_t,B^{t},A^{t-1},\bar{X}^{t-1})}{P_{h'}(\bar{X}_t|A_t,B^{t},A^{t-1},\bar{X}^{t-1})}\right)\label{eq:Z_{hh'}(n)_3}.
\end{align}
We now note that under the policy $\pi$, the probability of choosing arm $B_t$ at time $t$, based on the history up to time $t$, cannot be a function of the underlying odd arm location (which is unknown to $\pi$), and must therefore be the same under hypotheses $\mathcal{H}_h$ and $\mathcal{H}_{h'}$. Thus, the first term in \eqref{eq:Z_{hh'}(n)_0} and the expression in \eqref{eq:Z_{hh'}(n)_1} are $0$. Also, we note that $P_h(A_0|B_0)=P_{h'}(A_0|B_0)$, and for each $t$, $$P_h(A_t|B_t,A^{t-1},\bar{X}^{t-1})=P_{h'}(A_t|B_t,A^{t-1},\bar{X}^{t-1})$$ since $A_t$, the arm that is actually pulled at time $t$, is a function only of $B_t$ and is related to $B_t$ through \eqref{eq:trembling_hand_rule}. Therefore, given the history, the choice of $A_t$ is not a function of the odd arm location, and is the same under hypotheses $\mathcal{H}_h$ and $\mathcal{H}_{h'}$, implying that the second term in \eqref{eq:Z_{hh'}(n)_0} and the expression in \eqref{eq:Z_{hh'}(n)_2} are $0$. Finally, the probabilities in \eqref{eq:Z_{hh'}(n)_3} do not depend on the intended arm pulls $\{B_t\}$ since the state $\bar{X}_t$ observed on arm $A_t$ is a function only of the delay and the last observed state of arm $A_t$.
Letting $X_t^a$ denote the state of arm $A_t=a$, and defining
\begin{align}
	N(n,\underline{d},\underline{i},a)&\coloneqq\sum\limits_{t=K}^{n}\mathbb{I}_{\{\underline{d}(t)=\underline{d},\underline{i}(t)=\underline{i},A_t=a\}},\label{eq:N(n,d,i,a)}\\
	N(n,\underline{d},\underline{i},a,j)&\coloneqq \sum\limits_{t=K}^{n}\mathbb{I}_{\{\underline{d}(t)=\underline{d},\underline{i}(t)=\underline{i},A_t=a,X_t^a=j\}},\label{eq:N(n,d,i,a,j)}
\end{align}
for all $(\underline{d},\underline{i},a)\in\mathbb{S}\times\mathcal{A}$, and using the assumption that arm $1$ is selected at time $t=0$, arm $2$ at time $t=1$ and so on until arm $K$ at time $t=K-1$, we have
\begin{align}
	Z_{hh'}(n) &=\sum\limits_{a=1}^{K}\log \frac{P_h(X_{a-1}^a)}{P_{h'}(X_{a-1}^a)} + \sum\limits_{t=K}^{n}\log\frac{P_h(\bar{X}_t|A_t,B^{t},A^{t-1},\bar{X}^{t-1})}{P_{h'}(\bar{X}_t|A_t,B^{t},A^{t-1},\bar{X}^{t-1})}\nonumber\\
	&=\sum\limits_{a=1}^{K}\log \frac{P_h(X_{a-1}^a)}{P_{h'}(X_{a-1}^a)} + \sum\limits_{(\underline{d},\underline{i})\in\mathbb{S}}~
\sum\limits_{j\in\mathcal{S}}~\sum\limits_{a=1}^{K}~\sum\limits_{t=K}^{n}\mathbb{I}_{\{\underline{d}(t)=\underline{d},\underline{i}(t)=\underline{i},A_t=a,X_t^a=j\}}\,\log \frac{P_h(\bar{X}_t=j|A_t=a,B^{t},A^{t-1},\bar{X}^{t-1})}{P_{h'}(\bar{X}_t=j|A_t=a,B^{t},A^{t-1},\bar{X}^{t-1})} \nonumber\\
    &= \sum\limits_{a=1}^{K}\log \frac{P_h(X_{a-1}^a)}{P_{h'}(X_{a-1}^a)} + \sum\limits_{(\underline{d},\underline{i})\in\mathbb{S}}~
    \sum\limits_{j\in\mathcal{S}}~\sum\limits_{a=1}^{K}~\sum\limits_{t=K}^{n}\mathbb{I}_{\{\underline{d}(t)=\underline{d},\underline{i}(t)=\underline{i},A_t=a,X_t^a=j\}}\,\log \frac{P_h(X_{t}^a=j|A_t=a,X_{t-d_a}^a=i_a)}{P_{h'}(X_{t}^a=j|A_t=a,X_{t-d_a}^a=i_a)} \nonumber\\
    &=\sum\limits_{a=1}^{K}\log \frac{P_h(X_{a-1}^a)}{P_{h'}(X_{a-1}^a)} + \sum\limits_{(\underline{d},\underline{i})\in\mathbb{S}}~
    \sum\limits_{j\in\mathcal{S}}~\sum\limits_{a=1}^{K}~\sum\limits_{t=K}^{n}\mathbb{I}_{\{\underline{d}(t)=\underline{d},\underline{i}(t)=\underline{i},A_t=a,X_t^a=j\}}\,\log \frac{(P_h^a)^{d_a}(j|i_a)}{(P_{h'}^a)^{d_a}(j|i_a)} \nonumber\\
   	&=\sum\limits_{a=1}^{K}\log \frac{P_h(X_{a-1}^a)}{P_{h'}(X_{a-1}^a)}+\sum\limits_{(\underline{d},\underline{i})\in\mathbb{S}}~
     \sum\limits_{j\in\mathcal{S}}~\sum\limits_{a=1}^{K} N(n,\underline{d},\underline{i},a,j)\log\frac{(P_h^a)^{d_a}(j|i_a)}{(P_{h'}^a)^{d_a}(j|i_a)}\label{eq:Z_{hh'}(n)_temp_1}\\
     &=\sum\limits_{a=1}^{K}\log \frac{P_h(X_{a-1}^a)}{P_{h'}(X_{a-1}^a)}+\sum\limits_{(\underline{d},\underline{i})\in\mathbb{S}}~
     \sum\limits_{j\in\mathcal{S}}\left[N(n,\underline{d},\underline{i},h,j)\log\frac{P_1^{d_h}(j|i_h)}{P_2^{d_h}(j|i_h)}+N(n,\underline{d},\underline{i},h',j)\log\frac{P_2^{d_{h'}}(j|i_{h'})}{P_1^{d_{h'}}(j|i_{h'})}\right].\label{eq:Z_{hh'}(n)}
\end{align}
{\color{black} In the above set of equations, $P_h(X_{a-1}^a)$ denotes the law of the observation $X_{a-1}^a$ obtained from arm $a$ at time $a-1$ when the true hypothesis is $\mathcal{H}_h$; $P_{h'}(X_{a-1}^a)$ is defined similarly. Also, \eqref{eq:Z_{hh'}(n)} follows by noting that 
\begin{equation}
P_h^a = \begin{cases}
P_1, &a=h,\\
P_2, &a\neq h,
\end{cases} \qquad \qquad \qquad  P_{h'}^a = \begin{cases}
P_1, &a=h',\\
P_2, &a\neq h', \label{eq:P_h^a_and_P_{h'}^a}
\end{cases}
\end{equation}
and thus the only nonzero terms in the summation over the arms in \eqref{eq:Z_{hh'}(n)_temp_1} are those corresponding to $a=h$ and $a=h'$.}

To describe our policy, we first fix constants $\delta>0$ and $L>1$. These will be the parameters of our policy. Recall that the supremum in \eqref{eq:R_delta^*(h,P_1,P_2)} is over all SRS policies. By the definition of this supremum, we know that for any fixed hypothesis $\mathcal{H}_h$ and given $\delta>0$, there exists $\lambda(\cdot\mid\cdot)=\lambda_{h, \delta}(\cdot\mid\cdot)$ such that under the corresponding SRS policy $\pi^{\lambda_{h, \delta}}$, we have\begin{equation}
	\min\limits_{h'\neq h}\,\sum\limits_{(\underline{d},\underline{i})\in\mathbb{S}}~\sum\limits_{a=1}^{K} \nu^{\lambda_{h,\delta}}(\underline{d},\underline{i},a)\, \,\textcolor{black}{k_{hh'}(\underline{d}, \underline{i}, a)} \geq \frac{R^*(P_1,P_2)}{1+\delta}.\label{eq:policy_exists}
\end{equation}
Notice that $\lambda_{h, \delta}$ is, in general, a function of $\delta$ and the hypothesis $\mathcal{H}_h$ (the hypothesis that arm $h$ is the odd arm), although $R^*(P_1,P_2)$ itself is not a function $h$. 

Our policy, which we call $\pi^\star(L,\delta)$, is then as below.
\vspace{.1in}
\hrule

\vspace{.1in}

\noindent \textbf{\underline{\emph{Policy }$\pi^{\star}(L,\delta)$}}:\\
\noindent Fix $L>1$ and $\delta>0$. Let the parameter of the trembling hand be $\eta\in(0,1]$. Assume\footnote{If this is not the case, exercise arm pulls uniformly at random until each arm is selected at least once. It can be shown that this will only take finite time almost surely, and does not affect the asymptotic analysis of our policy.} that $A_0=1$, $A_1=2$, and so on until $A_{K-1}=K$. Let $M_h(n)=\min\limits_{h'\neq h}Z_{hh'}(n)$. Follow the below mentioned steps for each $n\geq K$.\\
\noindent (1) Let $\theta(n)=\arg\max\limits_{h\in\mathcal{A}}M_h(n)$; resolve ties at random.\\
\noindent (2) If $M_{\theta(n)}(n)\geq\log((K-1)L)$, stop further arm selections and declare $\theta(n)$ as the true index of the odd arm.\\
\noindent (3) If $M_{\theta(n)}(n)<\log((K-1)L)$, decide to pull arm $B_{n}$ according to the distribution $\lambda_{\theta(n),\delta}( \cdot \mid \underline{d}(n),\underline{i}(n))$.

\vspace{.1in}
\hrule
\vspace{0.1in}

In item (1) above, $\theta(n)$ denotes the guess of the odd arm at time $n$. In item (2), we check if the LLR of hypothesis $\mathcal{H}_{\theta(n)}$ with respect to each of its alternative hypotheses is separated sufficiently $(\geq \log(K-1)L)$. If this is the case, then the policy is confident that the true odd arm location is $\theta(n)$. The policy then terminates and outputs the index $\theta(n)$. If the condition in item (2) fails, then the policy picks the next arm to pull.

Recall that the supremum in \eqref{eq:R_delta^*(h,P_1,P_2)} is only over SRS policies. However, the policy $\pi^\star(L,\delta)$ described above is \emph{not} an SRS policy since the distribution in item (3) is a function of $\theta(n)$ that could potentially depend on the entire history of arm selections and observations up to time $n$. Yet, as we show below, its performance comes arbitrarily close to that of the lower bound.

\subsection{Performance of Policy $\pi^\star(L,\delta)$}
We now present results on the performance of our policy.

\begin{Lemma}\label{lem:liminf_Z_{hh'}(n)_strictly_positive}
	Fix $L>1$, $\delta>0$ and $h\in\mathcal{A}$, and suppose that $\mathcal{H}_h$ is the true hypothesis. Consider the non-stopping version of the policy $\pi^\star(L,\delta)$ which runs indefinitely (i.e., even if item (2) is true, it moves to item (3)). Under this policy, for every $h'\neq h$,
	\begin{equation}
		\liminf\limits_{n\to\infty}\frac{Z_{hh'}(n)}{n}>0\quad \text{almost surely}.\label{eq:liminf_Z_{hh'}(n)_strictly_positive}
	\end{equation}
\end{Lemma}
\begin{IEEEproof}
See Appendix \ref{appndx:proof_of_lem_liminf_Z_{hh'}(n)_strictly_positive}.	
\end{IEEEproof}
Thanks to Lemma \ref{lem:liminf_Z_{hh'}(n)_strictly_positive}, we have $\liminf\limits_{n\to\infty}M_h(n)/n >0$ almost surely under the true hypothesis $\mathcal{H}_h$. This implies that, almost surely,  $M_h(n)\geq \log((K-1)L)$ for all sufficiently large values of $n$, thus proving that the policy $\pi^\star(L,\delta)$ stops in finite time with probability $1$.

Next, we show that the probability of error of our policy may be controlled by setting the parameter $L$ suitably.
\begin{Lemma}\label{lem:pi_star(L)_in_Pi(epsilon)}
	Fix error probability $\epsilon>0$. If $L=1/\epsilon$, then for every $\delta>0$, $\pi^\star(L,\delta)\in\Pi(\epsilon)$. Here, $\Pi(\epsilon)$ is as defined in \eqref{eq:Pi(epsilon)}.
\end{Lemma}
\begin{IEEEproof}
The proof uses the fact that the policy stops in finite time with probability $1$. See Appendix \ref{appndx:proof_of_lem_pi_star(L)_in_Pi(epsilon)} for the details.	
\end{IEEEproof}

With the above ingredients in place, we state the main result of this section, which is that the expected stopping time of our policy satisfies an asymptotic upper bound that comes arbitrarily close to the lower bound in \eqref{eq:lower_bound}.
\begin{prop}\label{prop:upper_bound}
	Fix $h\in\mathcal{A}$ and $\delta>0$, and let $\mathcal{H}_h$ be the true hypothesis. The policy $\pi^\star(L, \delta)$ satisfies
	\begin{equation}
		\limsup\limits_{L\to\infty}\frac{E_h[\tau(\pi^\star(L, \delta))]}{\log L}\leq \frac{1+\delta}{R^*(P_1,P_2)}.\label{eq:upper_bound}
	\end{equation}
\end{prop}
\begin{IEEEproof}
In the proof, which we provide in Appendix \ref{appndx:proof_of_upper_bound}, we first show that as $L\to\infty$ (equivalently $\epsilon\downarrow 0$), the ratio $\tau(\pi^\star(L, \delta))/\log L$ satisfies an almost sure upper bound that matches with the right-hand side of \eqref{eq:upper_bound}. We then show that the family $\{\tau(\pi^\star(L, \delta))/\log L:L>1\}$ is uniformly integrable. Combining the almost sure upper bound with the uniform integrability result yields \eqref{eq:upper_bound}.
\end{IEEEproof}

\section{Main Result}\label{sec:main_result}
We are now ready to state the main result of this paper.

\begin{thrm}\label{prop:main_result}
	Consider a multi-armed bandit with $K\geq 3$ arms in which each arm is a time homogeneous and ergodic Markov process on the finite state space $\mathcal{S}$. Fix $h\in\mathcal{A}$, and suppose that $h$ is the odd arm. Let $P_1$ be the transition probability matrix of the Markov process of arm $h$. Further, for all $a\neq h$, let the transition probability matrix of arm $a$ be $P_2$, where $P_2\neq P_1$. Fix $\eta\in (0,1]$, and suppose that a decision maker who wishes to identify the odd arm has a trembling hand with parameter $\eta$. Assuming that $P_1$ and $P_2$ are known to the decision maker, the expected time required by the decision maker to identify the odd arm satisfies the asymptotic relation
	\begin{equation}
		\lim\limits_{\epsilon\downarrow 0}\inf\limits_{\pi\in\Pi(\epsilon)}\frac{E_h[\tau(\pi)]}{\log(1/\epsilon)}=\lim\limits_{\delta \downarrow 0}\lim\limits_{L\to\infty}\frac{E_h[\tau(\pi^\star(L,\delta))]}{\log L}= \frac{1}{R^*(P_1,P_2)}.\label{eq:main_result}
	\end{equation}
\end{thrm}

\begin{IEEEproof}
From Lemma \ref{lem:pi_star(L)_in_Pi(epsilon)}, we see that given any error tolerance parameter $\epsilon>0$, by setting $L=1/\epsilon$, we have $\pi^\star(L, \delta)\in \Pi(\epsilon)$ for all $\delta>0$. Therefore, it follows that for all $\epsilon, \delta>0$,  
\begin{equation}
\inf\limits_{\pi \in \Pi(\epsilon)} \frac{E_h[\tau(\pi)]}{\log \left(\frac{1}{\epsilon}\right)} \leq \frac{E_h[\tau(\pi^\star(L, \delta))]}{\log L}. 
\label{eq:proof_of_main_result_temp_1}
\end{equation}
Fixing $\delta>0$ and letting $\epsilon\downarrow 0$ (which is identical to letting $L\to \infty$) in \eqref{eq:proof_of_main_result_temp_1}, and using the upper bound in \eqref{eq:upper_bound}, we get
\begin{equation}
\limsup\limits_{\epsilon\downarrow 0}\inf\limits_{\pi\in\Pi(\epsilon)}\frac{E_h[\tau(\pi)]}{\log(1/\epsilon)} \leq \limsup\limits_{L\to\infty}\frac{E_h[\tau(\pi^\star(L, \delta))]}{\log L}\leq \frac{1+\delta}{R^*(P_1,P_2)}.
\label{eq:proof_of_main_result_temp_2}
\end{equation}
Letting $\delta\downarrow 0$ in \eqref{eq:proof_of_main_result_temp_2} and noting that the leftmost term in \eqref{eq:proof_of_main_result_temp_2} does not depend on $\delta$, we get 
\begin{equation}
\limsup\limits_{\epsilon\downarrow 0}\inf\limits_{\pi\in\Pi(\epsilon)}\frac{E_h[\tau(\pi)]}{\log(1/\epsilon)} \leq \lim\limits_{\delta \downarrow 0}\limsup\limits_{L\to\infty}\frac{E_h[\tau(\pi^\star(L, \delta))]}{\log L}\leq \frac{1}{R^*(P_1,P_2)}.
\label{eq:proof_of_main_result_temp_3}
\end{equation}
Combining the result in \eqref{eq:proof_of_main_result_temp_3} with the lower bound in \eqref{eq:lower_bound}, we get
\begin{equation}
\frac{1}{R^*(P_1,P_2)} \leq \liminf\limits_{\epsilon\downarrow 0}\inf\limits_{\pi\in\Pi(\epsilon)}\frac{E_h[\tau(\pi)]}{\log(1/\epsilon)} \leq \limsup\limits_{\epsilon\downarrow 0}\inf\limits_{\pi\in\Pi(\epsilon)}\frac{E_h[\tau(\pi)]}{\log(1/\epsilon)} \leq \lim\limits_{\delta \downarrow 0}\limsup\limits_{L\to\infty}\frac{E_h[\tau(\pi^\star(L, \delta))]}{\log L}\leq \frac{1}{R^*(P_1,P_2)}.
\label{eq:proof_of_main_result_temp_4}
\end{equation}
Thus, it follows that the limit infimum and the limit suprema in \eqref{eq:proof_of_main_result_temp_4} are indeed limits, thereby yielding \eqref{eq:main_result}. This completes the proof of the theorem.
\end{IEEEproof}

We thus see that the policy $\pi^\star(L,\delta)$ is asymptotically optimal. As noted in Lemma \ref{lem:pi_star(L)_in_Pi(epsilon)}, the parameter $L$ may be set appropriately so as to ensure that the policy meets the desired error probability at stoppage. Furthermore, the parameter $\delta$ may be set so as to ensure that the upper bound in \eqref{eq:upper_bound} is within a desired accuracy from the lower bound in \eqref{eq:lower_bound}. Finally, we emphasise here that our analysis of the lower and upper bounds crucially relies on the trembling hand parameter $\eta$ being strictly positive.

\section{The Case $\eta=0$}\label{sec:case_eta_=_0}
{\color{black} We now investigate the case $\eta=0$. Let us first recall that the key result of Lemma \ref{lem:pi_delta^lambda_is_an_SSRS}, which states that under every SRS policy the controlled Markov process $\{(\underline{d}(t), \underline{i}(t)):t\geq K\}$ is an ergodic Markov process, crucially relies on the trembling hand parameter $\eta$ being strictly positive. Such an ergodicity property may not be available when $\eta=0$. While, in principle, we may consider plugging $\eta=0$ in \eqref{eq:lower_bound} and treating the resulting expression as the lower bound for the case when $\eta=0$, it is not clear if this new lower bound can be approached asymptotically through a sequence of strategies (policies) in the sense of \eqref{eq:upper_bound}. 
Therefore, it is a priori not clear if the results of this paper extend directly to the case $\eta=0$.}


In what follows, we bring to light the following observations.
\begin{enumerate}
	\item Writing $R^*(P_1,P_2)$ of \eqref{eq:R_delta^*(h,P_1,P_2)} more explicitly as $R_\eta^*(P_1,P_2)$ for $\eta\in (0,1]$, we show that $\lim\limits_{\eta\downarrow 0}R_\eta^*(P_1,P_2)$ exists. This is based on a key monotonicity property which we elaborate upon in  Section \ref{subsec:subsec_key_monotonicity_property}.
	\item Writing $R_0^*(P_1,P_2)$ to denote the constant obtained by plugging $\eta=0$ in \eqref{eq:R_delta^*(h,P_1,P_2)}, we demonstrate that
		\begin{equation}
			\lim\limits_{\eta\downarrow 0}R_\eta^*(P_1,P_2)\leq R_0^*(P_1,P_2).\label{eq:R_eta^*_leq_R_0^*}
		\end{equation}
	 It is not clear if, in general, the inequality in \eqref{eq:R_eta^*_leq_R_0^*} is an equality.
	 \item We show in Section \ref{subsec:iid_observations_from_arms} and Section \ref{subsec:rested_markov_arms} that the lower bounds for the settings when either (a) each arm yields iid observations from a common finite alphabet, or (b) each arm yields Markov observations from a common finite state space and the arms are rested, may be recovered from \eqref{eq:R_delta^*(h,P_1,P_2)} by plugging $\eta=0$ in \eqref{eq:R_delta^*(h,P_1,P_2)}. Our proof of this is based on verifying that the hypotheses of the envelope theorem \cite[Theorem 2]{milgrom2002envelope} are satisfied for these settings. Thus, we show that the inequality in \eqref{eq:R_eta^*_leq_R_0^*} is an equality for each of the above settings, thereby implying that the lower bounds for these settings may be approached asymptotically through a sequence of ``trembling-hand'' based policies similar to that presented in this paper; the policies of \cite[Section II.B]{Vaidhiyan2017} and \cite[Section IV]{pnkarthik2019learning} are example cases in point. {\color{black} This demonstrates that our analysis of the setting of restless Markov arms carries over to the settings of the prior works with minor modifications.}      
\end{enumerate} 

\subsection{A Key Monotonicity Property}\label{subsec:subsec_key_monotonicity_property}
Fix $\eta\in (0,1]$, and assume that the decision maker possesses a trembling hand with parameter $\eta$. 
Let $\lambda=\lambda(\cdot\mid \cdot)$ be any conditional probability distribution on the arms, conditioned on the arm delays and the last observed states, as described in Section \ref{sec:MDP_preliminaries}, and let $\Lambda$ denote the set of all such conditional distributions. Define 
\begin{equation}
	\Lambda^\eta\coloneqq \left\lbrace \frac{\eta}{K}+(1-\eta)\,\lambda(\cdot\mid \cdot):\lambda(\cdot\mid \cdot)\in \Lambda \right\rbrace\label{eq:Lambda^eta}.
\end{equation}
Note that for any $\lambda(\cdot\mid \cdot)\in \Lambda$, the corresponding element of $\Lambda^\eta$ is the probability distribution according to which arms are \emph{actually} selected, when the decision maker \emph{intends} to pull the arms according to $\lambda(\cdot\mid \cdot)$. Notice that $\Lambda^\eta\subset \Lambda$ for all $\eta\in(0,1]$. 

The following Lemma shows that $\Lambda^\eta$ is non-decreasing as $\eta$ decreases.
\begin{Lemma}\label{lem:Lambda^eta_monotone}
	$\Lambda^\eta\subset  \Lambda^{\eta'}$ for all $0<\eta'< \eta\leq 1$.
\end{Lemma}
\begin{IEEEproof}
Fix $0<\eta'<\eta\leq 1$, and consider $\frac{\eta}{K}+(1-\eta)\,\lambda(\cdot\mid \cdot)\in \Lambda^\eta$ for some $\lambda(\cdot\mid \cdot)\in \Lambda$. Then, for all $(\underline{d},\underline{i},a)\in\mathbb{S}\times\mathcal{A}$,
\begingroup \allowdisplaybreaks\begin{align}
	\frac{\eta}{K}+(1-\eta)~\lambda(a|\underline{d},\underline{i})
	&=\frac{\eta'}{K}+\frac{\eta-\eta'}{K}+(1-\eta)~\lambda(a|\underline{d},\underline{i})\nonumber\\
	&=\frac{\eta'}{K}+(1-\eta')\left[\frac{\eta-\eta'}{1-\eta'}\cdot \frac{1}{K}+\frac{1-\eta}{1-\eta'}~\lambda(a|\underline{d},\underline{i})\right]\nonumber\\
	&=\frac{\eta'}{K}+(1-\eta')\left[\frac{\eta''}{K}+(1-\eta'')~\lambda(a|\underline{d},\underline{i})\right]\label{eq:lower_bound_8}\\
	&\in \Lambda^{\eta'},\label{eq:lower_bound_9}
\end{align}\endgroup
where in \eqref{eq:lower_bound_8}, $\eta''=\frac{\eta-\eta'}{1-\eta'}\in (0,1]$, and \eqref{eq:lower_bound_9} follows by noting that the term inside the square brackets in \eqref{eq:lower_bound_8} is a valid element of $\Lambda$. The relation in \eqref{eq:lower_bound_9} implies that every element of $\Lambda^\eta$ is also an element of $\Lambda^{\eta'}$ whenever $\eta'<\eta$. This completes the proof.
\end{IEEEproof}
Plugging $\eta=0$ in \eqref{eq:Lambda^eta}, and denoting the resulting set as $\Lambda^0$, we see that $\Lambda^{0}=\Lambda$. Thus, it follows from Lemma \ref{lem:Lambda^eta_monotone} that
\begin{equation}
	\bigcup\limits_{\eta\downarrow 0}\Lambda^\eta \subset \Lambda.\label{eq:lim_Lambda^eta_subset_of_Lambda^0}
\end{equation}
Let us now turn our attention to \eqref{eq:ergodic_state_action_occupancy_measure}, and note that the right-hand side of \eqref{eq:ergodic_state_action_occupancy_measure} represents the long-term probability of seeing the state $(\underline{d},\underline{i})$ and selecting arm $a$ subsequently with probability $\frac{\eta}{K}+(1-\eta)\,\lambda(a|\underline{d},\underline{i})$. Defining $\lambda^\eta(\cdot\mid\cdot) \coloneqq \frac{\eta}{K}+(1-\eta)\,\lambda(\cdot\mid\cdot)$, and writing $\nu^\lambda$ in \eqref{eq:ergodic_state_action_occupancy_measure} as $\nu^{\lambda^\eta}$, we may express the right-hand side of \eqref{eq:R_delta^*(h,P_1,P_2)} equivalently as 
\begin{equation}
	R_\eta^*(P_1,P_2)\coloneqq \sup\limits_{\lambda^\eta(\cdot\mid\cdot)\in \Lambda^\eta}~ \min\limits_{h'\neq h} ~\sum\limits_{(\underline{d},\underline{i})\in\mathbb{S}}~\sum\limits_{a=1}^{K}  \nu^{\lambda^\eta}(\underline{d},\underline{i},a) \,\textcolor{black}{k_{hh'}(\underline{d}, \underline{i}, a)}.\label{eq:alternative_expression_for_R^*(P_1,P_2)}
\end{equation}
It follows from Lemma \ref{lem:Lambda^eta_monotone} that $R_\eta^*(P_1,P_2)$ is non-decreasing in $\eta$; thus, $\lim\limits_{\eta\downarrow 0}R_\eta^*(P_1,P_2)$ exists.

Finally, denoting by $R_0^*(P_1,P_2)$ the quantity obtained by plugging $\eta=0$ in \eqref{eq:alternative_expression_for_R^*(P_1,P_2)}, it follows from \eqref{eq:lim_Lambda^eta_subset_of_Lambda^0} that \eqref{eq:R_eta^*_leq_R_0^*} holds. 

%

\subsection{IID Observations From The Arms}\label{subsec:iid_observations_from_arms}
We now show that when each arm yields iid observations coming from a finite alphabet common across the arms, the inequality in \eqref{eq:R_eta^*_leq_R_0^*} is indeed an equality. Fix $h\in \mathcal{A}$, and suppose that $\mathcal{H}_h$ is the true hypothesis. Let arm $h$ be associated with an iid process whose underlying law is $\nu_1$. Further, for all $h'\neq h$, let arm $h'$ be associated with an iid process whose law is $\nu_2$, where $\nu_2\neq \nu_1$. Assume that the iid process of any given arm is independent of the iid process of each of the remaining arms. Let $\nu_h^a$ denote the marginal law of the iid process of arm $a$ under the hypothesis $\mathcal{H}_h$, i.e.,
\begin{equation}
	\nu_h^a=\begin{cases}
		\nu_1,&a=h,\\
		\nu_2,&a\neq h.
	\end{cases}\label{eq:nu_h^a}
\end{equation}

Since any iid process is trivially a Markov process, with the state space of the Markov process being the alphabet of the iid process, we may let $P_1$ denote the transition probability matrix of arm $h$ and $P_2$ the transition probability matrix of each of the non-odd arms $h'\neq h$. Then, for all $i,j\in\mathcal{S}$ and $d\geq 1$, we have
\begin{equation}
	P_1^d(j|i)=\nu_1(j),\quad P_2^d(j|i)=\nu_2(j).\label{eq:correspondence_btw_P_and_nu}
\end{equation}
Thus, when each arm yields iid observations, the function $\textcolor{black}{k_{hh'}(\underline{d}, \underline{i}, a)}$ in \eqref{eq:k(a,d_a,i_a)} may be expressed as
\begin{equation}
	\textcolor{black}{k_{hh'}(\underline{d}, \underline{i}, a)}=
	\begin{cases}
		D(\nu_1\|\nu_2),&a=h,\\
		{D(\nu_2\|\nu_1)},&a=h',\\
		0, &\text{otherwise}.
	\end{cases}\label{eq:k(a,d_a,i_a)_iid}
\end{equation}
In other words, the function $k$ does not depend on either the arm delays or the last observed states. Noting that the right-hand side of \eqref{eq:k(a,d_a,i_a)_iid} may be written compactly as $D(\nu_h^a\|\nu_{h'}^a)$, and plugging this in \eqref{eq:alternative_expression_for_R^*(P_1,P_2)}, we get
\begingroup \allowdisplaybreaks
\begin{align}
R_\eta^*(P_1,P_2)&=\sup\limits_{\lambda^\eta(\cdot\mid\cdot)\in \Lambda^\eta}~ \min\limits_{h'\neq h} ~ \sum\limits_{a=1}^{K}~\sum\limits_{(\underline{d},\underline{i})\in\mathbb{S}}  \nu^{\lambda^\eta}(\underline{d},\underline{i},a) \,D(\nu_h^a\|\nu_{h'}^a)\nonumber\\
&\stackrel{(a)}{=}\sup\limits_{\lambda^\eta(\cdot\mid\cdot)\in \Lambda^\eta}~ \min\limits_{h'\neq h} ~\sum\limits_{a=1}^{K} \sum\limits_{(\underline{d},\underline{i})\in\mathbb{S}}\mu^{\lambda^\eta}(\underline{d},\underline{i})\,\lambda^\eta(a|\underline{d}, \underline{i})\,D(\nu_h^a\|\nu_{h'}^a)\nonumber\\
	&=\sup\limits_{\lambda(\cdot\mid\cdot)\in \Lambda}~ \min\limits_{h'\neq h} ~\sum\limits_{a=1}^{K} \sum\limits_{(\underline{d},\underline{i})\in\mathbb{S}}\mu^{\lambda^\eta}(\underline{d},\underline{i})\,\left[\frac{\eta}{K}+(1-\eta)\,\lambda(a|\underline{d},\underline{i})\right]\,D(\nu_h^a\|\nu_{h'}^a)\nonumber\\
	&\stackrel{(b)}{=}\sup\limits_{\lambda(\cdot\mid\cdot)\in \Lambda}~ \min\limits_{h'\neq h} ~\frac{\eta}{K}\,\sum\limits_{a=1}^{K} \,D(\nu_h^a\|\nu_{h'}^a)+(1-\eta)\,\sum\limits_{a=1}^{K} \sum\limits_{(\underline{d},\underline{i})\in\mathbb{S}}\mu^{\lambda^\eta}(\underline{d},\underline{i})\,\lambda(a|\underline{d},\underline{i})\,D(\nu_h^a\|\nu_{h'}^a)\nonumber\\
	&=\sup\limits_{\lambda\in \mathcal{P}(\mathcal{A})}~ \min\limits_{h'\neq h} ~\frac{\eta}{K}\,\sum\limits_{a=1}^{K} \,D(\nu_h^a\|\nu_{h'}^a)+(1-\eta)\,\sum\limits_{a=1}^{K} \lambda(a)\,D(\nu_h^a\|\nu_{h'}^a)\label{eq:D^*(delta,P_1,P_2)_iid_arms_case}\\
	&={\color{black} \sup\limits_{\lambda\in \mathcal{P}(\mathcal{A})}~\frac{\eta}{K}\left[D(\nu_1\|\nu_2)+D(\nu_2\|\nu_1)\right]+(1-\eta)\,\left[\lambda(h)\,D(\nu_1\|\nu_2)+\bigg(\min\limits_{h'\neq h} ~\lambda(h')\bigg)\,D(\nu_2\|\nu_1)\right],}
	\label{eq:D^*(delta,P_1,P_2)_iid_arms_case_1}
\end{align}\endgroup
{ \color{black} where in $(a)$ above, $\mu^{\lambda^\eta}$ is the long-term probability of observing the state $(\underline{d},\underline{i})$ when the arms are selected according to the distribution $\lambda^\eta(\cdot\mid \cdot)$, $(b)$ above follows by using the fact that $\nu^{\lambda^\eta}$ is a probability distribution on $\mathbb{S}\times \mathcal{A}$, and the term $\lambda(a)$ in \eqref{eq:D^*(delta,P_1,P_2)_iid_arms_case} is given by $$\lambda(a)=\sum\limits_{(\underline{d},\underline{i})\in\mathbb{S}}\mu^{\lambda^\eta}(\underline{d},\underline{i})\,\lambda(a|\underline{d},\underline{i}),\quad a\in\mathcal{A},$$ with $\mathcal{P}(\mathcal{A})$ in \eqref{eq:D^*(delta,P_1,P_2)_iid_arms_case} denoting the set of all probability distributions on the set $\mathcal{A}$. Lastly, \eqref{eq:D^*(delta,P_1,P_2)_iid_arms_case_1} follows by noting that
\begin{equation}
\nu_h^a=\begin{cases}
\nu_1, &a=h,\\
\nu_2, &a\neq h,
\end{cases} \qquad \qquad \qquad \nu_{h'}^a=\begin{cases}
\nu_1, &a=h',\\
\nu_2, &a\neq h',
\end{cases}
\end{equation}
and therefore the only non-zero terms in the summation over the arms in \eqref{eq:D^*(delta,P_1,P_2)_iid_arms_case} are those corresponding to $a=h$ and $a=h'$.}

{\color{black} We now note that for each $\lambda \in \mathcal{P}(\mathcal{A})$, the mapping 
$$  
\eta \longmapsto \frac{\eta}{K}\left[D(\nu_1\|\nu_2)+D(\nu_2\|\nu_1)\right]+(1-\eta)\,\left[\lambda(h)\,D(\nu_1\|\nu_2)+\bigg(\min\limits_{h'\neq h} ~\lambda(h')\bigg)\,D(\nu_2\|\nu_1)\right]
$$
is bounded and linear (hence absolutely continuous) for all $\eta\in[0,1]$. Using the envelope theorem \cite[Theorem 2]{milgrom2002envelope}, we get that the mapping $\eta \mapsto R_\eta^*(P_1, P_2)$ is absolutely continuous for all $\eta\in [0,1]$, thereby implying that $\lim\limits_{\eta \downarrow 0} R_\eta^*(P_1,P_2)=R_0^*(P_1, P_2)$. This establishes that the inequality in \eqref{eq:R_eta^*_leq_R_0^*} holds with equality.}

\subsection{Rested Markov Arms}\label{subsec:rested_markov_arms}
We now show that when each arm is a Markov process on a finite state space that is common across the arms, and the arms are rested, the inequality in \eqref{eq:R_eta^*_leq_R_0^*} is indeed an equality. Fix $h\in\mathcal{A}$, and suppose that $\mathcal{H}_h$ is the true hypothesis. Let each arm be associated with a time-homogeneous and ergodic discrete-time Markov process on a common, finite state space $\mathcal{S}$. Let $P_1$ be the transition probability matrix of the odd arm, and let $P_2$ be the transition probability matrix of each of the non-odd arms. Let $\mu_1$ and $\mu_2$ denote the unique stationary distributions of $P_1$ and $P_2$ respectively. Assume that the Markov process of any given arm is independent of the Markov process of each of the remaining arms.

Let $P_h^a$ denote the transition probability matrix of arm $a$ under the hypothesis $\mathcal{H}_h$, and let $\mu_h^a$ be the stationary distribution of $P_h^a$. It then follows that
\begin{equation}
	P_h^a=\begin{cases}
		P_1,& a=h,\\
		P_2,& a\neq h,
	\end{cases}\quad \quad  \mu_h^a=\begin{cases}
		\mu_1,& a=h,\\
		\mu_2,& a\neq h.
	\end{cases}\label{eq:P_h^a_and_mu_h^a}
\end{equation}

When the arms are {rested}, as noted at the beginning of this section, the delay parameter for every arm is identically equal to $1$, i.e., $d_a(t)\equiv 1$ for all $a\in\mathcal{A}$ and $t\geq K$. Thus, we may omit the summation over $\underline{d}$ in \eqref{eq:alternative_expression_for_R^*(P_1,P_2)}. Writing $\lambda(a|\underline{i})$ in place of $\lambda(a|\underline{d},\underline{i})$, writing $\nu^{\lambda^\eta}(\underline{i})$ in place of $\nu^{\lambda^\eta}(\underline{d},\underline{i})$, and the following the steps presented earlier for the case of iid observations, we have
\begingroup \allowdisplaybreaks\begin{align}
R_\eta^*(P_1,P_2)&=\sup\limits_{\lambda^\eta(\cdot\mid\cdot)\in \Lambda^\eta}~ \min\limits_{h'\neq h} ~ \sum\limits_{a=1}^{K}~\sum\limits_{\underline{i}\in\mathcal{S}^K}  \nu^{\lambda^\eta}(\underline{i},a) \,D(P_h^a(\cdot|i_a)\|P_{h'}^a(\cdot|i_a))\nonumber\\
	&=\sup\limits_{\lambda(\cdot\mid\cdot)\in \Lambda}~ \min\limits_{h'\neq h} ~\sum\limits_{a=1}^{K} \sum\limits_{\underline{i}\in\mathcal{S}^K}\mu^{\lambda^\eta}(\underline{i})\,\left[\frac{\eta}{K}+(1-\eta)\,\lambda(a|\underline{i})\right]\,D(P_h^a(\cdot|i_a)\|P_{h'}^a(\cdot|i_a))\nonumber\\
	&=\sup\limits_{\lambda(\cdot\mid\cdot)\in \Lambda}~ \min\limits_{h'\neq h} ~\bigg[\frac{\eta}{K}\,\sum\limits_{a=1}^{K}\sum\limits_{\underline{i}\in\mathcal{S}^K}\mu^{\lambda^\eta}(\underline{i}) \,D(P_h^a(\cdot|i_a)\|P_{h'}^a(\cdot|i_a))\nonumber\\
	&\hspace{5cm}+(1-\eta)\,\sum\limits_{a=1}^{K} \sum\limits_{\underline{i}\in\mathcal{S}^K}\mu^{\lambda^\eta}(\underline{i})\,\lambda(a|\underline{i})\,D(P_h^a(\cdot|i_a)\|P_{h'}^a(\cdot|i_a))\bigg]\nonumber\\
	&{\color{black} \stackrel{(a)}{=}\sup\limits_{\lambda(\cdot\mid\cdot)\in \Lambda}~ \min\limits_{h'\neq h} ~\bigg[\frac{\eta}{K}\,\sum\limits_{a=1}^{K}\sum\limits_{\underline{i}\in\mathcal{S}^K}\mu^{\lambda^\eta}(\underline{i}) \,D(P_h^a(\cdot|i_a)\|P_{h'}^a(\cdot|i_a))}\nonumber\\
	&\hspace{5cm}{\color{black}+(1-\eta)\,\sum\limits_{a=1}^{K}~ \sum\limits_{i_a\in \mathcal{S}}\left(\sum\limits_{\underline{i}^{-a}\in \mathcal{S}^{K-1}}~\mu^{\lambda^\eta}(\underline{i})\,\lambda(a|\underline{i})\right)\,D(P_h^a(\cdot|i_a)\|P_{h'}^a(\cdot|i_a))\bigg]}\nonumber\\
	&{\color{black} \stackrel{(b)}{=}\sup\limits_{\lambda(\cdot\mid\cdot)\in \Lambda}~ \min\limits_{h'\neq h} ~\bigg[\frac{\eta}{K}\,\sum\limits_{a=1}^{K}\sum\limits_{i_a\in\mathcal{S}}\mu^{\lambda^\eta}(i_a) \,D(P_h^a(\cdot|i_a)\|P_{h'}^a(\cdot|i_a))}\nonumber\\
	&\hspace{5cm}{\color{black} +(1-\eta)\,\sum\limits_{a=1}^{K} \sum\limits_{i_a\in\mathcal{S}}\mu^{\lambda^\eta}(i_a)\,\lambda(a|i_a)\,D(P_h^a(\cdot|i_a)\|P_{h'}^a(\cdot|i_a))\bigg],}\label{eq:D^*(delta,P_1,P_2)_rested_Markov_case_1}
\end{align}\endgroup
{\color{black} where in $(a)$ above, $\underline{i}^{-a}$ denotes the vector of last observed states excluding the component corresponding to arm $a$, and in $(b)$ above, $\mu^{\lambda^\eta}(i_a)$ denotes the marginal of $\mu^{\lambda^\eta}(\underline{i})$ corresponding to arm $a$. Further, in writing $(b)$, we use the simplification
\begin{equation}
\sum\limits_{\underline{i}^{-a}\in \mathcal{S}^{K-1}}\mu^{\lambda^\eta}(\underline{i})\,\lambda(a|\underline{i}) = \mu^{\lambda^\eta}(i_a)\,\lambda(a|i_a).
\end{equation}}

{\color{black} We now note that the product $\mu^{\lambda^\eta}(i_a)\,\lambda(a|i_a)$ represents the long-term probability of observing arm $a$ in state $i_a$ 
 and subsequently selecting arm $a$ according to the conditional distribution $\lambda(a|i_a)$. This may be interpreted as the long-term probability of first seeing a transition \emph{from} the state $i_a$ on arm $a$ and subsequently selecting arm $a$ based on the observed transition. Since the arms are rested, the long-term probability of seeing a transition \emph{from} the state $i_a$ on arm $a$ is equal to the long-term probability of seeing a transition \emph{to} the state $i_a$ on arm $a$. Due to the ergodic nature of each of the arms, these probabilities are in turn equal to the probability of observing the state $i_a$ on arm $a$ under its stationary distribution; we refer the reader to \cite[pp. 4336]{pnkarthik2019learning} for the details.}

{\color{black} Hence, under the hypothesis $\mathcal{H}_h$, we may write 
\begin{equation}
\mu^{\lambda^\eta}(i_a)\,\lambda(a|i_a)=\lambda(a) \cdot \mu_h^a(i_a),
\label{eq:rested_arms_inequality_holds_with_equality_1}
\end{equation}
where in \eqref{eq:rested_arms_inequality_holds_with_equality_1}, $\lambda(a)=\sum\limits_{i_a \in \mathcal{S}} \mu^{\lambda^\eta}(i_a)\,\lambda(a|i_a)$. Using \eqref{eq:rested_arms_inequality_holds_with_equality_1} in \eqref{eq:D^*(delta,P_1,P_2)_rested_Markov_case_1}, we have
\begin{align}
R_\eta^*(P_1,P_2)&=\sup\limits_{\lambda(\cdot\mid\cdot)\in \Lambda}~ \min\limits_{h'\neq h} ~\bigg[\frac{\eta}{K}\,\sum\limits_{a=1}^{K}\sum\limits_{i_a\in\mathcal{S}}\mu^{\lambda^\eta}(i_a) \,D(P_h^a(\cdot|i_a)\|P_{h'}^a(\cdot|i_a))\nonumber\\
	&\hspace{5cm}+(1-\eta)\,\sum\limits_{a=1}^{K} \sum\limits_{i_a\in\mathcal{S}}\lambda(a) \, \mu_h^a(i_a)\,D(P_h^a(\cdot|i_a)\|P_{h'}^a(\cdot|i_a))\bigg]\nonumber\\
	&\stackrel{(a)}{=}\sup\limits_{\lambda\in \mathcal{P}(\mathcal{A})}~ \min\limits_{h'\neq h} ~\bigg[\frac{\eta}{K}\,\sum\limits_{a=1}^{K} \,D(P_h^a(\cdot|\cdot )\|P_{h'}^a(\cdot|\cdot) | \mu_h^a)+(1-\eta)\,\sum\limits_{a=1}^{K} \lambda(a)\,D(P_h^a(\cdot|\cdot )\|P_{h'}^a(\cdot|\cdot) | \mu_h^a)\bigg]\nonumber\\
	&=\sup\limits_{\lambda\in \mathcal{P}(\mathcal{A})}~\bigg[\frac{\eta}{K}\bigg(D(P_1(\cdot|\cdot)\|P_2(\cdot|\cdot)\|\mu_1) + D(P_2(\cdot|\cdot)\|P_1(\cdot|\cdot)\|\mu_2)\bigg)\nonumber\\
	&\hspace{3cm}+(1-\eta)\bigg(\lambda(h)\, D(P_1(\cdot|\cdot)\|P_2(\cdot|\cdot)\|\mu_1) + \bigg(\min\limits_{h'\neq h} \lambda(h')\bigg)\,D(P_2(\cdot|\cdot)\|P_1(\cdot|\cdot)\|\mu_2)\bigg)\bigg]\
	\label{eq:D^*(delta,P_1,P_2)_rested_Markov_case_2}
\end{align}
where in $(a)$ above, 
$$
D(P_h^a(\cdot|\cdot )\|P_{h'}^a(\cdot|\cdot) | \mu_h^a) \coloneqq \sum\limits_{i_a\in\mathcal{S}} \mu_h^a(i_a)\,D(P_h^a(\cdot|i_a)\|P_{h'}^a(\cdot|i_a)),
$$
and \eqref{eq:D^*(delta,P_1,P_2)_rested_Markov_case_2} follows by noting that 
\begin{equation}
P_h^a=\begin{cases}
		P_1,& a=h,\\
		P_2,& a\neq h,
	\end{cases} \qquad  \mu_h^a=\begin{cases}
		\mu_1,& a=h,\\
		\mu_2,& a\neq h,
	\end{cases}\qquad 
P_{h'}^a=\begin{cases}
		P_1,& a=h',\\
		P_2,& a\neq h',
	\end{cases} \qquad  \mu_{h'}^a=\begin{cases}
		\mu_1,& a=h',\\
		\mu_2,& a\neq h',
	\end{cases}
\end{equation} 
hence, the only non-zero terms in the summation over the arms in $(a)$ above are those corresponding to $a=h$ and $a=h'$.}

{\color{black} Finally, we note that for each $\lambda\in \mathcal{P}(\mathcal{A})$, the mapping 
\begin{align*}
\eta &\longmapsto \frac{\eta}{K}\bigg(D(P_1(\cdot|\cdot)\|P_2(\cdot|\cdot)\|\mu_1) + D(P_2(\cdot|\cdot)\|P_1(\cdot|\cdot)\|\mu_2)\bigg)\\
	&\hspace{3cm}+(1-\eta)\bigg(\lambda(h)\, D(P_1(\cdot|\cdot)\|P_2(\cdot|\cdot)\|\mu_1) + \bigg(\min\limits_{h'\neq h} \lambda(h')\bigg)\,D(P_2(\cdot|\cdot)\|P_1(\cdot|\cdot)\|\mu_2)\bigg)
\end{align*}
is bounded and linear (hence absolutely continuous) for all $\eta\in [0,1]$. Using the envelope theorem \cite[Theorem 2]{milgrom2002envelope}, we get that the mapping $\eta \mapsto R_\eta^*(P_1, P_2)$ is absolutely continuous for all $\eta\in [0,1]$, thereby implying that $\lim\limits_{\eta \downarrow 0} R_\eta^*(P_1,P_2)=R_0^*(P_1, P_2)$. This establishes that the inequality in \eqref{eq:R_eta^*_leq_R_0^*} holds with equality.}

{\color{black} \subsection{A Subtle Remark on the Interpretation of $R_0^*(P_1, P_2)$}
Recall that $R_0^*(P_1,P_2)$ denotes the constant obtained by plugging $\eta=0$ in \eqref{eq:R_delta^*(h,P_1,P_2)}. The correct interpretation of this constant deserves some explanation, which is the content of this section. Recall that when $\eta>0$, under every SRS policy, the controlled Markov process $\{(\underline{d}(t), \underline{i}(t)): t\geq K\}$ is an ergodic Markov process (Lemma \ref{lem:pi_delta^lambda_is_an_SSRS}). This, in conjunction with Theorem  \ref{thrm:restriction_to_SRS_policies} of Appendix \ref{appndx:an_important_theorem}, leads to the supremum over the set of SRS policies in \eqref{eq:R_delta^*(h,P_1,P_2)}. However, it is important to note that Theorem  \ref{thrm:restriction_to_SRS_policies} crucially relies on the ergodicity property given by Lemma \ref{lem:pi_delta^lambda_is_an_SSRS}, the proof of which in turn holds only for the case $\eta>0$. Such an ergodicity property may not be available when $\eta=0$. Therefore, it is not clear how, after plugging $\eta=0$, the right hand side of \eqref{eq:R_delta^*(h,P_1,P_2)} is to be interpreted; for e.g., $\nu^\lambda$, the ergodic state-action occupancy measure under the SRS policy $\pi^\lambda$ when $\eta>0$, may no longer be interpreted so when $\eta=0$.
	
	In order to address the above mentioned issue, we appeal to the literature and note that a common assumption that appears in works that deal with controlled Markov processes is one of ``under every SRS policy, the underlying controlled Markov process is an ergodic Markov process''; see, for instance, \cite[pp. 58, Section II]{borkar1988control} or \cite{puterman2014markov}. Such an assumption readily holds for the case $\eta>0$. Thus,  $R_0^*(P_1,P_2)$ must be interpreted as the constant obtained by plugging $\eta=0$ in \eqref{eq:R_delta^*(h,P_1,P_2)}, under the assumption\footnote{Or any assumption that in turn guarantees ergodicity of the underlying controlled Markov process under every SRS policy.} that every SRS policy makes the underlying controlled Markov process an ergodic Markov process.}

\section{Concluding Remarks}\label{sec:conclusion}
We make several concluding remarks to end the paper.
\begin{enumerate}
\item  From \eqref{eq:main_result}, when the trembling hand parameter $\eta>0$, we see that
\begin{equation}
	\lim\limits_{\epsilon\downarrow 0} ~ \inf\limits_{\pi\in\Pi(\epsilon)} ~ \frac{E_h[\tau(\pi)]}{\log(1/\epsilon)} = \frac{1}{R_\eta^*(P_1,P_2)}.\label{eq:final-result}
\end{equation}
We have thus provided an answer to \eqref{eqn:objective} on the minimum growth rate of the expected time to identify the odd arm location as $\epsilon \downarrow 0$.

\item The asymptotically optimal $\lambda(\cdot|\cdot)$ in the restless case may depend on the history unlike that in the prior works \cite{Vaidhiyan2017, prabhu2017optimal, vaidhiyan2017learning, pnkarthik2019learning} where $\lambda(\cdot)$ did not depend on history, even in the rested Markov case. At first glance, this is surprising for the rested Markov case, but in retrospect, these features are apparent from an examination of the optimisation problem \eqref{eq:R_delta^*(h,P_1,P_2)} in these special cases. 

\item Computability of $R_\eta^*(P_1,P_2)$ may be an issue, and one must usually resort to $Q$-learning for restless Markov arms \cite{avrachenkov2020whittle} to arrive at good policies. The fact that $D(P_k^{d_a} (\cdot | i_a) || P_l^{d_a} (\cdot | i_a))$, $k,l\in\{1,2\}$, converges as $d_a \rightarrow \infty$ could enable restriction of the countable state space $\mathbb{S}$ to a finite set, and could lead to good approximations.


\item When the trembling hand parameter $\eta>0$, the ergodicity of the Markov process $(\underline{d}(t), \underline{i}(t))$ under any SRS policy ensures that time averages approach the ensemble averages. This is crucial to show achievability. Note also the use of uniqueness of the stationary distribution to show the converse. The trembling hand model may be viewed as a {\em regularisation} that gives stability of the aforementioned Markov process for free. If the trembling hand parameter $\eta$ were 0, one could deliberately add some regularisation parameterised by $\eta$, and let this parameter $\eta \downarrow 0$. $R_{0}^*(P_1,P_2)$ governs the lower bound, whereas $\lim\limits_{\eta \downarrow 0}R_{\eta}^*(P_1,P_2)$ governs the upper bound. The resulting lower and upper bounds on the growth rate may have a gap.

\item Open questions: The key difficulties when $\eta=0$ are (a) absence of ergodicity property, and (b) a formal verification of the envelope theorem. 
It would be interesting to study these. Another open question is the setting when $P_1$ and $P_2$ are unknown and have to be learnt along the way.
\end{enumerate}


\appendices
\section{Proof of Lemma \ref{lem:pi_delta^lambda_is_an_SSRS}}\label{appndx:proof_of_Lemma_pi_delta^lambda_is_an_SSRS}
Let $\eta\in(0,1]$ denote the trembling hand parameter. Fix $\pi^\lambda\in\Pi_{\textsf{SRS}}$ and $h\in\mathcal{A}$, and let $\mathcal{H}_h$ be the true hypothesis. Recall that under the SRS policy $\pi^\lambda$, the controlled Markov process  $\{(\underline{d}(t), \underline{i}(t)):t\geq K\}$ is a Markov process.

\begin{IEEEproof}[Proof of Irreducibility]
	Consider any two states $(\underline{d},\underline{i})\in\mathbb{S}$ and $(\underline{d}',\underline{i}')\in\mathbb{S}$, and suppose that the Markov process $\{(\underline{d}(t), \underline{i}(t)):t\geq K\}$ is in the state $(\underline{d},\underline{i})$ at some time $t=T_0$. We shall now demonstrate that there exists $N$ such that the state $(\underline{d}',\underline{i}')$ may be reached starting from the state $(\underline{d},\underline{i})$ after $N$ steps under $\pi^\lambda$. Recall that at any time $t$, the arm intended to be pulled is $B_t$, while the arm actually pulled at time $t$ is its trembled version $A_t$; the arms $A_t$ and $B_t$ are related through the trembling hand relation in \eqref{eq:trembling_hand_rule}. For any $a\in\mathcal{A}$, we have
	\begin{align}
		P(A_t=a\mid B^{t-1},A^{t-1},\bar{X}^{t-1})&=\sum\limits_{b=1}^{K}P(B_t=b,\,A_t=a|B^{t-1},A^{t-1},\bar{X}^{t-1})\nonumber\\
		&=\sum\limits_{b=1}^{K}P(B_t=b\mid B^{t-1},A^{t-1},\bar{X}^{t-1})\cdot P(A_t=a\mid B_t=b,B^{t-1},A^{t-1},\bar{X}^{t-1})\nonumber\\
		&\stackrel{(a)}{=}\sum\limits_{b=1}^{K}\lambda(b\mid\underline{d}(t),\underline{i}(t))\cdot \left(\frac{\eta}{K}+(1-\eta)\,\mathbb{I}_{\{a=b\}}\right)\nonumber\\
		&=\frac{\eta}{K}+(1-\eta)\,\lambda(a\mid\underline{d}(t),\underline{i}(t))\nonumber\\
		&\geq \frac{\eta}{K},\label{eq:P(A_t=a|B_t=b,B^{t-1},A^{t-1},bar{X}^{t-1}}
	\end{align}
	where $(a)$ above follows from \eqref{eq:trembling_hand_rule} and the fact that under $\pi^\lambda$, the intended arm $B_t$ is selected  according to $\lambda(\cdot\mid\cdot)$. 
	
Assume without loss of generality that $\underline{d}'$, the vector of arm delays in the destination state $(\underline{d}', \underline{i}')$, is such that $d_1'>d_2'>\cdots>d_K^\prime=1$. Noting that $P_1$ and $P_2$ are transition probability matrices on the finite set $\mathcal{S}$, we use \cite[Proposition 1.7]{levin2017markov} for finite state Markov processes to deduce that there exists an integer $M$ such that for all $m\geq M$,
\begin{equation}
	P_1^m(j|i)>0\text{ for all }i,j\in\mathcal{S},\quad P_2^m(j|i)>0\text{ for all }i,j\in\mathcal{S}.\label{eq:P_1^M_and_P_2^M_entries_strictly_pos_entries}
\end{equation}
Consider the sequence of actions and observations as follows: starting from the state $(\underline{d},\underline{i})$ at time $t=T_0$, let the Markov process $\{(\underline{d}(t), \underline{i}(t)):t \geq K\}$ evolve for $M-1$ time instants. Thereafter, let arm $1$ be selected at the $(T_0+M)$th time instant and let the state observed on arm $1$ be $i_1'$; let arm $2$ be selected at the $(T_0+M+d_1'-d_2')$th time instant and let the state observed on arm $2$ be $i_2'$, and so on. Finally, let arm $K$ be observed at the $(T_0+M+d_1'-d_K')$th time instant, and let the state observed on arm $K$ be $i_K'$. Additionally, let arm $1$ not be selected for all $T_0+M<t<T_0+M+d_1'$; let arm $2$ not be selected for all $T_0+M+d_1'-d_2'<t<T_0+M+d_1'$ and so on.

Clearly, the above sequence of actions and observations leads to the state $(\underline{d}',\underline{i}')$ after $M+d_1'-d_K'$ time instants. Thus, the probability of starting from the state $(\underline{d},\underline{i})$ and reaching the state $(\underline{d}',\underline{i}')$ may be lower bounded by the probability that the above sequence of actions and observations occur under $\pi^\lambda$ which, when the true hypothesis is $\mathcal{H}_h$, is given by
\begin{align}
	&\bigg(\prod_{a=1}^{K} P(A_{T_0+M+d_1'-d_a'}=a\mid B^{T_0+M+d_1'-d_a'-1}, A^{T_0+M+d_1'-d_a'-1}, \bar{X}^{T_0+M+d_1'-d_a'-1})\bigg)~\cdot ~\bigg(\prod\limits_{a=1}^{K}(P_h^a)^{M+d_1'-d_{a}'+d_a}(i_a'|i_a)\bigg) \nonumber\\ 
	&\hspace{3cm}\cdot \bigg(\prod\limits_{a=1}^{K-1}~\prod\limits_{t=T_0+M+d_1'-d_a'+1}^{T_0+M+d_1'-d_{a+1}'}P(A_t\notin \{1,\ldots,a\}\mid B^{t-1},A^{t-1},\bar{X}^{t-1})\bigg)\nonumber\\
	&\stackrel{(a)}{\geq}\bigg(\frac{\eta}{K}\bigg)^{K}\cdot  \left[\prod\limits_{a=1}^{K}(P_h^a)^{M+d_1'-d_a'+d_a}(i_a'|i_a)\right]\cdot \left[\prod\limits_{a=1}^{K-1}\,\prod\limits_{t=T_0+M+d_1'-d_a'+1}^{T_0+M+d_1'-d_{a+1}'}\frac{\eta(K-a)}{K}\right]\nonumber\\
	&\stackrel{(b)}{\geq} \bigg(\frac{\eta}{K}\bigg)^{K}\cdot \left[\prod\limits_{a=1}^{K}(P_h^a)^{M+d_1'-d_a'+d_a}(i_a'|i_a)\right]\cdot \left[\prod\limits_{a=1}^{K-1}\,\prod\limits_{t=T_0+M+d_1'-d_a'+1}^{T_0+M+d_1'-d_{a+1}'}\frac{\eta}{K}\right]\nonumber\\
	&>0,
\end{align}
where $(a)$ above follows from the observation that the right-hand side of \eqref{eq:P(A_t=a|B_t=b,B^{t-1},A^{t-1},bar{X}^{t-1}}, for each $t$, is $\geq \eta/K$ and the fact that $$P(A_t\notin \{1,\ldots,a\}\mid B^{t-1},A^{t-1}\bar{X}^{t-1})=\sum\limits_{a'= a+1}^{K} P(A_t=a'\mid B^{t-1},A^{t-1},\bar{X}^{t-1})\geq \frac{\eta(K-a)}{K},$$ and $(b)$ follows by noting that $K-a\geq 1$ for $a\in \{1,\ldots,K-1\}$. Setting $N=M+d_1'-d_K'$, we see that the Markov process $\{(\underline{d}(t),\underline{i}(t):t\geq K)\}$ is in the state $(\underline{d}',\underline{i}')$ at time $t=T_0+N$. This establishes irreducibility.
\end{IEEEproof}

\begin{IEEEproof}[Proof of Aperiodicity]
	Fix an arbitrary $(\underline{d},\underline{i})\in\mathbb{S}$. We shall now demonstrate that starting from the state $(\underline{d},\underline{i})$, there is a strictly positive probability of the Markov process $\{(\underline{d}(t),\underline{i}(t)):t\geq K\}$ returning back to the state $(\underline{d},\underline{i})$ after $M'$ steps as well as after $(M'+1)$ steps, where $M'$ is  sufficiently large and such that \eqref{eq:P_1^M_and_P_2^M_entries_strictly_pos_entries} holds for all $m\geq M'$. This will then establish the desired aperiodicity property since the period of the state $(\underline{d},\underline{i})$ is equal to the gcd of $M'$ and $M'+1$, which is $1$. 
	
Assume, without loss of generality, that $\underline{d}$ is such that $d_1>d_2>\cdots d_K=1$. Let $M$ be such that \eqref{eq:P_1^M_and_P_2^M_entries_strictly_pos_entries} holds for all $m\geq M$. Using arguments similar to those presented above in the proof of irreducibility, the probability of starting from the state $(\underline{d},\underline{i})$ at some time $t=T_0$ and returning back to the state $(\underline{d},\underline{i})$ after $M+d_1-d_K$ time instants may be lower bounded, under hypothesis $\mathcal{H}_h$, by 
	\begin{equation}
		\bigg(\frac{\eta}{K}\bigg)^{K}\cdot \left[\prod\limits_{a=1}^{K}(P_h^a)^{M+d_1}(i_a'|i_a)\right]\cdot \left[\prod\limits_{a=1}^{K-1}\,\prod\limits_{t=T_0+M+d_1-d_a+1}^{T_0+M+d_1-d_{a+1}}\frac{\eta}{K}\right]>0.\label{eq:aperiodicity}
	\end{equation}  
Setting $M'=M+d_1-d_K$ yields the desired result. 
\end{IEEEproof}

\begin{IEEEproof}[Proof of positive recurrence]
	 Let 
\begingroup \allowdisplaybreaks\begin{align}
	& p_\eta\coloneqq \frac{\eta}{K}\min\bigg\lbrace \min\{P_1^{M}(j|i):i,j\in\mathcal{S}\},\,\min\{P_2^{M}(j|i):i,j\in\mathcal{S}\} \bigg\rbrace;\label{eq:p_delta_strictly_positive}
\end{align}\endgroup
here, once again, $M$ is such that \eqref{eq:P_1^M_and_P_2^M_entries_strictly_pos_entries} holds for all $m\geq M$. Therefore, it follows that $p_\eta>0$. Let 
\begin{equation}
	r(\pi^\lambda)\coloneqq\min\{t> K:\underline{d}(t)=\underline{d}(K),\underline{i}(t)=\underline{i}(K)\}\label{eq:tau(pi_delta^lambda)}
\end{equation}
denote the first return time of the Markov process $\{(\underline{d}(t), \underline{i}(t)):t\geq K\}$ to its initial state (i.e., the state at time $t=K$) under $\pi^\lambda$. We may then upper bound $r(\pi^\lambda)$ as
\begin{equation}
	r(\pi^\lambda)\leq M\cdot K\cdot \tau_{\eta}\quad \text{almost surely},\label{eq:r(pi_delta^lambda)_as_upper_bound}
\end{equation}
where $\tau_\eta$ is a Geometric random variable with parameter $p_\eta$. In other words, $r(\pi^\lambda)$ may be almost surely upper bounded by the first return time of the process $\{(\underline{d}(t), \underline{i}(t)):t\geq K\}$ to its initial state measured only at time instants that are integer multiples of $M\cdot K$. It then follows that
\begin{align}
	E\left[r(\pi^\lambda)\right]&\leq M\cdot K\cdot E[\tau_\eta]\nonumber\\
	&=M\cdot K\cdot \frac{1}{p_\eta}\nonumber\\
	&<\infty,\label{eq:sup_first_moment_of_return_time_finite}
\end{align}
thus implying that the Markov process $\{(\underline{d}(t), \underline{i}(t)):t\geq K\}$ is positive recurrent under $\pi^\lambda$. This completes the proof of positive recurrence, and also the proof of the lemma.
\end{IEEEproof}

\section{Proof of Proposition \ref{prop:lower_bound}}\label{appndx:proof_of_prop_lower_bound}
This proof is organised as follows. Given $\epsilon>0$, we first obtain a lower bound for $E_h[Z_{hh'}(\tau(\pi)]$ for all $\pi\in\Pi(\epsilon)$ using a change of measure argument of Kaufmann et al. \cite{Kaufmann2016}. Following this, we obtain an upper bound for $E_h[Z_{hh'}(\tau(\pi)]$ in terms of $E_h[\tau(\pi)]$. Combining the upper and the lower bounds, and letting $\epsilon\downarrow 0$, we arrive at the desired result. The ergodicity property established in Lemma \ref{lem:pi_delta^lambda_is_an_SSRS} for SRS policies plays a crucial role in deriving the final lower bound of \eqref{eq:lower_bound}.

\subsection{A Lower Bound on $E_h[Z_{hh'}(\tau(\pi))]$ for $\pi\in\Pi(\epsilon)$}
As a first step towards deriving the lower bound, we use a result of Kaufmann et al. \cite{Kaufmann2016} to obtain a lower bound for $E_h[Z_{hh'}(\tau(\pi))]$ in terms of the error probability parameter $\epsilon$. This is based on a generalisation of \cite[Lemma 18]{Kaufmann2016}, a change of measure argument for iid observations from the arms, to the setting of restless arms with Markov observations. We present this generalisation in the following lemma. 

\begin{Lemma}\label{lem:change_of_measure}
	Fix $\pi\in\Pi(\epsilon)$, and let $\tau(\pi)$ be the stopping time of policy $\pi$. Let $\mathcal{F}_{\tau(\pi)}$ be the $\sigma$-algebra
	\begin{equation}
		\mathcal{F}_{\tau(\pi)}=\{E\in\mathcal{F}:E\cap \{\tau(\pi)= t\}\in\mathcal{F}_t\text{ for all }t\geq 0\},\label{eq:F_tau}
	\end{equation}
	where $\mathcal{F}_0=\sigma(\Omega, \emptyset)$ and $\mathcal{F}_t=\sigma(B^t,A^t,\bar{X}^t)$ for all $t\geq 1$. Then, for any $h,h'\in\mathcal{A}$ such that $h'\neq h$, the relation 
	\begin{equation}
		P_{h'}(E)=E_h[1_{E}\,\exp(-Z_{hh'}(\tau(\pi))]\label{eq:change_of_measure}
	\end{equation}
	holds for all $E\in\mathcal{F}_{\tau(\pi)}$.
\end{Lemma}

\begin{IEEEproof}[Proof of Lemma \ref{lem:change_of_measure}]
 We prove the Lemma by demonstrating, through mathematical induction, that the relation
 \begin{equation}
 	E_{h'}[g(B^t,A^t,\bar{X}^t)]=E_h[g(B^t,A^t,\bar{X}^t)\,\exp(-Z_{hh'}(t))]\label{eq:change_of_measure_equiv}
 \end{equation}
 holds for all $t\geq 0$ and for all measurable functions $g:\mathcal{A}^{t+1}\times\mathcal{A}^{t+1}\times\mathcal{S}^{t+1}\to\mathbb{R}$. The proof for the case $t=0$ may be obtained as follows. For any measurable $g:\mathcal{A}\times\mathcal{A}\times\mathcal{S}\to\mathbb{R}$, we have
 \begingroup \allowdisplaybreaks\begin{align}
 	E_{h'}[g(B_0,A_0,\bar{X}_0)]&=\sum\limits_{b=1}^K\sum\limits_{a=1}^{K}\sum\limits_{i\in\mathcal{S}}~g(b,a,i)~P_{h'}(B_0=b, A_0=a,\bar{X}_0=i)\nonumber\\
 	&=\sum\limits_{b=1}^K\sum\limits_{a=1}^{K}\sum\limits_{i\in\mathcal{S}}~g(b,a,i)~P_{h'}(B_0=b)~P_{h'}(A_0=a|B_0=b)~P_{h'}(\bar{X}_0=i|B_0=b,A_0=a)\nonumber\\
 	&\stackrel{(a)}{=}\sum\limits_{b=1}^K\sum\limits_{a=1}^{K}\sum\limits_{i\in\mathcal{S}}~g(b,a,i)~P_h(B_0=b)~P_h(A_0=a|B_0=b)~P_{h'}(\bar{X}_0=i|A_0=a)\nonumber\\
 	&=\sum\limits_{b=1}^K\sum\limits_{a=1}^{K}\sum\limits_{i\in\mathcal{S}}~g(b,a,i)~P_h(B_0=b)~P_h(A_0=a|B_0=b)~P_{h'}(X_0^a=i),\label{eq:change_of_measure_1}
 \end{align}\endgroup
 where $(a)$ follows using the facts that $P_h(B_0=b)=P_{h'}(B_0=b)$ and $P_h(A_0=a|B_0=b)=P_{h'}(A_0=a|B_0=b)$ (see Section \ref{sec:achievability}). Assuming that $X_0^a\sim \nu$, where $\nu$ is a probability distribution on $\mathcal{S}$, independent of the true hypothesis (which is not known to the policy $\pi$), we have
 \begingroup \allowdisplaybreaks\begin{align}
 	E_{h'}[g(B_0,A_0,\bar{X}_0)]&=\sum\limits_{b=1}^{K}\sum\limits_{a=1}^{K}\sum\limits_{i\in\mathcal{S}}~g(b,a,i)~P_h(B_0=b)~P_h(A_0=a|B_0=b)~\nu(i)\\
 	&=\sum\limits_{b=1}^{K}\sum\limits_{a=1}^{K}\sum\limits_{i\in\mathcal{S}}~g(b,a,i)~P_h(B_0=b)~P_h(A_0=a|B_0=b)~P_h(X_0^a=i|A_0=a)\\
 	&=\sum\limits_{b=1}^{K}\sum\limits_{a=1}^{K}\sum\limits_{i\in\mathcal{S}}~g(b,a,i)~P_h(B_0=b)~P_h(A_0=a|B_0=b)~P_h(X_0^a=i|A_0=a, B_0=b).\label{eq:change_of_measure_2}
 \end{align}\endgroup
 Also, we have (see Section \ref{sec:achievability})
 \begingroup \allowdisplaybreaks\begin{align}
 	Z_{hh'}(0)&=\log \frac{P_h(B_0,A_0,\bar{X}_0)}{P_{h'}(B_0,A_0,\bar{X}_0)}=0.\label{eq:change_of_measure_3}
 \end{align}\endgroup
 Combining \eqref{eq:change_of_measure_2} and \eqref{eq:change_of_measure_3}, we get $E_{h'}[g(B_0,A_0,\bar{X}_0)]=E_h[g(B_0,A_0,\bar{X}_0)\exp (-Z_{hh'}(0))]$, thus proving \eqref{eq:change_of_measure_equiv} for $t=0$.
 
 We now assume that \eqref{eq:change_of_measure_equiv} is true for some $t>0$, and demonstrate that it also true for $t+1$. By the law of iterated expectations,
 \begingroup \allowdisplaybreaks\begin{align}
 	E_{h'}[g(B^{t+1},A^{t+1},\bar{X}^{t+1})]=E_{h'}[E_{h'}[g(B^{t+1},A^{t+1},\bar{X}^{t+1})|\mathcal{F}_t]].\label{eq:change_of_measure_4}
 \end{align}\endgroup
Noting that $E_{h'}[g(B^{t+1},A^{t+1},\bar{X}^{t+1})|\mathcal{F}_t]$ is a measurable function of $(B^t,A^t,\bar{X}^t)$, by the induction hypothesis, we have
 \begingroup \allowdisplaybreaks\begin{align}
 	E_{h'}[g(B^{t+1},A^{t+1},\bar{X}^{t+1})|\mathcal{F}_t]&=E_h[E_{h'}[g(B^{t+1},A^{t+1},\bar{X}^{t+1})|\mathcal{F}_t]\,\exp(-Z_{hh'}(t))].\label{eq:change_of_measure_5}
 \end{align}\endgroup
 We now note that
 \begingroup \allowdisplaybreaks\begin{align}
 	& E_{h'}[g(B^{t+1},A^{t+1},\bar{X}^{t+1})|\mathcal{F}_t]\,\exp(-Z_{hh'}(t))\nonumber\\
 	&\stackrel{(a)}{=}E_{h'}[g(B^{t+1},A^{t+1},\bar{X}^{t+1})\,\exp(-Z_{hh'}(t))|\mathcal{F}_t]\,\nonumber\\
 	&=\sum\limits_{b=1}^{K}\sum\limits_{a=1}^{K}\sum\limits_{i\in\mathcal{S}}\bigg[g(B^t,A^t,\bar{X}^t,b, a, i)\cdot P_{h'}(B_{t+1}=b|B^t,A^t,\bar{X}^t)\nonumber\\
 	&\hspace{3cm} \cdot P_{h'}(A_{t+1}=a|B^{t+1}=b,B^t,A^t,\bar{X}^t)\cdot P_{h'}(\bar{X}_{t+1}=i|B^{t+1}=b,A_{t+1}=a,B^t,A^t,\bar{X}^t)\cdot \exp(-Z_{hh'}(t))\bigg]\nonumber\\
 	&\stackrel{(b)}{=}\sum\limits_{b=1}^{K}\sum\limits_{a=1}^{K}\sum\limits_{i\in\mathcal{S}}\bigg[g(B^t,A^t,\bar{X}^t,b, a, i)\cdot P_h(B_{t+1}=b|B^t,A^t,\bar{X}^t)\nonumber\\
 	&\hspace{3cm} \cdot P_h(A_{t+1}=a|B^{t+1}=b,B^t,A^t,\bar{X}^t)\cdot P_{h'}(\bar{X}_{t+1}=i|A_{t+1}=a,A^t,\bar{X}^t)\cdot \exp(-Z_{hh'}(t))\bigg],\label{eq:change_of_measure_6}
 \end{align}\endgroup
 where $(a)$ above is due to the fact that $Z_{hh'}(t)$ is a measurable function of $(B^t, A^t,\bar{X}^t)$, and in writing $(b)$, we use the following facts: for any $t$,
 \begin{itemize}
 	\item $P_{h'}(B_{t+1}=b|B^t,A^t,\bar{X}^t)=P_h(B_{t+1}=b|B^t,A^t,\bar{X}^t)$,
 	\item $P_{h'}(A_{t+1}=a|B_{t+1}=b,B^t,A^t,\bar{X}^t)=P_h(A_{t+1}=a|B_{t+1}=b,B^t,A^t,\bar{X}^t)$, and
 	\item $P_{h'}(\bar{X}_{t+1}=i|B_{t+1}=b,A_{t+1}=a,B^t,A^t,\bar{X}^t)=P_{h'}(\bar{X}_{t+1}=i|A_{t+1}=a,A^t,\bar{X}^t)$.
 \end{itemize}
 See Section \ref{sec:achievability} for a justification of why the above facts are true. It then follows that
 \begingroup \allowdisplaybreaks\begin{align}
 	& \sum\limits_{i\in\mathcal{S}}P_{h'}(\bar{X}_{t+1}=i|A_{t+1}=a,A^t,\bar{X}^t)\,\exp(-Z_{hh'}(t))\nonumber\\
 	&=\sum\limits_{i\in\mathcal{S}}\frac{P_{h'}(\bar{X}_{t+1}=i|A_{t+1}=a,A^t,\bar{X}^t)}{P_h(\bar{X}_{t+1}=i|A_{t+1}=a,A^t,\bar{X}^t)}\,\exp(-Z_{hh'}(t))\,P_h(\bar{X}_{t+1}=i|A_{t+1}=a,A^t,\bar{X}^t)\nonumber\\
 	&=\sum\limits_{i\in\mathcal{S}}\exp(-Z_{hh'}(t+1,a,i))\,P_h(\bar{X}_{t+1}=i|A_{t+1}=a,A^t,\bar{X}^t).\label{eq:change_of_measure_7}
 \end{align}\endgroup
 where in \eqref{eq:change_of_measure_7}, the quantity $Z_{hh'}(t+1,a,i)$ is defined as $$Z_{hh'}(t+1,a,i)\coloneqq Z_{hh'}(t)+\log\frac{P_h(\bar{X}_{t+1}=i|A_{t+1}=a,A^t,\bar{X}^t)}{P_{h'}(\bar{X}_{t+1}=i|A_{t+1}=a,A^t,\bar{X}^t)}.$$
 Substituting \eqref{eq:change_of_measure_7} in \eqref{eq:change_of_measure_6} and simplifying, we get
 \begingroup \allowdisplaybreaks\begin{align}
& E_{h'}[g(B^{t+1},A^{t+1},\bar{X}^{t+1})|\mathcal{F}_t]\,\exp(-Z_{hh'}(t))\nonumber\\
&=\sum\limits_{b=1}^{K}\sum\limits_{a=1}^{K}\sum\limits_{i\in\mathcal{S}} \bigg[g(B^t,A^t,\bar{X}^t, b, a, i)\cdot P_h(B_{t+1}=b|B^t,A^t,\bar{X}^t)\nonumber\\
&\hspace{1cm}\cdot P_h(A_{t+1}=a|B_{t+1}=b,B^t,A^t,\bar{X}^t)\cdot P_h(\bar{X}_{t+1}=i|B_{t+1}=b,A_{t+1}=a,B^t,A^t,\bar{X}^t)\cdot \exp(-Z_{hh'}(t+1,a,i))\bigg]\\
&=E_h[g(B^{t+1},A^{t+1},\bar{X}^{t+1})~\exp(-Z_{hh'}(t+1))|\mathcal{F}_t].\label{eq:change_of_measure_8}
 \end{align}\endgroup
 Applying $E_h[\cdot]$ to both sides of \eqref{eq:change_of_measure_8}, we arrive at the desired relation. This proves \eqref{eq:change_of_measure_equiv} for all $t\geq 0$.
 
 Finally, for any $E\in\mathcal{F}_{\tau(\pi)}$, we have
 \begingroup \allowdisplaybreaks\begin{align}
 	P_{h'}(E)&=E_{h'}[1_{E}]\nonumber\\
 	&=E_{h'}\left[\sum\limits_{t\geq 0} 1_{E\cap \{\tau(\pi)= t\}}\right]\nonumber\\
 	&\stackrel{(a)}{=}\sum\limits_{t\geq 0}E_{h'}\left[1_{E\cap \{\tau(\pi)= t\}}\right]\nonumber\\
 	&\stackrel{(b)}{=}\sum\limits_{t\geq 0}E_h\left[1_{E\cap \{\tau(\pi)= t\}}\,\exp(-Z_{hh'}(t))\right]\nonumber\\
 	&=\sum\limits_{t\geq 0}E_h\left[1_{E\cap \{\tau(\pi)= t\}}\,\exp(-Z_{hh'}(\tau(\pi)))\right]\nonumber\\
 	&=E_h\left[1_{E}\,\exp(-Z_{hh'}(\tau(\pi)))\right],\label{eq:change_of_measure_9}
 \end{align}\endgroup
 where $(a)$ is due to monotone convergence theorem, and $(b)$ above follows from \eqref{eq:change_of_measure_equiv} and the fact that $E\cap \{\tau(\pi)=t\}\in\mathcal{F}_t$ for all $t\geq 0$ since $E\in\mathcal{F}_{\tau(\pi)}$. This completes the proof of the lemma.
\end{IEEEproof}

Lemma \ref{lem:change_of_measure}, in conjunction with \cite[Lemma 19]{Kaufmann2016}, yields the following inequality for all policies $\pi\in \Pi(\epsilon)$ and all $h'\neq h$:
\begin{equation}
	E_h[Z_{hh'}(\tau(\pi))]\geq \sup\limits_{E\in\mathcal{F}_{\tau(\pi)}} d(P_h(E),P_{h'}(E)),\label{eq:Kaufmann_DPI_bound}
\end{equation} 
where for any $x,y\in [0,1]$, $$d(x,y)\coloneqq x\log (x/y)+(1-x)\log ((1-x)/(1-y))$$ is the binary relative entropy function. As noted in \cite{Kaufmann2016}, $x\mapsto d(x,y)$ is monotone increasing for $x<y$ and the $y\mapsto d(x,y)$ is monotone decreasing for any fixed $x$. Also, for any $\pi\in\Pi(\epsilon)$, we have $$P_h(\theta(\pi)=h)\geq 1-\epsilon,\quad P_{h'}(\theta(\pi)=h)\leq \epsilon$$ for all $h'\neq h$. Combining the aforementioned facts, we get
\begin{equation}
	\min\limits_{h'\neq h} E_h[Z_{hh'}(\tau(\pi))]\geq d(\epsilon,1-\epsilon).\label{eq:Kaufmann_DPI_bound_epsilon}
\end{equation}
for all $\pi\in\Pi(\epsilon)$.
\subsection{An Upper Bound for  $E_h[Z_{hh'}(\tau(\pi))]$ in Terms of $ E_h[\tau(\pi)]$}

We now obtain an upper bound for the left-hand side of \eqref{eq:Kaufmann_DPI_bound_epsilon}. Fix $\pi\in\Pi(\epsilon)$ and $h'\neq h$ arbitrarily. Then, from \eqref{eq:Z_{hh'}(n)},
\begingroup \allowdisplaybreaks\begin{align}
	& E_h[Z_{hh'}(\tau(\pi))]\nonumber\\
	&=E_h\bigg[\sum\limits_{a=1}^{K}\log\frac{P_h(X_{a-1}^a)}{P_{h'}(X_{a-1}^a)}\bigg]+E_h\bigg[\sum\limits_{(\underline{d},\underline{i})\in\mathbb{S}}~\sum\limits_{a=1}^{K}\sum\limits_{j\in\mathcal{S}}N(\tau(\pi),\underline{d},\underline{i},a,j)\log\frac{(P_h^{a})^{d_a}(j|i_a)}{(P_{h'}^{a})^{d_a}(j|i_a)}\bigg].\label{eq:lower_bound_1}
\end{align}\endgroup
To simplify the second expectation term on the right-hand side of \eqref{eq:lower_bound_1}, we use the following Lemma.
\begin{Lemma}\label{lem:relation_between_N(tau,d,i,j,a)_and_N(tau,d,i,a)}
	{\color{black} Fix $h\in \mathcal{A}$. For every $(\underline{d},\underline{i})\in \mathbb{S}$, $a\in\mathcal{A}$ and $j\in \mathcal{S}$,}
	\begin{equation}
		E_h[E_h[N(\tau(\pi),\underline{d},\underline{i},a,j)|X_{a-1}^a]|\tau(\pi)]=E_h[E_h[N(\tau(\pi),\underline{d},\underline{i},a)|X_{a-1}^a]|\tau(\pi)]\,(P_h^a)^{d_a}(j|i_a).\label{eq:relation_between_N(tau,d,i,j,a)_and_N(tau,d,i,a)}
	\end{equation}
\end{Lemma}

\begin{IEEEproof}[Proof of Lemma \ref{lem:relation_between_N(tau,d,i,j,a)_and_N(tau,d,i,a)}]
 Substituting $n=\tau(\pi)$ in \eqref{eq:N(n,d,i,a,j)}, we have
 \begingroup \allowdisplaybreaks\begin{align}
 	E_h[E_h[N(\tau(\pi),\underline{d},\underline{i},a,j)|X_{a-1}^a]|\tau(\pi)]&=E_h\bigg[E_h\bigg[\sum\limits_{t=K}^{\tau(\pi)} 1_{\{\underline{d}(t)=\underline{d},\underline{i}(t)=\underline{i},A_{t}=a,X_{t}^a=j\}}\bigg|X_{a-1}^a\bigg]\bigg|\tau(\pi)\bigg]\nonumber\\
 	&=E_h\bigg[\sum\limits_{t=K}^{\tau(\pi)}P_h(\underline{d}(t)=\underline{d},\underline{i}(t)=\underline{i},A_{t}=a,X_{t}^a=j|X_{a-1}^a)\,\bigg|\tau(\pi)\bigg].\label{eq:relation_btw_two_qty_1}
 \end{align}\endgroup
 For each $t$ in the range of the summation in \eqref{eq:relation_btw_two_qty_1}, the conditional probability term for $t$ may be expressed as
 \begingroup \allowdisplaybreaks\begin{align}
 	& P_h(\underline{d}(t)=\underline{d},\underline{i}(t)=\underline{i},A_{t}=a,X_{t}^a=j|X_{a-1}^a)\nonumber\\
 	&=P_h(\underline{d}(t)=\underline{d},\underline{i}(t)=\underline{i},A_{t}=a|X_{a-1}^a)\cdot P_h(X_{t}^a=j|A_{t}=a,\underline{d}(t)=\underline{d},\underline{i}(t)=\underline{i},X_{a-1}^a)\nonumber\\
 	&=P_h(\underline{d}(t)=\underline{d},\underline{i}(t)=\underline{i},A_{t}=a|X_{a-1}^a)\cdot (P_h^a)^{d_a}(j|i_a).\label{eq:relation_btw_two_qty_2}
 \end{align}\endgroup
 Plugging \eqref{eq:relation_btw_two_qty_2} back in \eqref{eq:relation_btw_two_qty_1} and simplifying, we arrive at the desired relation in \eqref{eq:relation_between_N(tau,d,i,j,a)_and_N(tau,d,i,a)}.
\end{IEEEproof}

Using Lemma \ref{lem:relation_between_N(tau,d,i,j,a)_and_N(tau,d,i,a)}, the second expectation term on the right-hand side of \eqref{eq:lower_bound_1} can be simplified as follows.
\begingroup \allowdisplaybreaks\begin{align}
	& E_h\bigg[\sum\limits_{(\underline{d},\underline{i})\in\mathbb{S}}~\sum\limits_{a=1}^{K}\sum\limits_{j\in\mathcal{S}}N(\tau(\pi),\underline{d},\underline{i},a,j)\log\frac{(P_h^{a})^{d_a}(j|i_a)}{(P_{h'}^{a})^{d_a}(j|i_a)}\bigg]\nonumber\\
	&=E_h\bigg[\sum\limits_{(\underline{d},\underline{i})\in\mathbb{S}}~\sum\limits_{a=1}^{K}\sum\limits_{j\in\mathcal{S}}N(\tau(\pi),\underline{d},\underline{i},a,j)\log\frac{(P_h^{a})^{d_a}(j|i_a)}{(P_{h'}^{a})^{d_a}(j|i_a)}\bigg]\nonumber\\
 &=E_h\left[\sum\limits_{(\underline{d},\underline{i})\in\mathbb{S}}~\sum\limits_{a=1}^{K}\sum\limits_{j\in\mathcal{S}}E_h[E_h[N(\tau(\pi),\underline{d},\underline{i},a,j)|X_{a-1}^a]|\tau(\pi)]\log\frac{(P_h^{a})^{d_a}(j|i_a)}{(P_{h'}^{a})^{d_a}(j|i_a)}\right]\nonumber\\
 &\stackrel{(a)}{=} E_h\left[\sum\limits_{(\underline{d},\underline{i})\in\mathbb{S}}~\sum\limits_{a=1}^{K}\sum\limits_{j\in\mathcal{S}}E_h[E_h[N(\tau(\pi),\underline{d},\underline{i},a)|X_{a-1}^a]|\tau(\pi)]\cdot (P_h^a)^{d_a}(j|i)\cdot\log\frac{(P_h^{a})^{d_a}(j|i_a)}{(P_{h'}^{a})^{d_a}(j|i_a)}\right]\nonumber\\
 &= E_h\left[\sum\limits_{(\underline{d},\underline{i})\in\mathbb{S}}~\sum\limits_{a=1}^{K}E_h[E_h[N(\tau(\pi),\underline{d},\underline{i},a)|X_{a-1}^a]|\tau(\pi)]\cdot D((P_h^a)^{d_a}(\cdot|i_a)\|(P_{h'}^a)^{d_a}(\cdot|i_a))\right]\nonumber\\
 &=\sum\limits_{(\underline{d},\underline{i})\in\mathbb{S}}~\sum\limits_{a=1}^{K}E_h[N(\tau(\pi),\underline{d},\underline{i},a)]\cdot D((P_h^a)^{d_a}(\cdot|i_a)\|(P_{h'}^a)^{d_a}(\cdot|i_a)),\label{eq:lower_bound_2}
\end{align}\endgroup
where in the above set of equations, $(a)$ follows from Lemma \ref{lem:relation_between_N(tau,d,i,j,a)_and_N(tau,d,i,a)},
and \eqref{eq:lower_bound_2} is due to monotone convergence theorem and the fact that $$E_h[E_h[E_h[N(\tau(\pi),\underline{d},\underline{i},a)|X_{a-1}^a]|\tau(\pi)]]=E_h[N(\tau(\pi),\underline{d},\underline{i},a)].$$

Plugging \eqref{eq:lower_bound_2} back in \eqref{eq:lower_bound_1}, 
we get
\begingroup \allowdisplaybreaks\begin{align}
	& E_h[Z_{hh'}(\tau(\pi))]\nonumber\\
	&=E_h\left[\sum\limits_{a=1}^{K} \log\frac{P_h(X_{a-1}^a)}{P_{h'}(X_{a-1}^a)}\right]
	+\sum\limits_{(\underline{d},\underline{i})\in\mathbb{S}}~\sum\limits_{a=1}^{K}E_h[N(\tau(\pi),\underline{d},\underline{i},a)]\cdot D((P_h^a)^{d_a}(\cdot|i_a)\|(P_{h'}^a)^{d_a}(\cdot|i_a)).\label{eq:lower_bound_3}
\end{align}\endgroup
Noting that 
\begingroup \allowdisplaybreaks\begin{align}
	\sum\limits_{(\underline{d},\underline{i})\in\mathbb{S}}~\sum\limits_{a=1}^{K}~E_h[N(\tau(\pi),\underline{d},\underline{i},a)]&\stackrel{(a)}{=}E_h\bigg[\sum\limits_{(\underline{d},\underline{i})\in\mathbb{S}}~\sum\limits_{a=1}^{K}N(\tau(\pi),\underline{d},\underline{i},a)\bigg]\nonumber\\
	&=E_h\bigg[\sum\limits_{(\underline{d},\underline{i})\in\mathbb{S}}~\sum\limits_{a=1}^{K}\sum\limits_{t=K}^{\tau(\pi)}1_{\{\underline{d}(t)=\underline{d},\underline{i}(t)=\underline{i},A_{t}=a\}}\bigg]\nonumber\\
	&=E_h\bigg[\sum\limits_{t=K}^{\tau(\pi)}1\bigg]\\
	&=E_h[\tau(\pi)-K+1],
\end{align}\endgroup
where $(a)$ above is due to monotone convergence theorem,
we write \eqref{eq:lower_bound_3} as
\begingroup \allowdisplaybreaks\begin{align}
	& E_h[Z_{hh'}(\tau(\pi))]\nonumber\\
	&=E_h\left[\sum\limits_{a=1}^{K} \log\frac{P_h(X_{a-1}^a)}{P_{h'}(X_{a-1}^a)}\right]+\bigg(E_h[\tau(\pi)-K+1]\bigg)\cdot \sum\limits_{(\underline{d},\underline{i})\in\mathbb{S}}~\sum\limits_{a=1}^{K}\frac{E_h[N(\tau(\pi),(\underline{d},\underline{i}),a)]}{E_h[\tau(\pi)-K+1]}\cdot D((P_h^a)^{d_a}(\cdot|i_a)\|(P_{h'}^a)^{d_a}(\cdot|i_a)).\label{eq:lower_bound_4}
\end{align}\endgroup
Combining \eqref{eq:Kaufmann_DPI_bound_epsilon} and \eqref{eq:lower_bound_4}, and noting that \eqref{eq:lower_bound_4} holds for all $h'\neq h$, we get
\begingroup \allowdisplaybreaks\begin{align}
	d(\epsilon,1-\epsilon)&\leq \min\limits_{h'\neq h}\bigg\lbrace E_h\left[\sum\limits_{a=1}^{K} \log\frac{P_h(X_{a-1}^a)}{P_{h'}(X_{a-1}^a)}\right]\nonumber\\
	&\hspace{2.5cm}+ \bigg(E_h[\tau(\pi)-K+1]\bigg)\cdot\,\sum\limits_{(\underline{d},\underline{i})\in\mathbb{S}}~\sum\limits_{a=1}^{K}\frac{E_h[N(\tau(\pi),\underline{d},\underline{i},a)]}{E_h[\tau(\pi)-K+1]}\cdot D((P_h^a)^{d_a}(\cdot|i_a)\|(P_{h'}^a)^{d_a}(\cdot|i_a))\bigg\rbrace\nonumber\\
	&\leq \sup\limits_{\nu}\,\min\limits_{h'\neq h}\bigg\lbrace E_h\left[\sum\limits_{a=1}^{K} \log\frac{P_h(X_{a-1}^a)}{P_{h'}(X_{a-1}^a)}\right]\nonumber\\
	&\hspace{2.5cm}+\bigg(E_h[\tau(\pi)-K+1]\bigg)\cdot\,\sum\limits_{(\underline{d},\underline{i})\in\mathbb{S}}~\sum\limits_{a=1}^{K}\nu(\underline{d},\underline{i},a)\, D((P_h^a)^{d_a}(\cdot|i_a)\|(P_{h'}^a)^{d_a}(\cdot|i_a))\bigg\rbrace,\label{eq:lower_bound_5}
	\end{align}\endgroup
where the supremum in \eqref{eq:lower_bound_5} is over all state-action occupancy measures satisfying 
\begin{align}
	\sum\limits_{a=1}^{K}\nu(\underline{d}',\underline{i}',a)&=\sum\limits_{(\underline{d},\underline{i})\in\mathbb{S}}~\sum\limits_{a=1}^{K}\,\nu(\underline{d},\underline{i},a)\,Q(\underline{d}',\underline{i}'|\underline{d},\underline{i},a)\quad \text{for all }(\underline{d}',\underline{i}')\in\mathbb{S},\label{eq:lower_bound_6_1}\\
	&\sum\limits_{(\underline{d},\underline{i})\in\mathbb{S}}~\sum\limits_{a=1}^{K}\,\nu(\underline{d},\underline{i},a)=1,\label{eq:lower_bound_6_2}\\
	&\nu(\underline{d},\underline{i},a)\geq 0\quad \text{for all }(\underline{d},\underline{i},a)\in\mathbb{S}\times\mathcal{A}.\label{eq:lower_bound_6_3}
\end{align}
Recall that $Q$ in \eqref{eq:lower_bound_6_1} denotes the transition probability matrix given by \eqref{eq:MDP_transition_probabilities_A_t}. The left-hand side of \eqref{eq:lower_bound_6_1} represents the long-term probability of leaving the state $(\underline{d},\underline{i})$, while the right-hand side of \eqref{eq:lower_bound_6_2} represents the long-term probability of entering into the state $(\underline{d},\underline{i})$. Thus, \eqref{eq:lower_bound_6_1} is the \emph{global balance equation} for the controlled Markov process $\{(\underline{d}(t),\underline{i}(t)):t\geq K\}$. Equations \eqref{eq:lower_bound_6_2} and \eqref{eq:lower_bound_6_3} together imply that $\nu$ is a probability measure on $\mathbb{S}\times\mathcal{A}$.
 
As outlined in Section \ref{sec:MDP_preliminaries}, the controlled Markov process $\{(\underline{d}(t),\underline{i}(t)):t\geq K\}$, together with the sequence $\{B_t:t\geq 0\}$ of intended arm selections (or equivalently the sequence $\{A_t:t\geq 0\}$ of actual arm selections), defines a Markov decision problem (MDP) with state space $\mathbb{S}$ and action space $\mathcal{A}$. From Lemma \ref{lem:pi_delta^lambda_is_an_SSRS}, we know that $\{(\underline{d}(t), \underline{i}(t)): t\geq K\}$ is an ergodic Markov process under every SRS policy. This suffices to apply  Theorem \ref{thrm:restriction_to_SRS_policies} of Appendix \ref{appndx:an_important_theorem} to deduce a one-one correspondence between feasible solutions to \eqref{eq:lower_bound_6_1}-\eqref{eq:lower_bound_6_3} and policies in $\Pi_{\textsf{SRS}}$. In other words,  Theorem \ref{thrm:restriction_to_SRS_policies} implies that for any given $\nu$ satisfying \eqref{eq:lower_bound_6_1}-\eqref{eq:lower_bound_6_3}, we can find an SRS policy $\pi^\lambda\in\Pi_{\textsf{SRS}}$ such that $\nu^\lambda(\underline{d},\underline{i},a)=\nu(\underline{d},\underline{i},a)$ for all $(\underline{d},\underline{i},a)\in\mathbb{S}\times\mathcal{A}$. Recall that under the SRS policy $\pi^\lambda$, the stationary distribution of the Markov process $\{(\underline{d}(t),\underline{i}(t)):t\geq K\}$ is $\mu^\lambda$. The associated ergodic state occupancy measure, $\nu^\lambda$, is then defined according to \eqref{eq:ergodic_state_action_occupancy_measure}.

On account of Theorem \ref{thrm:restriction_to_SRS_policies}, we may replace the supremum in \eqref{eq:lower_bound_5} by a supremum over all SRS policies. Doing so leads us to the relation
\begin{align}
	d(\epsilon,1-\epsilon)&\leq \sup\limits_{\pi^\lambda\in \Pi_{\textsf{SRS}}}\,\min\limits_{h'\neq h}\bigg\lbrace E_h\left[\sum\limits_{a=1}^{K} \log\frac{P_h(X_{a-1}^a)}{P_{h'}(X_{a-1}^a)}\right]\nonumber\\
	&\hspace{2.5cm}+\bigg(E_h[\tau(\pi)-K+1]\bigg)\cdot\,\sum\limits_{(\underline{d},\underline{i})\in\mathbb{S}}~\sum\limits_{a=1}^{K}\nu^\lambda(\underline{d},\underline{i},a)\, D((P_h^a)^{d_a}(\cdot|i_a)\|(P_{h'}^a)^{d_a}(\cdot|i_a))\bigg\rbrace.\label{eq:lower_bound_7}
\end{align}
for all $\pi\in\Pi(\epsilon)$. Observe that the constant term multiplying $E_h[\tau(\pi)-K+1]$ in \eqref{eq:lower_bound_7} is finite; further, it is not a function of either $\epsilon$ or of $\pi\in\Pi(\epsilon)$. The finiteness of this constant follows from the following observation: denote by $\mu_h^a$ the stationary distribution of the transition probability matrix $P_h^a$ (i.e., $\mu_h^a=\mu_1$ for $a=h$ and $=\mu_2$ for all $a\neq h$). An application of the ergodic theorem to the Markov process of arm $a$ yields 
	    \begin{equation}
	    	D((P_h^a)^{d_a}(\cdot|i_a)\|(P_{h'}^a)^{d_a}(\cdot|i_a))\longrightarrow D(\mu_h^a\|\mu_{h'}^a)<\infty \quad \text{as }d_a\to\infty.
	    \end{equation}
	    Since every convergent sequence is bounded, we may write $D((P_h^a)^{d_a}(\cdot|i_a)\|(P_{h'}^a)^{d_a}(\cdot|i_a))\leq C$ for all $(\underline{d},\underline{i},a)\in\mathbb{S}\times\mathcal{A}$, where $0<C<\infty$. Using \eqref{eq:lower_bound_6_2}, it follows that the constant term multiplying $E_h[\tau(\pi)-K+1]$ in \eqref{eq:lower_bound_7} is bounded above by $C$.
	    
Let us also note that the first term inside the braces in \eqref{eq:lower_bound_7} does not depend on $\epsilon$. Since $d(\epsilon,1-\epsilon)\to d(0,1)=+\infty$ as $\epsilon\downarrow 0$, the boundedness of $R^*(P_1,P_2)$ shows that $\epsilon\downarrow 0$ is equivalent to $E_h[\tau(\pi)]\to \infty$ for all $\pi\in\Pi(\epsilon)$.
Letting $\epsilon\downarrow 0$, and using  $d(\epsilon,1-\epsilon)/\log(1/\epsilon)\longrightarrow 1$ as $\epsilon\downarrow 0$, we arrive at the lower bound in \eqref{eq:lower_bound}. This completes the proof of the proposition.

\section{Proof of Lemma \ref{lem:liminf_Z_{hh'}(n)_strictly_positive}}\label{appndx:proof_of_lem_liminf_Z_{hh'}(n)_strictly_positive}
Observe that the key ingredient in the proof of Lemma \ref{lem:pi_delta^lambda_is_an_SSRS} is  the strict positivity of the probability term in \eqref{eq:P(A_t=a|B_t=b,B^{t-1},A^{t-1},bar{X}^{t-1}} when the trembling hand parameter $\eta>0$. Clearly, this is satisfied even under the policy $\pi^\star(L,\delta)$. We leverage this to first show that under the policy $\pi^\star(L, \delta)$,
\begin{equation}
	\liminf\limits_{n\to\infty}\frac{N(n,\underline{d},\underline{i})}{n}>0\quad \text{almost surely}\label{eq:lim_N(n,d,i)}
\end{equation}
for every $(\underline{d},\underline{i})\in\mathbb{S}$, where for each $n\geq K$,
\begin{equation}
	N(n,\underline{d},\underline{i})\coloneqq \sum\limits_{t=K}^{n}\mathbb{I}_{\{\underline{d}(t)=\underline{d},~\underline{i}(t)=\underline{i}\}}\label{eq:N(n,d,i)}
\end{equation}
denotes the number of times the controlled Markov process $\{(\underline{d}(t),\underline{i}(t)):t\geq K\}$ visits the state $(\underline{d},\underline{i})$. Noting that $P_1$ and $P_2$ are transition probability matrices on the finite set $\mathcal{S}$, we use \cite[Proposition 1.7]{levin2017markov} for finite state Markov processes to deduce that there exists an integer $M$ such that for all $m\geq M$,
\begin{equation}
P_1^m(j|i)>0\text{ for all }i,j\in\mathcal{S},\quad P_2^m(j|i)>0\text{ for all }i,j\in\mathcal{S}.
\end{equation}
Fix an arbitrary $(\underline{d}, \underline{i})\in\mathbb{S}$, and assume without loss of generality that $\underline{d}$ is such that $d_1>d_2>\cdots>d_K=1$. Also assume, again without loss of generality, that the controlled Markov process $\{(\underline{d}(t), \underline{i}(t)):t\geq K\}$ starts in the state $(\underline{d}, \underline{i})$, i.e., $\underline{d}(K)=\underline{d}$, $\underline{i}(K)=\underline{i}$. Let $p(\underline{d}, \underline{i})$ denote the probability of the process  $\{(\underline{d}(t),\underline{i}(t)):t\geq K\}$ starting in the state $(\underline{d},\underline{i})$ and returning back to the state $(\underline{d},\underline{i})$. The analysis presented in Appendix \ref{appndx:proof_of_Lemma_pi_delta^lambda_is_an_SSRS} shows that this probability is lower bounded by the probability of returning after $M+d_1$ time instants given by \eqref{eq:aperiodicity}. Since \eqref{eq:aperiodicity} is strictly positive, it follows that $p(\underline{d},\underline{i})>0$. 

Clearly, then, the term $N(n,\underline{d},\underline{i})$ may be lower bounded almost surely by the number of visits to the state $(\underline{d},\underline{i})$ measured only at times $t=K+M+d_1, K+2(M+d_1),K+3(M+d_1)$ and so on until time $t=n$. Note that at each of these time instants, the probability that the process $\{(\underline{d}(t),\underline{i}(t)):t\geq K\}$ is in the state $(\underline{d},\underline{i})$ is equal to $p(\underline{d},\underline{i})$. Thus, we have
\begin{align}
N(n,\underline{d},\underline{i})\geq \text{Bin}\left(\frac{n-K+1}{M+d_1},~~p(\underline{d}, \underline{i})\right)\quad \text{almost surely},\label{eq:N(n,d,i)_lower_bounded_by_binomial}
\end{align}
where the notation $\text{Bin}(m,q)$ denotes a Binomial random variable with parameters $m$ and $q$. It then follows that, almost surely,
\begin{align}
\liminf\limits_{n\to\infty}\frac{N(n,\underline{d},\underline{i})}{n}&\geq \liminf\limits_{n\to\infty}\frac{\text{Bin}\left(\frac{n-K+1}{M+d_1},~~p(\underline{d}, \underline{i})\right)}{n}\nonumber\\
&=\liminf\limits_{n\to\infty} \frac{\text{Bin}\left(\frac{n-K+1}{M+d_1},~~p(\underline{d}, \underline{i})\right)}{\frac{n-K+1}{M+d_1}}\cdot \frac{n-K+1}{n}\cdot \frac{1}{M+d_1}\nonumber\\
&\stackrel{(a)}{=}\frac{p(\underline{d}, \underline{i})}{M+d_1}\nonumber\\
&>0,\label{eq:liminf_N(n,d,i)_strictly positive}
\end{align}
where $(a)$ above is due to the strong law of large numbers. This establishes \eqref{eq:lim_N(n,d,i)}.

We now show that for all $(\underline{d}, \underline{i})\in \mathbb{S}$ and $a\in\mathcal{A}$,
\begin{equation}
	\liminf\limits_{n\to\infty}\frac{N(n,\underline{d},\underline{i},a)}{n}>0\quad \text{almost surely}.\label{eq:liminf_N(n,d,i_a)_strictly_positive}
\end{equation}
We shall then use \eqref{eq:liminf_N(n,d,i_a)_strictly_positive} to establish \eqref{eq:liminf_Z_{hh'}(n)_strictly_positive}. Fix an arbitrary $a\in\mathcal{A}$, and define
\begin{equation}
	S(n,\underline{d},\underline{i},a)\coloneqq \sum\limits_{t=K}^{n}\bigg[\mathbb{I}_{\{A_t=a,\underline{d}(t)=\underline{d},\underline{i}(t)=\underline{i}\}}-P(A_t=a,\underline{d}(t)=\underline{d},\underline{i}(t)=\underline{i}|B^{t-1},A^{t-1},\bar{X}^{t-1})\bigg].\label{eq:S(n,d,i.a)}
\end{equation}
For each $t\geq K$, since $|\mathbb{I}_{\{A_t=a,\underline{d}(t)=\underline{d},\underline{i}(t)=\underline{i}\}}-P(A_t=a,\underline{d}(t)=\underline{d},\underline{i}(t)=\underline{i}|B^{t-1},A^{t-1},\bar{X}^{t-1})|\leq 2$ almost surely, and $$E[\mathbb{I}_{\{A_t=a,\underline{d}(t)=\underline{d},\underline{i}(t)=\underline{i}\}}-P(A_t=a,\underline{d}(t)=\underline{d},\underline{i}(t)=\underline{i}|B^{t-1},A^{t-1},\bar{X}^{t-1})|B^{t-1}, A^{t-1},\bar{X}^{t-1}]=0,$$ the collection $\{\mathbb{I}_{\{A_t=a,\underline{d}(t)=\underline{d},\underline{i}(t)=\underline{i}\}}-P(A_t=a,\underline{d}(t)=\underline{d},\underline{i}(t)=\underline{i}|B^{t-1},A^{t-1},\bar{X}^{t-1})\}_{t\geq K}$ is a bounded martingale difference sequence. Using the concentration result \cite[Theorem 1.2A]{victor1999general} for bounded martingale difference sequences and subsequently applying the Borel-Cantelli lemma, we get that
\begin{equation}
	\frac{S(n,\underline{d},\underline{i},a)}{n}\longrightarrow 0\quad \text{as }n\to\infty,\quad \text{almost surely}.
\end{equation}
 This implies that for every choice of $\varepsilon>0$, there exists $N_\varepsilon$ sufficiently large such that
\begin{equation}
	\frac{N(n,\underline{d},\underline{i},a)}{n}\geq \frac{1}{n}\sum\limits_{t=K}^{n}P(A_t=a,\underline{d}(t)=\underline{d},\underline{i}(t)=\underline{i}|B^{t-1},A^{t-1},\bar{X}^{t-1})-\varepsilon \quad\text{ for all }n\geq N_\varepsilon \text{ almost surely}.\label{eq:frac{N(n,d,i,a)}{n}_is_lower_bounded}
\end{equation}
Now, for each $t\geq K$,
\begin{align}
	&P(A_t=a,\underline{d}(t)=\underline{d},\underline{i}(t)=\underline{i}|B^{t-1},A^{t-1},\bar{X}^{t-1})\nonumber\\
	&=P(A_t=a|\underline{d}(t)=\underline{d},\underline{i}(t)=\underline{i}, B^{t-1}, A^{t-1}, \bar{X}^{t-1})\cdot P(\underline{d}(t)=\underline{d},\underline{i}(t)=\underline{i}|B^{t-1},A^{t-1},\bar{X}^{t-1})\nonumber\\
	&=\left[\frac{\eta}{K}+(1-\eta)\,\lambda_{\theta(t),\delta}(a\mid \underline{d},\underline{i})\right]P(\underline{d}(t)=\underline{d},\underline{i}(t)=\underline{i}|B^{t-1},A^{t-1},\bar{X}^{t-1})\nonumber\\
	&\geq \frac{\eta}{K}\cdot \mathbb{I}_{\{(\underline{d}(t)=\underline{d},\underline{i}(t)=\underline{i}\}},\label{eq:lower_bound_P(A_t=a,d,i)}
\end{align}
where \eqref{eq:lower_bound_P(A_t=a,d,i)} follows from the fact that $\underline{d}(t)$ and $\underline{i}(t)$ are measurable with respect to the history $(B^{t-1}, A^{t-1},\bar{X}^{t-1})$.
Plugging \eqref{eq:lower_bound_P(A_t=a,d,i)} in \eqref{eq:frac{N(n,d,i,a)}{n}_is_lower_bounded}, we get
\begin{align}
	\frac{N(n,\underline{d},\underline{i},a)}{n} &\geq \frac{\eta}{K}\cdot \frac{N(n,\underline{d},\underline{i})}{n}-\varepsilon\quad \forall~ n\geq N_\varepsilon \quad \text{almost surely}.\label{eq:frac{N(n,d,i,a)}{n}_is_lower_bounded_1}
\end{align}
Using \eqref{eq:liminf_N(n,d,i)_strictly positive} in \eqref{eq:frac{N(n,d,i,a)}{n}_is_lower_bounded_1}, we get
\begin{equation}
	\frac{N(n,\underline{d},\underline{i},a)}{n-K+1} \geq \frac{\eta}{K}\cdot \frac{p(\underline{d}, \underline{i})}{2(M+d_1)}-\varepsilon\label{eq:frac{N(n,d,i,a)}{n}_is_lower_bounded_2}
\end{equation}
for all sufficiently large values of $n$, almost surely.
Setting $\varepsilon=\frac{\eta}{2K}\cdot \frac{p(\underline{d}, \underline{i})}{2(M+d_1)}$ establishes \eqref{eq:liminf_N(n,d,i_a)_strictly_positive}. 

\begin{IEEEproof}[Proof of Lemma \ref{lem:liminf_Z_{hh'}(n)_strictly_positive}]
For any $h'\neq h$, we have
\begin{align}
	\frac{1}{n}Z_{hh'}(n)&=\sum\limits_{(\underline{d},\underline{i})\in\mathbb{S}}\sum\limits_{j\in \mathcal{S}}\frac{N(n,\underline{d},\underline{i},h,j)}{n}\,\log\frac{P_1^{d_h}(j|i_h)}{P_2^{d_h}(j|i_h)}+\frac{N(n,\underline{d},\underline{i},h',j)}{n}\,\log\frac{P_1^{d_{h'}}(j|i_{h'})}{P_2^{d_{h'}}(j|i_{h'})}.\label{eq:Z_{hh'}(n)/n_1}
\end{align}
Since $N(n,\underline{d},\underline{i},a)\to\infty$ almost surely as $n\to\infty$ (this follows from the fact that $\liminf\limits_{n\to\infty}N(n,\underline{d},\underline{i},a)/n>0$ almost surely) for every $a\in\mathcal{A}$, we apply the Ergodic theorem to deduce that
\begin{equation}
	\frac{N(n,\underline{d},\underline{i},a,j)}{N(n,\underline{d},\underline{i},a)}\longrightarrow (P_h^a)^{d_a}(j|i_a)\quad \text{as }n\to\infty \quad \text{almost surely}.\label{eq:limit_for_N(n,d,i,a,j)/N(n,d,i,a)}
\end{equation}
Using \eqref{eq:limit_for_N(n,d,i,a,j)/N(n,d,i,a)} in \eqref{eq:Z_{hh'}(n)/n_1}, we get that for every choice of $\varepsilon$, there exists $N_\varepsilon$ sufficiently large such that for all $n\geq N_\varepsilon$, almost surely,
\begin{align}
	\frac{1}{n}Z_{hh'}(n)&\geq \sum\limits_{(\underline{d},\underline{i})\in\mathbb{S}}\sum\limits_{j\in \mathcal{S}} \frac{N(n,\underline{d},\underline{i},h)}{n}(P_1^{d_h}(j|i_h)+\varepsilon)\log P_1^{d_h}(j|i_h)\nonumber\\
	&+\sum\limits_{(\underline{d},\underline{i})\in\mathbb{S}}\sum\limits_{j\in \mathcal{S}} \frac{N(n,\underline{d},\underline{i},h)}{n}(P_1^{d_h}(j|i_h)-\varepsilon)\log \frac{1}{P_2^{d_h}(j|i_h)}\nonumber\\
	&+\sum\limits_{(\underline{d},\underline{i})\in\mathbb{S}}\sum\limits_{j\in \mathcal{S}} \frac{N(n,\underline{d},\underline{i},h')}{n}(P_2^{d_{h'}}(j|i_{h'})+\varepsilon)\log P_2^{d_{h'}}(j|i_{h'})\nonumber\\
	&+\sum\limits_{(\underline{d},\underline{i})\in\mathbb{S}}\sum\limits_{j\in \mathcal{S}} \frac{N(n,\underline{d},\underline{i},h')}{n}(P_2^{d_{h'}}(j|i_{h'})-\varepsilon)\log \frac{1}{P_1^{d_{h'}}(j|i_{h'})}\nonumber\\
	&=\sum\limits_{(\underline{d},\underline{i})\in\mathbb{S}}\frac{N(n,\underline{d},\underline{i},h)}{n}\,D(P_1^{d_h}(\cdot|i_h)\|P_2^{d_h}(\cdot|i_h))+\frac{N(n,\underline{d},\underline{i},h')}{n}\,D(P_2^{d_{h'}}(\cdot|i_{h'})\|P_1^{d_{h'}}(\cdot|i_{h'}))\nonumber\\
	&+\varepsilon\left[\sum\limits_{(\underline{d},\underline{i})\in\mathbb{S}}\frac{N(n,\underline{d},\underline{i},h)}{n}\left(\sum\limits_{j\in \mathcal{S}} \log P_1^{d_h}(j|i_h)\,P_2^{d_h}(j|i_h)\right)+\frac{N(n,\underline{d},\underline{i},h')}{n}\left(\sum\limits_{j\in \mathcal{S}} \log P_1^{d_{h'}}(j|i_{h'})\,P_2^{d_{h'}}(j|i_{h'})\right)\right].\label{eq:Z_{hh'}(n)/n_lower_bound_1}
\end{align}
As a consequence of the convergence theorem for finite state Markov processes \cite[Theorem 4.9]{levin2017markov}, we have 
\begin{align}
	P_1^{d}(j|i)&\longrightarrow \mu_1(j)>0\quad \text{ as }d\to\infty\nonumber\\
	P_2^{d}(j|i)&\longrightarrow \mu_2(j)>0\quad \text{ as }d\to\infty\label{eq:convergence_theorem_convergences}
\end{align}
for all $i,j\in\mathcal{S}$. This implies that
 the term inside the square brackets in \eqref{eq:Z_{hh'}(n)/n_lower_bound_1} is bounded from below (say by a constant $C<0$). We then have
 \begin{align}
 	\frac{1}{n}Z_{hh'}(n)&\geq \sum\limits_{(\underline{d},\underline{i})\in\mathbb{S}}\frac{N(n,\underline{d},\underline{i},h)}{n}\,D(P_1^{d_h}(\cdot|i_h)\|P_2^{d_h}(\cdot|i_h))+\frac{N(n,\underline{d},\underline{i},h')}{n}\,D(P_2^{d_{h'}}(\cdot|i_{h'})\|P_1^{d_{h'}}(\cdot|i_{h'})) + C\varepsilon\nonumber\\
 	&\geq \frac{N(n,\underline{d},\underline{i},h)}{n}\,D(P_1^{d_h}(\cdot|i_h)\|P_2^{d_h}(\cdot|i_h))+\frac{N(n,\underline{d},\underline{i},h')}{n}\,D(P_2^{d_{h'}}(\cdot|i_{h'})\|P_1^{d_{h'}}(\cdot|i_{h'})) + C\varepsilon\label{eq:Z_{hh'}(n)/n_lower_bound_2}
 \end{align} 
for all $(\underline{d},\underline{i})\in\mathbb{S}$ and
for all $n\geq N_\varepsilon$, almost surely. Now, fix an arbitrary $(\underline{d},\underline{i})\in\mathbb{S}$ such that $d_1>d_2>\cdots>d_K=1$. From \eqref{eq:frac{N(n,d,i,a)}{n}_is_lower_bounded_1}, we know that there exist constants $N_h, N_{h'}$ sufficiently large such that
\begin{equation}
	\frac{N(n,\underline{d},\underline{i},h)}{n}\geq \frac{\eta}{K}\cdot \frac{p(\underline{d},\underline{i})}{2(M+d_1)}-\varepsilon,\quad \frac{N(n,\underline{d},\underline{i},h')}{n}\geq \frac{\eta}{K}\cdot \frac{p(\underline{d},\underline{i})}{2(M+d_1)}-\varepsilon\label{eq:N(n,d,i,h)/n_and_N(n,d,i,h')/n_lower_bound_2}
\end{equation}
for all $n\geq \max\{N_h,N_{h'},N_\varepsilon\}$, almost surely. 
Combining \eqref{eq:N(n,d,i,h)/n_and_N(n,d,i,h')/n_lower_bound_2} and \eqref{eq:Z_{hh'}(n)/n_lower_bound_2}, we may choose $\varepsilon>0$ appropriately so that the right-hand side of \eqref{eq:Z_{hh'}(n)/n_lower_bound_2} is strictly positive. This establishes the desired result.
\end{IEEEproof}

\section{Proof of Lemma \ref{lem:pi_star(L)_in_Pi(epsilon)}}\label{appndx:proof_of_lem_pi_star(L)_in_Pi(epsilon)}
 The policy $\pi^\star(L,\delta)$ commits error if one of the following events is true:
\begin{enumerate}
	\item The policy never stops in finite time.
	\item The policy stops in finite time and declares $h'\neq h$ as the true index of the odd arm.
\end{enumerate}
The event in item $1$ above has zero probability, thanks to Lemma \ref{lem:liminf_Z_{hh'}(n)_strictly_positive}.
Thus, the probability of error of policy $\pi=\pi^\star(L,\delta)$ may be evaluated as follows: suppose $\mathcal{H}_h$ is the true hypothesis. Then,
\begin{align}
	P_h(\theta(\tau(\pi))\neq h)
	=P_h\bigg(\exists~ n\text{ and }~h'\neq h\text{ such that }
	\theta(\tau(\pi))=h'\text{ and } \tau(\pi)=n\bigg).\label{eq:P_e_partial_1}
\end{align}
We now let
\begin{align}
	\mathcal{R}_{h'}(n)\coloneqq\{\omega:\tau(\pi)(\omega)=n,\,\theta(\tau(\pi))(\omega)=h'\}\label{eq:R_{h'}(n)}
\end{align}
denote the set of all sample paths for which the policy stops at time $n$ and declares $h'\neq h$ as the true index of the odd arm. Clearly, $\{\mathcal{R}_{h'}(n):h'\neq h,\,n\geq 0\}$ is a collection of mutually disjoint sets. Therefore, we have
\begin{align}
P_h(\theta(\tau(\pi))\neq h) &=P_h\left(\bigcup\limits_{h'\neq h}\,\bigcup\limits_{n=0}^{\infty}\mathcal{R}_{h'}(n)\right)\nonumber\\
&= \sum\limits_{h'\neq h}\sum\limits_{n=0}^{\infty}P_h(\tau(\pi)=n,~\theta(\tau(\pi))=h')\nonumber\\
&= \sum\limits_{h'\neq h}\sum\limits_{n=0}^{\infty}~\int\limits_{\mathcal{R}_{h'}(n)}\,dP_h(\omega)\nonumber\\
&\stackrel{(a)}{=}\sum\limits_{h'\neq h}\sum\limits_{n=0}^{\infty}~\int\limits_{\mathcal{R}_{h'}(n)}\exp(Z_h(n)(\omega))\quad d(B^n(\omega),A^n(\omega),\bar{X}^n(\omega))\nonumber\\
&\stackrel{(b)}{=}\sum\limits_{h'\neq h}\sum\limits_{n=0}^{\infty}~\int\limits_{\mathcal{R}_{h'}(n)}\exp({-Z_{h'h}(n)(\omega)})\quad {\exp}(Z_{h'}(n)(\omega))\quad d(B^n(\omega),A^n(\omega),\bar{X}^n(\omega))\nonumber\\
&\stackrel{(c)}{\leq} \sum\limits_{h'\neq h}\sum\limits_{n=0}^{\infty}~\bigg\lbrace\int\limits_{\mathcal{R}_{h'}(n)}\frac{1}{(K-1)L}~dP_{h'}(\omega)\bigg\rbrace\nonumber\\
&=\sum\limits_{h'\neq h}\frac{1}{(K-1)L}~P_{h'}\left(\bigcup\limits_{n=0}^{\infty}\mathcal{R}_{h'}(n)\right){\leq}~ \frac{1}{L},
\end{align}
where in $(a)$ above, $$Z_{h}(n)\coloneqq \log P_h(B^n,A^n,\bar{X}^n)$$ denotes the log-likelihood of all the intended arm pulls, the actual arm pulls and the observations up to time $n$ under the hypothesis $\mathcal{H}_h$, $(b)$ above follows by noting that $Z_{hh'}(n)=Z_{h}(n)-Z_{h'}(n)=-Z_{h'h}(n)$, and $(c)$ follows from the fact that when $\mathcal{H}_{h'}$ is the true hypothesis, the condition $M_{h'}(n)\geq \log((K-1)L)$ is satisfied when the policy $\pi=\pi^\star(L, \delta)$ stops at time $\tau(\pi)=n$, which in particular implies that $Z_{h'h}(n)\geq \log((K-1)L)$. Finally, setting $L=1/\epsilon$ yields the desired result. This completes the proof of the lemma.

\section{Proof of Proposition \ref{prop:upper_bound}}\label{appndx:proof_of_upper_bound}
This section is organised as follows. First, we show in Proposition \ref{prop:Z_{hh'}(n)_has_the_right_drift} that under the policy $\pi^\star(L,\delta)$, the test statistic $M_h(n)$ has the correct drift, one that comes from the ergodic occupancy measure corresponding to $\pi^{\lambda_{h,\delta}}$ when $\mathcal{H}_h$ is the true hypothesis. We then show in Lemma \ref{Lemma:stopping_time_of_policy_goes_to_infinity} that the stopping time of the policy $\pi^\star(L,\delta)$ grows with $L$ (i.e., lower probability of error implies more time required to stop and declare the odd arm location correctly with high confidence). More specifically, we show in Lemma \ref{Lemma:almost_sure_upper_bound_for_policy_pi_star} that ratio $\tau(\pi)/\log L$ has, in the limit as $L\to\infty$, an almost sure upper bound that matches with the right-hand side of \eqref{eq:upper_bound}. Finally, we prove in Proposition \ref{prop:uniform_integrability} that the family $\{\tau(\pi)/\log L:L>1\}$ is uniformly integrable. The almost sure upper bound of Lemma \ref{Lemma:almost_sure_upper_bound_for_policy_pi_star} combined with uniform integrability result of Proposition \ref{prop:uniform_integrability} yields the desired upper bound in \eqref{eq:upper_bound}.

\begin{prop}\label{prop:Z_{hh'}(n)_has_the_right_drift}
	Fix an arbitrary $L>1$, $\delta>0$ and $h\in\mathcal{A}$, and let $\mathcal{H}_h$ be the true hypothesis. For every $h'\neq h$, under the non-stopping version of policy $\pi^\star(L,\delta)$, we have, almost surely,
	\begin{equation}
		\lim\limits_{n\to\infty}\frac{Z_{hh'}(n)}{n}=\sum\limits_{(\underline{d},\underline{i})\in\mathbb{S}}\nu^{\lambda_{h,\delta}}(\underline{d},\underline{i},h)\,D(P_1^{d_h}(\cdot|i_h)\|P_2^{d_{h}}(\cdot|i_h))+\nu^{\lambda_{h,\delta}}(\underline{d},\underline{i},h')\,D(P_2^{d_{h'}}(\cdot|i_{h'})\|P_1^{d_{h'}}(\cdot|i_{h'})).\label{eq:limit_Z_{hh'}(n)/n}
	\end{equation}
	Consequently, it follows that almost surely,
	\begin{equation}
		\lim\limits_{n\to\infty}\frac{M_h(n)}{n}=\min\limits_{h'\neq h}\sum\limits_{(\underline{d},\underline{i})\in\mathbb{S}}\nu^{\lambda_{h,\delta}}(\underline{d},\underline{i},h)\,D(P_1^{d_h}(\cdot|i_h)\|P_2^{d_{h}}(\cdot|i_h))+\nu^{\lambda_{h,\delta}}(\underline{d},\underline{i},h')\,D(P_2^{d_{h'}}(\cdot|i_{h'})\|P_1^{d_{h'}}(\cdot|i_{h'})).\label{eq:limit_M_h(n)/n}
	\end{equation}
\end{prop}

\begin{IEEEproof}[Proof of Proposition \ref{prop:Z_{hh'}(n)_has_the_right_drift}]
From Lemma \ref{lem:liminf_Z_{hh'}(n)_strictly_positive}, it follows that when $\mathcal{H}_h$ is the true hypothesis,
\begin{equation}
	\liminf\limits_{n\to\infty}\frac{M_h(n)}{n}=\liminf\limits_{n\to\infty}\,\min\limits_{h'\neq h}\frac{Z_{hh'}(n)}{n}>0\quad \text{almost surely}.
\end{equation}
This in turn implies that $\liminf\limits_{n\to\infty}M_h(n)>0$ almost surely. An immediate consequence of this is that for any $h'\neq h$, almost surely,
\begin{align}
	\limsup\limits_{n\to\infty}M_{h'}(n)&=\limsup\limits_{n\to\infty}\min\limits_{a\neq h'}Z_{h'a}(n)\nonumber\\
	&\leq \limsup\limits_{n\to\infty}Z_{h'h}(n)\nonumber\\
	&=\limsup\limits_{n\to\infty}-Z_{hh'}(n)\nonumber\\
	&= -\liminf\limits_{n\to\infty}Z_{hh'}(n)\nonumber\\
	&\leq -\liminf\limits_{n\to\infty}M_{h}(n)\nonumber\\
	&<0.\label{eq:limsup_M_{h'}(n)_less_than_0}
\end{align}
The above set of inequalities imply the following important result: suppose $\mathcal{H}_h$ is the true hypothesis. Then, for any $L>1$ and $\delta>0$, under the non-stopping version of policy $\pi^\star(L,\delta)$, we have
\begin{equation}
	\theta(n)=h\quad \text{for all sufficiently large values of }n,\text{ almost surely}. \label{eq:theta(n)=h_for_all_n_large}
\end{equation}
The condition in \eqref{eq:theta(n)=h_for_all_n_large} implies that for all $(\underline{d},\underline{i},a)\in\mathbb{S}\times\mathcal{A}$,
\begin{align}
	\lim\limits_{n\to\infty} P(A_n=a|\underline{d}(n)=\underline{d},\underline{i}(n)=\underline{i},\{(\underline{d}(t),\underline{i}(t)):K\leq t< n\})&=\lim\limits_{n\to\infty}\frac{\eta}{K}+(1-\eta)\,\lambda_{\theta(n),\delta}(a|\underline{d},\underline{i})\nonumber\\
	&=\frac{\eta}{K}+(1-\eta)\,\lambda_{h,\delta}(a|\underline{d},\underline{i}).
\end{align}
Thus, we observe that because the arms are selected according to $\lambda_{\theta(n), \delta}(\cdot\mid \cdot)$ in the beginning, the non-stopping version of policy $\pi^\star(L, \delta)$ may not be regarded as an SRS policy (since $\theta(n)$ is, in general, a function of the entire history up to time $n$). However, $\theta(n)=h$ for all sufficiently large values of $n$, and therefore the non-stopping version of policy $\pi^\star(L, \delta)$ eventually turns into an SRS policy. As an immediate consequence of this, we have the following almost sure convergences as $n\to\infty$:
\begin{align}
	\frac{N(n,\underline{d},\underline{i},a)}{N(n,\underline{d},\underline{i})}&\longrightarrow \frac{\eta}{K}+(1-\eta)\,\lambda_{h,\delta}(a|\underline{d},\underline{i}),\label{eq:limit_N(n,d,i,a)/N(n,d,i)}\\
	\frac{N(n,\underline{d},\underline{i})}{n}&\longrightarrow \mu^{\lambda_{h,\delta}}(\underline{d},\underline{i}).\label{eq:limit_N(n,d,i)/n}
\end{align}
It now follows that for any $h'\neq h$, almost surely,
\begin{align}
	&\lim\limits_{n\to\infty}\frac{Z_{hh'}(n)}{n}\nonumber\\
	&=\lim\limits_{n\to\infty}\sum\limits_{(\underline{d},\underline{i})\in\mathbb{S}}\sum\limits_{j\in\mathcal{S}}\frac{N(n,\underline{d},\underline{i},h,j)}{n}\,\log\frac{P_1^{d_h}(j|i_h)}{P_2^{d_h}(j|i_h)}+\frac{N(n,\underline{d},\underline{i},h',j)}{n}\,\log\frac{P_2^{d_{h'}}(j|i_{h'})}{P_1^{d_{h'}}(j|i_{h'})}\nonumber\\
	&=\lim\limits_{n\to\infty}\sum\limits_{(\underline{d},\underline{i})\in\mathbb{S}}\sum\limits_{j\in\mathcal{S}}\left(\frac{N(n,\underline{d},\underline{i})}{n}\right)\left(\frac{N(n,\underline{d},\underline{i},h)}{N(n,\underline{d},\underline{i})}\right)\left(\frac{N(n,\underline{d},\underline{i},h,j)}{N(n,\underline{d},\underline{i},h)}\right)\,\log\frac{P_1^{d_h}(j|i_h)}{P_2^{d_h}(j|i_h)}\nonumber\\
	&\hspace{2cm}+\lim\limits_{n\to\infty}\sum\limits_{(\underline{d},\underline{i})\in\mathbb{S}}\sum\limits_{j\in\mathcal{S}}\left(\frac{N(n,\underline{d},\underline{i})}{n}\right)\left(\frac{N(n,\underline{d},\underline{i},h')}{N(n,\underline{d},\underline{i})}\right)\left(\frac{N(n,\underline{d},\underline{i},h',j)}{N(n,\underline{d},\underline{i},h')}\right)\,\log\frac{P_2^{d_{h'}}(j|i_{h'})}{P_1^{d_{h'}}(j|i_{h'})}.\label{eq:lim_Z_{hh'}(n)/n_1}
\end{align}
Note that in each of the logarithmic terms in \eqref{eq:lim_Z_{hh'}(n)/n_1}, when either the numerator or the denominator is equal to $0$, the corresponding coefficient term is also equal to $0$. Thus, we may assume without loss of generality that each term inside the summations in \eqref{eq:lim_Z_{hh'}(n)/n_1} is nonzero for all values of the summation indices. Under this assumption, it follows from the convergences in  \eqref{eq:convergence_theorem_convergences} that the logarithmic terms in \eqref{eq:lim_Z_{hh'}(n)/n_1} are bounded.
Using the dominated convergence theorem to pass the limit inside the summation in each of the terms, and using the results in \eqref{eq:limit_for_N(n,d,i,a,j)/N(n,d,i,a)}, \eqref{eq:limit_N(n,d,i,a)/N(n,d,i)} and \eqref{eq:limit_N(n,d,i)/n}, we arrive at the desired result.
\end{IEEEproof}

We now show that the stopping time of policy $\pi^\star(L,\delta)$ grows with $L$. 
\begin{Lemma}\label{Lemma:stopping_time_of_policy_goes_to_infinity}
	Fix $h\in\mathcal{A}$ and $\delta>0$, and suppose that $\mathcal{H}_h$ is the true hypothesis. Then, under policy $\pi=\pi^{\star}(L,\delta)$, we have
	\begin{equation}
		\liminf\limits_{L\to\infty}\tau(\pi)=\infty\text{ almost surely.}\label{eq:stopping_time_of_policy_goes_to_infinity}
	\end{equation}
\end{Lemma}
\begin{IEEEproof}[Proof of Lemma \ref{Lemma:stopping_time_of_policy_goes_to_infinity}]
Assume without loss of generality that the policy $\pi=\pi^{\star}(L,\delta)$ pulls arm $1$ at time $t=0$, arm $2$ at time $t=1$ and so on until arm $K$ at time $t=K-1$. In order to prove the Lemma, we note that it suffices to prove the following statement:
\begin{equation}
	\text{for each $m\geq K$,}\quad \lim\limits_{L\to\infty}P_h(\tau(\pi)\leq m)=0.
\end{equation}
Fix $m\geq K$, and note that
\begingroup\allowdisplaybreaks\begin{align}
	\limsup\limits_{L\to\infty}\,P_h(\tau(\pi)\leq m)
	&=\limsup\limits_{L\to\infty}\,P_h\bigg(\exists~K\leq n\leq m \text{ and }\tilde{h}\in\mathcal{A}
	\text{ such that }M_{\tilde{h}}(n)>\log((K-1)L)\bigg)\nonumber\\
	&\leq \limsup\limits_{L\to\infty}\sum\limits_{\tilde{h}\in\mathcal{A}}\sum\limits_{n=K}^{m}P_h(M_{\tilde{h}}(n)>\log((K-1)L))\nonumber\\
	&\leq \limsup\limits_{L\to\infty}\frac{1}{\log((K-1)L)}\sum\limits_{\tilde{h}\in\mathcal{A}}\sum\limits_{n=K}^{m}E_h[M_{\tilde{h}}(n)],\label{eq:stop_time_goes_to_infty_1}
\end{align}\endgroup
where the first inequality above follows from the union bound, and the second inequality is due to Markov's inequality.

We now show that for each $n\in\{K,\ldots,m\}$, the expectation term inside the summation in \eqref{eq:stop_time_goes_to_infty_1} is finite. This will then imply that the limit supremum on the right-hand side of \eqref{eq:stop_time_goes_to_infty_1} is equal to $0$, thus proving the desired result. Note that
\begingroup\allowdisplaybreaks\begin{align}
	M_{\tilde{h}}(n)=\min\limits_{h'\neq \tilde{h}}Z_{\tilde{h}h'}(n)\leq Z_{\tilde{h}h'}(n)\text{ for all }h'\neq \tilde{h}.\label{eq:mod_glr_upper_bounded_by_glr}
\end{align}\endgroup
Fix an arbitrary $h'\neq \tilde{h}$. Then, almost surely,
\begin{align}
	Z_{\tilde{h}h'}(n)&=\sum\limits_{(\underline{d},\underline{i})\in\mathbb{S}}\sum\limits_{j\in\mathcal{S}}{N(n,\underline{d},\underline{i},\tilde{h},j)}\,\log\frac{P_1^{d_{\tilde{h}}}(j|i_{\tilde{h}})}{P_2^{d_{\tilde{h}}}(j|i_{\tilde{h}})}+{N(n,\underline{d},\underline{i},h',j)}\,\log\frac{P_2^{d_{h'}}(j|i_{h'})}{P_1^{d_{h'}}(j|i_{h'})}\nonumber\\
	&\leq n\max\bigg\lbrace\max\left\lbrace \log\frac{P_1^{d}(j|i)}{P_2^{d}(j|i)}:d\in\mathbb{N},i,j\in\mathcal{S}\right\rbrace, \max\left\lbrace \log\frac{P_2^{d}(j|i)}{P_1^{d}(j|i)}:d\in\mathbb{N},i,j\in\mathcal{S}\right\rbrace\bigg\rbrace.\label{eq:Z_{h_tilde_h'}(n)_upper_bound}
\end{align}
From to the convergences in \eqref{eq:convergence_theorem_convergences}, we note that the coefficient of $n$ in  \eqref{eq:Z_{h_tilde_h'}(n)_upper_bound} is finite. Thus, it follows that $E[M_{\tilde{h}}(n)]\leq E[Z_{\tilde{h}h'}(n)]\leq nC$ for all $h'\neq \tilde{h}$, where $C<\infty$ represents the constant multiplying $n$ in \eqref{eq:Z_{h_tilde_h'}(n)_upper_bound}.
\end{IEEEproof}

Going further, let $R_{\lambda_{h,\delta}}$ denote the right-hand side of \eqref{eq:limit_M_h(n)/n}.
\begin{Lemma}\label{Lemma:almost_sure_upper_bound_for_policy_pi_star}
Fix $h\in\mathcal{A}$ and $\delta>0$, and suppose that $\mathcal{H}_h$ is the true hypothesis. Then, under policy $\pi=\pi^*(L,\delta)$, we have
\begingroup\allowdisplaybreaks\begin{align}
	\limsup\limits_{L\to\infty}\, \frac{\tau(\pi)}{\log L}\leq  \frac{1}{R_{\lambda_{h,\delta}}}\quad \text{almost surely}.\label{eq:almost_sure_upper_bound_for_policy_pi_star}
\end{align}\endgroup	
\end{Lemma}

\begin{IEEEproof}[Proof of Lemma \ref{Lemma:almost_sure_upper_bound_for_policy_pi_star}]
Note that as a consequence of Proposition \ref{prop:Z_{hh'}(n)_has_the_right_drift} and Lemma \ref{Lemma:stopping_time_of_policy_goes_to_infinity}, we have 
\begin{equation}
		\lim\limits_{L\to\infty}\frac{M_{h}(\tau(\pi))}{\tau(\pi)}=
		R_{\lambda_{h,\delta}}\quad \text{almost surely}.\label{eq:M_h(N(pi))/N(pi)_has_almost_correct_drift}
	\end{equation}
We now show that for any $h'\neq h$ and $n\geq K$, the increment $Z_{hh'}(n)-Z_{hh'}(n-1)$ is bounded almost surely. Observe that, almost surely,
\begin{align}
	& Z_{hh'}(n)-Z_{hh'}(n-1)\nonumber\\
	&=\log\frac{P_h(A^n,\bar{X}^n)}{P_{h'}(A^n,\bar{X}^n)}-\log\frac{P_h(A^{n-1},\bar{X}^{n-1})}{P_{h'}(A^{n-1},\bar{X}^{n-1})}\nonumber\\
	&=\log\frac{P_h^{A_n}(\bar{X}_n|A^{n-1},\bar{X}^{n-1})}{P_{h'}^{A_n}(\bar{X}_n|A^{n-1},\bar{X}^{n-1})}\nonumber\\
	&=\sum\limits_{(\underline{d},\underline{i})\in\mathbb{S}}~\sum\limits_{a=1}^{K}\sum\limits_{j\in\mathcal{S}}\mathbb{I}_{\{\underline{d}(n)=\underline{d},\underline{i}(n)=\underline{i},A_n=a,X_n^a=j\}}\,\log\frac{(P_h^a)^{d_a}(j|i_a)}{(P_{h'}^a)^{d_a}(j|i_a)}\nonumber\\
	&=\sum\limits_{(\underline{d},\underline{i})\in\mathbb{S}}\sum\limits_{j\in\mathcal{S}}\bigg[\mathbb{I}_{\{\underline{d}(n)=\underline{d},\underline{i}(n)=\underline{i},A_n=h,X_n^h=j\}}\,\log\frac{P_1^{d_h}(j|i_h)}{P_2^{d_h}(j|i_h)}+\mathbb{I}_{\{\underline{d}(n)=\underline{d},\underline{i}(n)=\underline{i},A_n=h',X_n^{h'}=j\}}\,\log\frac{P_2^{d_{h'}}(j|i_{h'})}{P_1^{d_{h'}}(j|i_{h'})}\bigg].\label{eq:Z_{hh'}(n)-Z_{hh'}(n-1)_upper_bound}
\end{align}
We now note that whenever either the numerator or the denominator of the logarithmic terms in \eqref{eq:Z_{hh'}(n)-Z_{hh'}(n-1)_upper_bound} is equal to $0$, then the corresponding indicator function is also equal to $0$. This, together with the convergences in \eqref{eq:convergence_theorem_convergences}, implies that the right-hand side of \eqref{eq:Z_{hh'}(n)-Z_{hh'}(n-1)_upper_bound} is bounded. This, together with the collection  $\{Z_{hh'}(n)-Z_{hh'}(n-1):1\leq n\leq K-1\}$ of finitely many terms, each of which is finite almost surely, establishes the almost sure boundedness of the increments $Z_{hh'}(n)-Z_{hh'}(n-1)$ for all $n\geq 1$ and all $h'\neq h$. 

When $\mathcal{H}_h$ is the true hypothesis, we note from the definition of stopping time $\tau(\pi)$ that $M_{h}(\tau(\pi)-1)<\log ((K-1)L)$, which implies that there exists $h''
\neq h$ such that $Z_{hh''}(\tau(\pi)-1)<\log((K-1)L)$. Using this, we have
\begingroup\allowdisplaybreaks\begin{align}
	\limsup\limits_{L\to\infty}\frac{M_{h}(\tau(\pi))}{\log L}&=\limsup\limits_{L\to\infty}\min\limits_{h'\neq h}\frac{Z_{hh'}(\tau(\pi))}{\log L}\nonumber\\
	&\leq \limsup\limits_{L\to\infty}\frac{Z_{hh''}(\tau(\pi))}{\log L}\nonumber\\
	&\stackrel{(a)}{=}\limsup\limits_{L\to\infty}\frac{Z_{hh''}(\tau(\pi)-1)}{\log L}\nonumber\\
	&\leq \limsup\limits_{L\to\infty}\frac{\log((K-1)L)}{\log L}\nonumber\\
	&=1\quad \text{almost surely},\label{eq:limsup_M_hh'(tau)/log_L}
\end{align}\endgroup	
where $(a)$ above is due to the almost sure boundedness of the increments established earlier. Then, using \eqref{eq:M_h(N(pi))/N(pi)_has_almost_correct_drift} along with \eqref{eq:limsup_M_hh'(tau)/log_L} yields
\begingroup\allowdisplaybreaks\begin{align}
	\limsup\limits_{L\to\infty}\frac{\tau(\pi)}{\log L}&=\limsup\limits_{L\to\infty}\bigg\lbrace\left(\frac{\tau(\pi)}{M_{h}(\tau(\pi))}\right)\left(\frac{M_{h}(\tau(\pi))}{\log L}\right)\bigg\rbrace \nonumber\\
	&=\left(\lim\limits_{L\to\infty}\frac{\tau(\pi)}{M_{h}(\tau(\pi))}\right)\left(\limsup\limits_{L\to\infty}\frac{M_{h}(\tau(\pi))}{\log L}\right)\nonumber\\
	&\leq \frac{1}{R_{\lambda_{h,\delta}}}\quad \text{almost surely},
\end{align}\endgroup
thus completing the proof of the lemma.
\end{IEEEproof}
Since, by definition, $R_{\lambda_{h,\delta}}>\frac{R^*(P_1,P_2)}{1+\delta}$, it follows that 
\begin{equation}
	\limsup\limits_{L\to\infty}\frac{\tau(\pi)}{\log L}\leq \frac{1+\delta}{R^*(P_1,P_2)}\quad \text{almost surely}.\label{eq:a.s._upper_bound_for_tau/logL}
\end{equation}

We now prove that the family $\{\tau(\pi^\star(L, \delta))/\log L:L>1\}$ is uniformly integrable for all $\delta>0$. This, along with the almost sure upper bound of \eqref{eq:a.s._upper_bound_for_tau/logL} yields the desired upper bound of \eqref{eq:upper_bound}.

\begin{prop}\label{prop:uniform_integrability}
	For any fixed $\delta>0$, the family of random variables $\{\tau(\pi^\star(L,\delta))/\log L:L>1\}$ is uniformly integrable.
\end{prop}

\begin{IEEEproof}[Proof of Proposition \ref{prop:uniform_integrability}]
Fix $h\in\mathcal{A}$, and suppose that $\mathcal{H}_h$ is the true hypothesis. Then, in order to establish the desired uniform integrability, it suffices to show that
	\begin{equation}
	\limsup\limits_{L\to\infty}E_h\bigg[\exp\bigg(\frac{\tau(\pi)}{\log L}\bigg)\bigg]<\infty.
\end{equation}
Towards this, let us first define
\begin{equation}
	D_{hh'}\coloneqq \sum\limits_{(\underline{d},\underline{i})\in\mathbb{S}}\,\,\sum\limits_{a=1}^{K}\nu^{\lambda_{h,\delta}}(\underline{d},\underline{i},a)\,D((P_h^a)^{d_a}(\cdot\mid i_a)\|(P_{h'}^a)^{d_a}(\cdot\mid i_a)).\label{eq:D_{hh'}}
\end{equation}
Let
\begin{equation}
	\tilde{n}(L)\coloneqq \frac{4\log((K-1)L)}{D_{hh'}}+K-1,\label{eq:n_tilde}
\end{equation}
and let
\begin{equation}
	u(L)\coloneqq \exp\left(\frac{1+\tilde{n}(L)}{\log L}\right).\label{eq:u(L)}
\end{equation}
Let $\pi^\star_h=\pi^\star_h(L,\delta)$ denote the version of policy $\pi^{\star}(L,\delta)$ that stops only upon declaring $h$ as the index of the odd arm. Clearly, $\tau(\pi^\star_h)\geq \tau(\pi)$ a.s.. Then,
\begingroup\allowdisplaybreaks\begin{align}
	\limsup\limits_{L\to\infty}E_h\bigg[\exp\bigg(\frac{\tau(\pi)}{\log L}\bigg)\bigg]&=\limsup\limits_{L\to\infty}\int\limits_{0}^{\infty}P_h\bigg(\frac{\tau(\pi)}{\log L}>\log x\bigg)\,dx\nonumber\\
	&\leq \limsup\limits_{L\to\infty}\int\limits_{0}^{\infty}P_h\bigg({\tau(\pi^\star_h)}\geq \lceil(\log x)({\log L})\rceil\bigg)\,dx\nonumber\\
	&\stackrel{(a)}{\leq} \limsup\limits_{L\to\infty}\bigg\lbrace u(L)+\int\limits_{u(L)}^{\infty}P_h\bigg({\tau(\pi^\star_h)}\geq \lceil(\log x)({\log L})\rceil\bigg)\,dx\bigg\rbrace\nonumber\\
	&= \exp\left(\frac{4}{D_{hh'}}\right)+\limsup\limits_{L\to\infty}\sum\limits_{n\geq \tilde{n}(L)}\exp\bigg(\frac{n+1}{\log L}\bigg)\,P_h(M_h(n)<\log((K-1)L)),\label{eq:uniform_integrability_1}
\end{align}\endgroup
where $(a)$ above follows by upper bounding the probability term by $1$ for all $x\leq u(L)$. In Lemma \ref{lem:exp_upper_bound}, we show that the probability term in \eqref{eq:uniform_integrability_1} has an exponential upper bound. It then follows that this exponential upper bound results in the finiteness of the right-hand side of \eqref{eq:uniform_integrability_1}, thus completing the proof of the proposition. 
\end{IEEEproof}

\section{An Exponential Upper Bound for $P_h(M_h(n)<\log((K-1)L))$} \label{appndx:exp_upper_bound_for_a_certain_term}
We now demonstrate the stated exponential upper bound used in \eqref{eq:uniform_integrability_1}.

\begin{Lemma}\label{lem:exp_upper_bound}
	Fix $\delta>0$ and $h\in\mathcal{A}$, and suppose that $\mathcal{H}_h$ is the true hypothesis. There exist constants $B>0$ and $0<\theta<\infty$ independent of $L$ such for all $n\geq \tilde{n}(L)$, 
	\begin{equation}
		P_h(M_h(n)<\log((K-1)L))\leq Be^{-n\theta}.\label{eq:exp_upper_bound} 
	\end{equation}
\end{Lemma}

\begin{IEEEproof}[Proof of Lemma \ref{lem:exp_upper_bound}]
Since
\begingroup\allowdisplaybreaks\begin{align}
	P_h(M_h(n)<\log((K-1)L))
	&=P_h\left(\min\limits_{h'\neq h}Z_{hh'}(n)<\log((K-1)L)\right)\nonumber\\
	&\leq \sum\limits_{h'\neq h}P_h\left(Z_{hh'}(n)<\log((K-1)L)\right);\label{eq:exp_bound_1}
\end{align}\endgroup
the last line above follows from the union bound. In order to prove the lemma, it suffices to show that each term inside the summation in \eqref{eq:exp_bound_1} is exponentially bounded.  

Fix $h'\neq h$. Recall that under the hypothesis $\mathcal{H}_h$, the transition probability matrix of arm $h$ is $P_1$, while that of arm $h'$ is $P_2$, where $P_2\neq P_1$. The latter condition of $P_2\neq P_1$ implies that there exists $i^*\in\mathcal{S}$ such that $P_1(\cdot|i^*)\neq P_2(\cdot|i^*)$. Equivalently, we have $$D(P_1(\cdot|i^*)\|P_2(\cdot|i^*))>0, \quad D(P_2(\cdot|i^*)\|P_1(\cdot|i^*))>0.$$ Going further, let us fix an arbitrary $(\underline{d}^*,\underline{i}^*)\in\mathbb{S}$ such that $d_h^*=1$ and $i_h^*=i^*$, where $i^*$ is as defined above.

For $n\geq K$, let
\begin{align}
	\Delta Z_{hh'}(n)&\coloneqq Z_{hh'}(n)-Z_{hh'}(n-1)\nonumber\\
					 &=\sum\limits_{(\underline{d},\underline{i})\in\mathbb{S}}\sum\limits_{a=1}^{K}\sum\limits_{j\in\mathcal{S}}\mathbb{I}_{\{\underline{d}(n)=\underline{d},\underline{i}(n)=\underline{i},A_n=a,X^a_n=j\}}\log\frac{(P_h^a)^{d_a}(j|i_a)}{(P_{h'}^a)^{d_a}(j|i_a)}\label{eq:delta_Z_{hh'}(n)}
\end{align}
denote the increment of the log-likelihood process of all the intended arm pulls, actual arm pulls and observations under hypothesis $\mathcal{H}_h$ with respect to those under hypothesis $\mathcal{H}_{h'}$;
 note that $\Delta Z_{h'h}(n)=-\Delta Z_{hh'}(n)$. We then have the following key property satisfied by $\Delta Z_{h'h}(n)$.
 \begin{Lemma}\label{lem:mgf_delta_Z_{hh'}(n)_strictly_less_than_1}
 	For any $(\underline{d},\underline{i})\in \mathbb{S}$, $a\in\mathcal{A}$ and $0<s<1$, we have
 	\begin{equation}
 		E_h\left[e^{s\Delta Z_{h'h}(n)}\bigg|A_n=a,\underline{d}(n)=\underline{d},\underline{i}(n)=\underline{i}\right]\leq 1\quad \forall n,\label{eq:mgf_delta_Z_{hh'}(n)_strictly_less_than_1} 
  	\end{equation}
 	with strict inequality in \eqref{eq:mgf_delta_Z_{hh'}(n)_strictly_less_than_1} if $(\underline{d},\underline{i})=(\underline{d}^*,\underline{i}^*)$ and $a=h$.
 \end{Lemma}
 \begin{IEEEproof}[Proof of Lemma \ref{lem:mgf_delta_Z_{hh'}(n)_strictly_less_than_1}]
 Note that
 \begin{align}
 	E_h\left[e^{s\Delta Z_{h'h}(n)}\bigg|A_n=h,\underline{d}(n)=\underline{d},\underline{i}(n)=\underline{i}\right]&=\sum\limits_{j\in\mathcal{S}} \left(\frac{(P_{h'}^a)^{d_a}(j|i_a)}{(P_h^a)^{d_a}(j|i_a)}\right)^s\,P_h(X_n^h=j|A_n=a,\underline{d}(n)=\underline{d},\underline{i}(n)=\underline{i})\nonumber\\
 	&=\sum\limits_{j\in\mathcal{S}} \left(\frac{(P_{h'}^a)^{d_a}(j|i_a)}{(P_h^a)^{d_a}(j|i_a)}\right)^s\,(P_h^a)^{d_a}(j|i_a)\nonumber\\
 	&=\sum\limits_{j\in\mathcal{S}} ((P_h^a)^{d_a}(j|i_a))^{1-s}\,\,((P_{h'}^a)^{d_a}(j|i_a))^s\nonumber\\
 	&\stackrel{(a)}{\leq} \left(\sum\limits_{j\in\mathcal{S}}(P_h^a)^{d_a}(j|i_a)\right)^{1-s}\cdot \left(\sum\limits_{j\in\mathcal{S}}(P_{h'}^a)^{d_a}(j|i_a)\right)^s\nonumber\\
 	&=1,
 \end{align}
 where $(a)$ above is due to H\"{o}lder's inequality, and the last line follows from the fact that $(P_h^a)^{d_a}(\cdot|i_a)$ and $(P_{h'}^a)^{d_a}(\cdot|i_a)$ are probability distributions on $\mathcal{S}$. When $(\underline{d},\underline{i})=(\underline{d}^*,\underline{i}^*)$ and $a=h$, the inequality in $(a)$ is a strict inequality since $(P_h^a)^{d_a}(\cdot|i_a)=P_1(\cdot|i^*)$ and $(P_{h'}^a)^{d_a}(\cdot|i_a)=P_2(\cdot|i^*)$, and since by the definition of $i^*$, $P_1(\cdot|i^*)\neq P_2(\cdot|i^*)$.
 \end{IEEEproof}
 
As an immediate consequence of Lemma \ref{lem:mgf_delta_Z_{hh'}(n)_strictly_less_than_1}, we have the following result.
\begin{Lemma}\label{lem:mgf_delta_Z_{hh'}(n)_without_A_n_conditioning_strictly_less_than_1}
For any $(\underline{d},\underline{i})\in \mathbb{S}$, $a\in\mathcal{A}$ and $0<s<1$, we have
 	\begin{equation}
 		E_h\left[e^{s\Delta Z_{h'h}(n)}\mid \mathcal{F}_{n-1}\right]\,\mathbb{I}_{\{\underline{d}(n)=\underline{d},\underline{i}(n)=\underline{i}\}}\leq 1\quad \forall n\quad \text{almost surely},\label{eq:mgf_delta_Z_{hh'}(n)_without_A_n_conditioning_strictly_less_than_1} 
  	\end{equation}
 	with strict inequality in \eqref{eq:mgf_delta_Z_{hh'}(n)_strictly_less_than_1} if $(\underline{d},\underline{i})=(\underline{d}^*,\underline{i}^*)$ and $a=h$.
\end{Lemma}
\begin{IEEEproof}[Proof of Lemma \ref{lem:mgf_delta_Z_{hh'}(n)_without_A_n_conditioning_strictly_less_than_1}]
We have, almost surely,
\begin{align}
	&E_h\left[e^{s\Delta Z_{h'h}(n)}\mid \mathcal{F}_{n-1}\right]\,\mathbb{I}_{\{\underline{d}(n)=\underline{d},\underline{i}(n)=\underline{i}\}}=E_h\left[e^{s\Delta Z_{h'h}(n)}\mid\underline{d}(n)=\underline{d},\underline{i}(n)=\underline{i},\mathcal{F}_{n-1}\right]\nonumber\\
	&=\sum\limits_{a=1}^{K} P(A_n=a\mid\underline{d}(n)=\underline{d},\underline{i}(n)=\underline{i},\mathcal{F}_{n-1})\cdot E_h\left[e^{s\Delta Z_{h'h}(n)}\mid A_n=a,\underline{d}(n)=\underline{d},\underline{i}(n)=\underline{i},\mathcal{F}_{n-1}\right]\nonumber\\
	&\stackrel{(a)}=P(A_n=h\mid\underline{d}(n)=\underline{d},\underline{i}(n)=\underline{i},\mathcal{F}_{n-1})\cdot E_h\left[e^{s\Delta Z_{h'h}(n)}\mid A_n=h,\underline{d}(n)=\underline{d},\underline{i}(n)=\underline{i}\right]\nonumber\\
	&\hspace{3cm}+\sum\limits_{a\neq h}P(A_n=a\mid \underline{d}(n)=\underline{d},\underline{i}(n)=\underline{i},\mathcal{F}_{n-1})\cdot E_h\left[e^{s\Delta Z_{h'h}(n)}\mid A_n=a,\underline{d}(n)=\underline{d},\underline{i}(n)=\underline{i}\right]\nonumber\\
	&\stackrel{(b)}{\leq}P(A_n=h\mid \underline{d}(n)=\underline{d},\underline{i}(n)=\underline{i},\mathcal{F}_{n-1})\cdot E_h\left[e^{s\Delta Z_{h'h}(n)}\mid A_n=h,\underline{d}(n)=\underline{d},\underline{i}(n)=\underline{i}\right]\nonumber\\
	&\hspace{8cm}+(1-P(A_n=h\mid \underline{d}(n)=\underline{d},\underline{i}(n)=\underline{i},\mathcal{F}_{n-1}))\nonumber\\
	&=P(A_n=h\mid \underline{d}(n)=\underline{d},\underline{i}(n)=\underline{i},\mathcal{F}_{n-1})\cdot \bigg(E_h\left[e^{s\Delta Z_{h'h}(n)}\mid A_n=h,\underline{d}(n)=\underline{d},\underline{i}(n)=\underline{i}\right]-1\bigg) + 1\nonumber\\
	&\stackrel{(c)}{\leq} \frac{\eta}{K}\,\,\bigg(E_h\left[e^{s\Delta Z_{h'h}(n)}\mid A_n=h,\underline{d}(n)=\underline{d},\underline{i}(n)=\underline{i}\right]-1\bigg) + 1\label{eq:temp_1}\\
	&\stackrel{(d)}{\leq} 1,\label{eq:temp_2}
\end{align}
where $(a)$ above follows by noting that $$E_h\left[e^{s\Delta Z_{h'h}(n)}\mid A_n=a,\underline{d}(n)=\underline{d},\underline{i}(n)=\underline{i},\mathcal{F}_{n-1}\right]=E_h\left[e^{s\Delta Z_{h'h}(n)}\mid A_n=a,\underline{d}(n)=\underline{d},\underline{i}(n)=\underline{i}\right],$$ $(b)$ uses the result of Lemma \ref{lem:mgf_delta_Z_{hh'}(n)_strictly_less_than_1}, $(c)$ follows from the fact that for any $n\geq K$, under the policy $\pi^\star(L,\delta)$,
\begin{align*}
	P(A_n=h\mid\underline{d}(n)=\underline{d},\underline{i}(n)=\underline{i},\mathcal{F}_{n-1})&=\frac{\eta}{K}+(1-\eta)\,\lambda_{\theta(n),\delta}(h|\underline{d},\underline{i})\\
	&\geq \frac{\eta}{K},
\end{align*}
and $(d)$ is straightforward.
Clearly, the inequalities in $(b)$, $(c)$ and $(d)$ above are strict when $(\underline{d},\underline{i})=(\underline{d}^*,\underline{i}^*)$ and $a=h$. 
\end{IEEEproof}

Going further, let $c$ denote the constant on the right-hand side of \eqref{eq:temp_1} when $(\underline{d},\underline{i})=(\underline{d}^*,\underline{i}^*)$. From the arguments above, we have $c<1$. Then,
\begin{align}
	&E_h\left[e^{s\Delta Z_{h'h}(n)} \mid \mathcal{F}_{n-1}\right]\nonumber\\
	&=\sum\limits_{(\underline{d},\underline{i})\in\mathbb{S}}E_h\left[e^{s\Delta Z_{h'h}(n)} \mid \mathcal{F}_{n-1}\right]\cdot \mathbb{I}_{\{\underline{d}(n)=\underline{d},\underline{i}(n)=\underline{i}\}}\nonumber\\
	&=c\,\,\mathbb{I}_{\{\underline{d}(n)=\underline{d}^*,\underline{i}(n)=\underline{i}^*\}}+\sum\limits_{(\underline{d},\underline{i})\neq (\underline{d}^*,\underline{i}^*)}E_h\left[e^{s\Delta Z_{h'h}(n)} \mid \underline{d}(n)=\underline{d},\underline{i}(n)=\underline{i},\mathcal{F}_{n-1}\right]\cdot \mathbb{I}_{\{\underline{d}(n)=\underline{d},\underline{i}(n)=\underline{i}\}}\nonumber\\
	&=\begin{cases}
		c,&\underline{d}(n)=\underline{d}^*,\underline{i}(n)=\underline{i}^*,\\
		\leq 1,&\text{otherwise}.
	\end{cases}\label{eq:temp_3}
\end{align}
The above set of inequalities immediately lead us to the following important result.
\begin{Lemma}\label{lem:exp_upper_bound_for_mgf_of_Z_{h'h}(n)}
	For $0<s<1$, 
	\begin{equation}
		E_h\left[e^{s Z_{h'h}(n)}\right]\leq B_1\,e^{-\theta_1 n},\label{eq:exp_upper_bound_for_mgf_of_Z_{h'h}(n)}
	\end{equation}
	where $B_1>0$  and $\theta_1>0$ are constants which depend on $h$, $h'$ and $s$.
\end{Lemma}
\begin{IEEEproof}[Proof of Lemma \ref{lem:exp_upper_bound_for_mgf_of_Z_{h'h}(n)}]
We have
\begin{align}
	E_h\left[e^{s Z_{h'h}(n)}\right]&=E_h[e^{s Z_{h'h}(n-1)}\,E_h[e^{s \Delta Z_{h'h}(n)}\mid \mathcal{F}_{n-1}]]\nonumber\\
	&\stackrel{(a)}{\leq}E_h\left[c^{N(n,\underline{d}^*,\underline{i}^*)}\right]\nonumber\\
	&\stackrel{(b)}{=}E_h\left[c^{N(n,\underline{d}^*,\underline{i}^*)}\,\,;\,\,N(n,\underline{d}^*,\underline{i}^*)>\frac{n\mu^{\lambda_{h,\delta}}(\underline{d}^*,\underline{i}^*)}{2}\right]+E_h\left[c^{N(n,\underline{d}^*,\underline{i}^*)}\,\,;\,\,N(n,\underline{d}^*,\underline{i}^*)\leq\frac{n\mu^{\lambda_{h,\delta}}(\underline{d}^*,\underline{i}^*)}{2}\right]\nonumber\\
	&\leq c^{n\frac{\mu^{\lambda_{h,\delta}}(\underline{d}^*,\underline{i}^*)}{2}}+P_h\left(N(n,\underline{d}^*,\underline{i}^*)\leq\frac{n\mu^{\lambda_{h,\delta}}(\underline{d}^*,\underline{i}^*)}{2}\right).\label{eq:temp_4}
\end{align}
In the above set of equations, $(a)$ follows from by repeatedly applying \eqref{eq:temp_3}, the notation $E[X;A]$ in $(b)$ stands for $E[X\,\mathbb{I}_A]$, and the last line follows by noting that $c^{n\frac{\mu^{\lambda_{h,\delta}}(\underline{d}^*,\underline{i}^*)}{2}}\leq 1$ almost surely. We now note that $\{N(n,\underline{d}^*,\underline{i}^*)-N(K,\underline{d}^*,\underline{i}^*):n\geq K\}$ is a bounded martingale. Using the Azuma-Hoeffding inequality, we then have
\begin{align}
	P_h\left(N(n,\underline{d}^*,\underline{i}^*)\leq\frac{n\mu^{\lambda_{h,\delta}}(\underline{d}^*,\underline{i}^*)}{2}\right)&=P_h\left(N(n,\underline{d}^*,\underline{i}^*)-N(K,\underline{d}^*,\underline{i}^*)\leq\frac{n\mu^{\lambda_{h,\delta}}(\underline{d}^*,\underline{i}^*)}{2}-N(K,\underline{d}^*,\underline{i}^*)\right)\nonumber\\
	&\leq P_h\left(N(n,\underline{d}^*,\underline{i}^*)-N(K,\underline{d}^*,\underline{i}^*)\leq\frac{n\mu^{\lambda_{h,\delta}}(\underline{d}^*,\underline{i}^*)}{2}\right)\nonumber\\
	&\leq \exp\left(-\frac{n(\mu^{\lambda_{h,\delta}}(\underline{d}^*,\underline{i}^*))^2}{8}\right).\label{eq:temp_5}
\end{align}
Plugging \eqref{eq:temp_5} back in \eqref{eq:temp_4}, and noting that $c$ is a function of $s$, we arrive at \eqref{eq:exp_upper_bound_for_mgf_of_Z_{h'h}(n)}.
\end{IEEEproof}

As a consequence of Lemma \ref{lem:exp_upper_bound_for_mgf_of_Z_{h'h}(n)}, we have the following result.
\begin{Lemma}\label{lem:exp_upper_bound_for_min_Z_{hh'}(n)<R}
	Fix an arbitrary $h\in\mathcal{A}$, and suppose that $\mathcal{H}_h$ is the true hypothesis. Consider the non-stopping version of the policy $\pi=\pi^\star(L,\delta)$. There exist constants $C_R$ and $\gamma>0$ such that
	\begin{equation}
		P_h\left(\min\limits_{h'\neq h}Z_{hh'}(n)<R\right)\leq C_R\,\,e^{-\gamma n}.\label{eq:exp_upper_bound_for_min_Z_{hh'}(n)<R}
	\end{equation}
	In \eqref{eq:exp_upper_bound_for_min_Z_{hh'}(n)<R}, $C_R$ is independent of $h$ but $\gamma$ depends on $h$.
\end{Lemma}
\begin{IEEEproof}[Proof of Lemma \ref{lem:exp_upper_bound_for_min_Z_{hh'}(n)<R}]
Observe that 
\begin{align}
	P_h\left(\min\limits_{h'\neq h}Z_{hh'}(n)<R\right)&=P_h\left(\max\limits_{h'\neq h}Z_{h'h}(n)>-R\right)\nonumber\\
	&\leq \sum\limits_{h'\neq h}P_h\left(Z_{h'h}(n)>-R\right)\nonumber\\
	&=\sum\limits_{h'\neq h}P_h\left(sZ_{h'h}(n)>-sR\right)\quad \forall\,\, 0<s<1\nonumber\\
	&\stackrel{(a)}{\leq} \sum\limits_{h'\neq h}e^{sR}\,\,E_h\left[e^{s Z_{h'h}(n)}\right]\nonumber\\
	&\stackrel{(b)}{\leq} e^{sR}\sum\limits_{h'\neq h}B_1\,\,e^{-\theta n}\nonumber\\
	&\leq e^{sR}\cdot (K-1)\cdot\max\limits_{h'\neq h} \,B_1\,e^{-\theta n}\nonumber\\
	&\leq C_R\,\,e^{-\gamma n},
\end{align}
where $\max\limits_{h'\neq h} \,B_1\,e^{-\theta n}=e^{-\gamma}$ and $C_R=Ke^{sR}$. In the above set of equations, $(a)$ is due to Chernoff's bound for $0<s<1$, and $(b)$ is due to Lemma \ref{lem:exp_upper_bound_for_mgf_of_Z_{h'h}(n)}.
\end{IEEEproof}

From \eqref{eq:theta(n)=h_for_all_n_large}, we know that under the non-stopping version of the policy $\pi^\star(L,\delta)$, the guess of the odd arm $\theta(n)$ eventually settles at $h$ with probability $1$ under the hypothesis $\mathcal{H}_h$. Indeed, we now show using Lemma \ref{lem:exp_upper_bound_for_min_Z_{hh'}(n)<R} that something stronger holds. Towards this, fix $h\in\mathcal{A}$, and suppose that $\mathcal{H}_h$ is the true hypothesis. Let
\begin{equation}
	T_h\coloneqq \inf\{n:\theta(n')=h\text{ for all }n'\geq n\}.\label{eq:T_h}
\end{equation}
We have the following result for $T_h$.
\begin{Lemma}\label{lem:exp_upper_bound_for_T_h}
	Fix an arbitrary $h\in\mathcal{A}$, and suppose that $\mathcal{H}_h$ is the true hypothesis. Consider the non-stopping version of the policy $\pi^\star(L,\delta)$. There exist constants $C>0$ and $b>0$, both finite and possibly depending on $h$, such that
	\begin{equation}
		P_h\left(T_h>n\right)\leq C\,e^{-bn}.
	\end{equation}
\end{Lemma}
\begin{IEEEproof}[Proof of Lemma \ref{lem:exp_upper_bound_for_T_h}]
	We have
	\begin{align}
		P_h\left(T_h>n\right)&\leq P_h\left(\exists\text{ }n'\geq n\text{ such that }\theta(n')\neq h\right)\nonumber\\
		&\leq \sum\limits_{n'\geq n}P_h(\theta(n')\neq h)\nonumber\\
		&= \sum\limits_{n'\geq n}P_h\left(\exists\text{ }h'\neq h\text{ such that }\theta(n')=h'\right)\nonumber\\
		&\leq \sum\limits_{n'\geq n}P_h\left(M_{h'}(n')> \max\limits_{h''\neq h'}M_{h''}(n)\right)\nonumber\\
		&\leq \sum\limits_{n'\geq n}P_h\left(M_{h}(n')-M_{h'}(n')< 0\right).\label{eq:temp_6}
	\end{align}
	We now note that, almost surely,
	\begin{align}
		M_h(n')-M_{h'}(n')&=M_h(n')-\min\limits_{h''\neq h'}Z_{h'h''}(n')\nonumber\\
		&\geq M_{h}(n')-Z_{h'h}(n')\nonumber\\
		&=M_{h}(n')+Z_{hh'}(n')\nonumber\\
		&\geq 2\min\limits_{h'\neq h}Z_{hh'}(n').\label{eq:temp_7}
	\end{align}
	Using \eqref{eq:temp_7} in \eqref{eq:temp_6}, we get
	\begin{align}
		P_h\left(T_h>n\right)&\leq \sum\limits_{n'\geq n}P_h\left(\min\limits_{h'\neq h}Z_{hh'}(n)< 0\right).
	\end{align}
	The result now follows from Lemma \ref{lem:exp_upper_bound_for_min_Z_{hh'}(n)<R}.
\end{IEEEproof}

We now use the results presented above to derive the desired exponential upper bound for each term of the summation in \eqref{eq:exp_bound_1} to finish the proof of Lemma \ref{lem:exp_upper_bound}. 
Note that for any $\epsilon'>0$, we have
\begin{align}
	&P_h\left(Z_{hh'}(n)<\log((K-1)L)\right)\nonumber\\
	&=P_h\left(\sum\limits_{k=K}^{n}\Delta Z_{hh'}(k)<\log((K-1)L)\right)\nonumber\\
	&=P_h\bigg(\sum\limits_{k=K}^n(\Delta Z_{hh'}(k)-E_h[\Delta Z_{hh'}(k)|\mathcal{F}_{k-1}]+\epsilon')\nonumber\\
	&\hspace{3cm}+\sum\limits_{k=K}^n (E_h[\Delta Z_{hh'}(k)|\mathcal{F}_{k-1}]-D_{hh'}+\epsilon')\nonumber\\
	&\hspace{5cm}+(n-K+1)\,(D_{hh'}-2\epsilon')<\log((K-1)L)\bigg)\nonumber\\
	&\leq P_h\bigg(\sum\limits_{k=K}^n(\Delta Z_{hh'}(k)-E_h[\Delta Z_{hh'}(k)|\mathcal{F}_{k-1}]+\epsilon')<0\bigg)+P_h\bigg(\sum\limits_{k=K}^n (E_h[\Delta Z_{hh'}(k)|\mathcal{F}_{k-1}]-D_{hh'}+\epsilon')<0\bigg)\nonumber\\
	&\hspace{6cm}+P_h\bigg((n-K+1)\,(D_{hh'}-2\epsilon')<\log((K-1)L)\bigg).\label{eq:temp_8}
\end{align}
We first choose $\epsilon'$ such that $$(n-K+1)\,(D_{hh'}-2\epsilon')\geq \log((K-1)L)\quad \forall n\geq \tilde{n}(L).$$ {\color{black} In particular, it suffices to set $\epsilon'=D_{hh'}/4$}. Let us fix this value of $\epsilon'$ for the rest of the proof, and note that this choice of $\epsilon'$ ensures that the third probability term in \eqref{eq:temp_8} is equal to $0$. We now focus on the first probability term in \eqref{eq:temp_8}, and note that each term inside the summation has strictly positive mean. Thus, from Chernoff's bounding technique \cite[Lemma 2]{Chernoff1959}, we get that there exists $b(\epsilon')$ such that 
\begin{equation}
	P_h\bigg(\sum\limits_{k=K}^n(\Delta Z_{hh'}(k)-E_h[\Delta Z_{hh'}(k)|\mathcal{F}_{k-1}]+\epsilon')<0\bigg)\leq e^{-(n-K+1)\,b(\epsilon')}.
\end{equation}
It thus remains to show that the second probability term in \eqref{eq:temp_8} is bounded above exponentially. To do so, we use the proof technique of Vaidhiyan et al. \cite[pp. 4793-4794]{Vaidhiyan2017} and adapt it to our setting of restless arms. 

Let 
\begin{align}
	\tilde{C}&\coloneqq \inf\limits_{(\underline{d},\underline{i})\in\mathbb{S},\,\,a\in\mathcal{A}}E_h[\Delta Z_{hh'}(n)\mid A_n=a,\underline{d}(n)=\underline{d},\underline{i}(n)=\underline{i}]-D_{hh'} \nonumber\\
	&= \inf\limits_{(\underline{d},\underline{i})\in\mathbb{S},\,\,a\in\mathcal{A}}~D((P_h^a)^{d_a}(\cdot\mid i_a)\|(P_{h'}^a)^{d_a}(\cdot\mid i_a))-D_{hh'}.\label{eq:C_tilde}
\end{align}
Note that $\tilde{C}\leq 0$ by the definition of $D_{hh'}$. Choose $\epsilon''$ such that $$\tilde{\epsilon}\coloneqq \epsilon'+\epsilon'' \tilde{C}>0;$$ here, $\epsilon'=D_{hh'}/4$ as chosen earlier. We may then write the second probability in \eqref{eq:temp_8} as follows:
\begin{align}
	&P_h\bigg(\sum\limits_{k=K}^n (E_h[\Delta Z_{hh'}(k)|\mathcal{F}_{k-1}]-D_{hh'}+\epsilon')<0\bigg)\nonumber\\
	&= P_h\bigg(\sum\limits_{k=K}^n (E_h[\Delta Z_{hh'}(k)|\mathcal{F}_{k-1}]-D_{hh'}+\epsilon')<0,~T_h\leq n\epsilon''\bigg)\nonumber\\
	&\hspace{5cm}+P_h\bigg(\sum\limits_{k=K}^n (E_h[\Delta Z_{hh'}(k)|\mathcal{F}_{k-1}]-D_{hh'}+\epsilon')<0,~T_h>n\epsilon''\bigg)\nonumber\\
	&\leq  P_h\bigg(\sum\limits_{k=K}^n (E_h[\Delta Z_{hh'}(k)|\mathcal{F}_{k-1}]-D_{hh'}+\epsilon')<0,~T_h\leq n\epsilon''\bigg)+P_h\bigg(T_h>n\epsilon''\bigg).\label{eq:temp_9}
\end{align}
From Lemma \ref{lem:exp_upper_bound_for_T_h}, the second probability term in \eqref{eq:temp_9} is bounded above exponentially. The first probability term in \eqref{eq:temp_9} may be upper bounded as
\begin{align}
	&P_h\bigg(\sum\limits_{k=K}^n (E_h[\Delta Z_{hh'}(k)|\mathcal{F}_{k-1}]-D_{hh'}+\epsilon')<0,~T_h\leq n\epsilon''\bigg)\nonumber\\
	&= P_h\bigg(\sum\limits_{k=K}^{\lfloor n\epsilon''\rfloor} (E_h[\Delta Z_{hh'}(k)|\mathcal{F}_{k-1}]-D_{hh'}+{\epsilon'})+ \sum\limits_{k=\lfloor n\epsilon''\rfloor+1}^n (E_h[\Delta Z_{hh'}(k)|\mathcal{F}_{k-1}]-D_{hh'}+{\epsilon'})<0,~T_h\leq n\epsilon''\bigg)\nonumber\\
	&\stackrel{(a)}{\leq} P_h\bigg((\lfloor n\epsilon''\rfloor-K+1)(\tilde{C}+\epsilon') + \sum\limits_{k=\lfloor n\epsilon''\rfloor+1}^n (E_h[\Delta Z_{hh'}(k)|\mathcal{F}_{k-1}]-D_{hh'}+{\epsilon'})<0,~T_h\leq n\epsilon''\bigg)\nonumber\\
	&=P_h\bigg((\lfloor n\epsilon''\rfloor-K+1)(\tilde{C}+\epsilon') +(n-\lfloor n\epsilon''\rfloor)(\epsilon'-\tilde{\epsilon})+ \sum\limits_{k=\lfloor n\epsilon''\rfloor+1}^n (E_h[\Delta Z_{hh'}(k)|\mathcal{F}_{k-1}]-D_{hh'}+\tilde{\epsilon})<0,~T_h\leq n\epsilon''\bigg)\nonumber\\
	&\stackrel{(b)}{\leq} P_h\bigg(\lfloor n\epsilon''\rfloor(\epsilon''\tilde{C}+\epsilon') -(K-1)\epsilon'+ \sum\limits_{k=\lfloor n\epsilon''\rfloor+1}^n (E_h[\Delta Z_{hh'}(k)|\mathcal{F}_{k-1}]-D_{hh'}+\tilde{\epsilon})<0,~T_h\leq n\epsilon''\bigg)\nonumber\\
	&\stackrel{(c)}{\leq} P_h\bigg(\sum\limits_{k=\lfloor n\epsilon''\rfloor+1}^n (E_h[\Delta Z_{hh'}(k)|\mathcal{F}_{k-1}]-D_{hh'}+\tilde{\epsilon})<0,~T_h\leq n\epsilon''\bigg)\nonumber\\
	&=\tilde{P}_h\bigg(\sum\limits_{k=\lfloor n\epsilon''\rfloor+1}^n (E_h[\Delta Z_{hh'}(k)|\mathcal{F}_{k-1}]-D_{hh'}+\tilde{\epsilon})<0\bigg),\label{eq:temp_10}
\end{align}
where in writing $(a)$, we use the fact that for each $k\geq K$, we have $E_h[\Delta Z_{hh'}(k)|\mathcal{F}_{k-1}]\geq \tilde{C}$, $(b)$ follows by noting that 
\begin{align}
&(\lfloor n\epsilon''\rfloor-K+1)(\tilde{C}+\epsilon') +(n-\lfloor n\epsilon''\rfloor)(\epsilon'-\tilde{\epsilon})\nonumber\\
&=(\lfloor n\epsilon''\rfloor-K+1)(\tilde{C}+\epsilon') -(n-\lfloor n\epsilon''\rfloor)\,\epsilon''\tilde{C}\nonumber\\
&=\lfloor n\epsilon''\rfloor(\epsilon' + \epsilon''\tilde{C})+\tilde{C}(\lfloor n\epsilon''\rfloor-n\epsilon'')-(K-1)(\tilde{C}+\epsilon')\nonumber\\
&\geq \lfloor n\epsilon''\rfloor(\epsilon''\tilde{C}+\epsilon') -(K-1)\epsilon'
\end{align}
since $\tilde{C}\leq 0$, $(c)$ and the equality in \eqref{eq:temp_10} hold for all $n$ such that $\lfloor n\epsilon''\rfloor(\epsilon''\tilde{C}+\epsilon') -(K-1)\epsilon' \geq 0$, and in
  \eqref{eq:temp_10}, $\tilde{P}_h$ is a new probability measure under which at each time instant, an arm is selected according to the policy $\pi^\star(L,\delta)$ but assuming that the guess of the odd arm $\theta(k)=h$ for all $k$. 



We now 
note that under the measure $\tilde{P}_h$,
\begin{equation}
	\tilde{E}_h[E_h[\Delta Z_{hh'}(k)|\mathcal{F}_{k-1}]]=\sum\limits_{(\underline{d},\underline{i})\in\mathbb{S}}\sum\limits_{a=1}^{K} \tilde{P}_h(\underline{d}(k)=\underline{d},\underline{i}(k)=\underline{i})\,\left(\frac{\eta}{K}+(1-\eta)\lambda_{h,\delta}(a|\underline{d},\underline{i})\right)\,D((P_h^a)^{d_a}(\cdot\mid i_a)\|(P_{h'}^a)^{d_a}(\cdot\mid i_a)),\label{eq:temp_11}
\end{equation}
where $\tilde{E}_h$ in \eqref{eq:temp_11} denotes expectation under the measure $\tilde{P}_h$. We claim that under the measure $\tilde{P}_h$, the collection $\{(\underline{d}(k),\underline{i}(k)):k\geq \lfloor n\epsilon''\rfloor+1\}$ is a Markov process. Indeed, for all $k\geq \lfloor n\epsilon''\rfloor+1$,
\begin{align}
	&\tilde{P}_h(\underline{d}(k+1)=\underline{d}',\underline{i}(k+1)=\underline{i}'\mid (\underline{d}(t),\underline{i}(t)),~\lfloor n\epsilon''\rfloor+1\leq t\leq k)\nonumber\\
	&=\begin{cases}
		\left(\frac{\eta}{K}+(1-\eta)\,\lambda_{h,\delta}(a|\underline{d}(k),\underline{i}(k))\right)\,(P_h^a)^{d_a(k)}(i_a'|i_a(k)),&\text{if }d_a'=1\text{ and }d_b'=d_b(k)+1\text{ for all }b\neq a,\\
		&i_b'=i_b(k)\text{ for all }b\neq a,\\
		0,&\text{otherwise}.
	\end{cases}
\end{align}

Fix an integer $M>> K-1$  such that \eqref{eq:P_1^M_and_P_2^M_entries_strictly_pos_entries} holds, and let $\underline{d}'=(K,K-1,\ldots,1)$ and $\underline{i}'=(i,\ldots,i)$ for some $i\in \mathcal{S}$. In what follows, we demonstrate that $(\underline{d}', \underline{i}')$ may be reached starting from any $(\underline{d}, \underline{i})$, with a strictly positive probability. 
Indeed, given any $(\underline{d}, \underline{i})$, assume that the Markov process $\{(\underline{d}(t), \underline{i}(t):t\geq K)\}$ is in the state $(\underline{d}, \underline{i})$ at some time $T_0\geq \lfloor n\epsilon'' \rfloor + 1$. Consider the following sequence of arm selections and observations: pull arm $1$ at time $t=T_0+1$, arm $2$ at time $t=T_0+2$ and so on until arm $K$ at time $t=T_0+K$. Thereafter, pull arm $1$ at time $t=T_0+M+1$ and observe it in state $i$. Pull arm $2$ at time $t=T_0+M+2$ and observe it in state $i$. Continuing this way, finally pull arm $K$ at time $t=T_0+M+K$ and observe it in state $i$. Notice that we do not specify the states of the arms as observed at times $T_0+1, \ldots, T_0+K$. For computational purposes, let these states be $s_1, \ldots, s_K$ from arms $1, \ldots, K$ respectively. 

Clearly, at time $t=T_0+M+2K+1$, we have $\underline{d}(t)=\underline{d}'$, $ \underline{i}(t)=\underline{i}'$, and 
\begin{align}
	\tilde{P}_h(\underline{d}(T_0+M+2K+1)=\underline{d}',\underline{i}(T_0+M+2K+1)=\underline{i}'\mid \underline{d}(T_0)=\underline{d},\underline{i}(T_0)=\underline{i})\geq \bigg(\frac{\eta}{K}\bigg)^{2K}\cdot \left[\prod\limits_{a=1}^{K}(P_h^a)^{M}(i|s_a)\right]\label{eq:temp_12}.
\end{align}
Denoting the right-hand side of \eqref{eq:temp_12} by $\alpha$, and noting that $\alpha>0$ and independent of $(\underline{d}, \underline{i})$, we have
\begin{align}
(\tilde{P}_h)^{M+2K+1}((\underline{d}'',\underline{i}'')\mid \underline{d},\underline{i})\geq \alpha\,\, \mathbb{I}_{\{(\underline{d}'',\underline{i}'')=(\underline{d}',\underline{i}')\}} \quad \text{for all } (\underline{d},\underline{i}),~(\underline{d}'',\underline{i}'')\in\mathbb{S}. \label{eq:temp_13}
\end{align}
The condition in \eqref{eq:temp_13} is referred to as the ``Doeblin's minorisation condition''  \cite[Eq. (5)]{kontoyiannis2005relative}. Noting that (a) the Markov process $\{(\underline{d}(k),\underline{i}(k)):k\geq \lfloor n\epsilon''\rfloor+1\}$ is ergodic under the measure $\tilde{P}_h$, with $\mu^{\lambda_{h,\delta}}$ as its unique stationary distribution, (b) \eqref{eq:temp_13} holds, and (c) the increment  $\Delta Z_{hh'}(k)$ is almost surely bounded for each $k$ as demonstrated in \eqref{eq:Z_{hh'}(n)-Z_{hh'}(n-1)_upper_bound}, we apply \cite[Theorem 1]{kontoyiannis2005relative} to deduce that the second probability term in \eqref{eq:temp_8} is bounded above exponentially. This establishes the lemma.
\end{IEEEproof}

\section{An Infinite-Dimensional Linear Programming Problem}\label{subsec:infinite_LPP} \label{appndx:infinite_dimensional_LP}
In order to better appreciate the usefulness of taking into account the arm delays and last observed states of \emph{all} the arms in deriving the lower bound, we present below a {\color{black} proof sketch} of a possibly weaker lower bound in which we first fix an arm $a$ and consider only its delay and last observed state in the subsequent calculations. Fix arm $a\in\mathcal{A}$. Given an integer $d\geq 1$, $i,j\in\mathcal{S}$ and a policy $\pi$, let
\begin{equation}
	N(\tau(\pi),d,i,a,j)\coloneqq \sum\limits_{t=K}^{\tau(\pi)}\mathbb{I}_{\{d_a(t)=d,i_a(t)=i,A_t=a,X_t^a=j\}}\label{eq:N(tau(pi),d,i,a,j)}.
\end{equation}
Recall that $\tau(\pi)$ denotes the stopping time of policy $\pi$. Following the earlier approaches of \cite{Vaidhiyan2017, vaidhiyan2017learning, prabhu2017optimal, pnkarthik2019learning}, and using the data processing inequality, one arrives at\footnote{For the gentle reader interested in the details, this can be obtained by following the chain of equalities leading up to \eqref{eq:lower_bound_2} in Appendix \ref{appndx:proof_of_prop_lower_bound}, with the inner summation over $(\underline{d},\underline{i})$ now replaced by a summation over $(d,i)\in\{1,2,\ldots\}\times \mathcal{S}$.}
\begin{equation}
	\sum\limits_{a=1}^{K}~\sum\limits_{d=1}^{\infty}~\sum\limits_{i\in\mathcal{S}}E_h[N(\tau(\pi),d,i,a)]\,D((P_h^a)^{d}(\cdot|i)\|(P_{h'}^a)^{d}(\cdot|i)),\label{eq:equiv_of_lower_bound_2}
\end{equation}
where $N(\tau(\pi),d,i,a)$ in \eqref{eq:equiv_of_lower_bound_2} is simply the summation over all $j\in\mathcal{S}$ of the right-hand side of \eqref{eq:N(tau(pi),d,i,a,j)}. 

\begin{figure}
	\includegraphics[width=\textwidth]{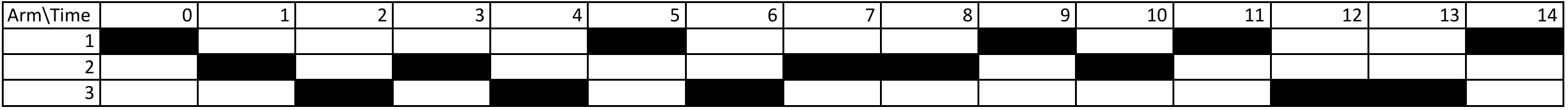}
	\caption{A schematic representation of arm selections over time for $K=3$ arms. In this schematic, an arm selected at any given time is indicated by a black box. Note that arm $1$ is selected at time $t=0$, arm $2$ at time $t=1$ and arm $3$ at time $t=2$. Thereafter, for $t\geq 3$, arm $1$ is selected at certain time instants and is not selected at certain other time instants. Whenever arm $1$ is not selected, \emph{some} other arm is selected, as a consequence of which the delay of arm $1$ increases, and it is this fact that must be captured as a constraint on the delays of arm $1$. Similar constraints apply for each of the other arms.}
	\label{fig:arm_selections}
\end{figure}

From the exposition in Section \ref{sec:preliminaries}, we know that at any given time $t\geq K$, the vector $\underline{d}(t)$ must satisfy the following constraint: exactly one component of $\underline{d}(t)$ is equal to $1$, and all the other components are strictly greater than $1$. Let us now express this constraint mathematically.  Recall the assumption that the policy $\pi$ selects, without loss of generality, arm $1$ at time $t=0$, arm $2$ at time $t=1$ and so on until arm $K$ at time $t=K-1$. From time $t=K$ onwards, arm $a$ may or may not be selected at all time instants, and whenever it is not selected, \emph{some} arm $b\neq a$ is selected. It is this observation (that some arm is selected at every time instant until the stopping time of the policy) that must be modelled as a constraint mathematically. Figure \ref{fig:arm_selections} depicts the selection of arms at various time instants for the case when $K=3$.

Assume without loss of generality that under the policy $\pi$, arm $a$ is selected at time $t=\tau(\pi)$. Then, it follows that 
\begin{equation}
	(a-1)+\sum\limits_{i\in\mathcal{S}}~\sum\limits_{d=1}^{\infty}~d\,N(\tau(\pi),d,i,a) + 1 = \tau(\pi)+1;\label{eq:constraint}
\end{equation}
in \eqref{eq:constraint}, the term $(a-1)$ on the left-hand side denotes the number of time instants that have passed before arm $a$ is selected for the first time. The second term on the left-hand side of \eqref{eq:constraint} denotes the total number of time instants that have passed, starting from time $t=K$, until the final selection time instant of arm $a$. The last term on the left-hand side of \eqref{eq:constraint} counts the final selection instant of arm $a$. Thus, the total value of the left-hand side of \eqref{eq:constraint} is equal to the total number of time instants that have passed from $t=0$ to $t=\tau(\pi)$ (both inclusive), which is precisely the quantity on the right-hand side of \eqref{eq:constraint}. Applying $E_h[\cdot]$ to both sides of \eqref{eq:constraint}, and using the monotone convergence theorem, we arrive at the following relation after some rearrangement:
\begin{equation}
	\sum\limits_{i\in\mathcal{S}}~\sum\limits_{d=1}^{\infty} d~\frac{E_h[N(\tau(\pi),d,i,a)]}{E_h[\tau(\pi)]}+\frac{a-1}{E_h[\tau(\pi)]}=1.\label{eq:constraint_1}
\end{equation}
In fact, it is easy to see that \eqref{eq:constraint}, and therefore \eqref{eq:constraint_1}, holds for every arm, whether or not the arm is selected at time $t=\tau(\pi)$. Mimicking the steps in Appendix \ref{appndx:proof_of_prop_lower_bound}, and using  \eqref{eq:equiv_of_lower_bound_2} in place of \eqref{eq:lower_bound_2} in Appendix \ref{appndx:proof_of_prop_lower_bound} along with the constraint in \eqref{eq:constraint_1}, we arrive at the following relation in place of \eqref{eq:lower_bound_5}:
\begin{align}
	d(\epsilon,1-\epsilon)&\leq \sup\limits_{\kappa}\,\min\limits_{h'\neq h}\bigg\lbrace E_h\left[\sum\limits_{a=1}^{K}\log\frac{P_h(X_{a-1}^{a})}{P_{h'}(X_{a-1}^{a})}\right]\nonumber\\
	&\hspace{2.5cm}+\bigg(E_h[\tau(\pi)-K+1]\bigg)\cdot\,\sum\limits_{a=1}^{K}~\sum\limits_{d=1}^{\infty}~\sum\limits_{i\in\mathcal{S}}\kappa(d,i,a)\, D((P_h^a)^{d}(\cdot|i)\|(P_{h'}^a)^{d}(\cdot|i))\bigg\rbrace,\label{eq:equiv_lower_bound_5}
\end{align}
where $d(\epsilon, 1-\epsilon)$ is the relative entropy between a Bernoulli distribution with parameter $\epsilon$ and a Bernoulli distribution with parameter $1-\epsilon$, and the supremum in \eqref{eq:equiv_lower_bound_5} is over all probability distributions $\kappa$ on $\{1,2,\ldots\}\times\mathcal{S}\times\mathcal{A}$ that satisfy the constraint
\begin{equation}
	\sum\limits_{i\in\mathcal{S}}~\sum\limits_{d=1}^{\infty} d~\kappa(d,i,a)=1\quad \text{ for all }a\in\mathcal{A}.\label{eq:constraint_2}
\end{equation}
The constraint in \eqref{eq:constraint_2} may be obtained from \eqref{eq:constraint_1} by letting $E_h[\tau(\pi)]\to \infty$ (which is the same as $\epsilon\downarrow 0$) and replacing the fractional term on the left-hand side of \eqref{eq:constraint_1} by $\kappa(d,i,a)$; here, $\kappa(d,i,a)$ represents the long-term joint probability of observing arm $a$ to have a delay $d$ and last observed state $i$, and subsequently selecting arm $a$. 

Dividing both sides of \eqref{eq:equiv_lower_bound_5} by $d(\epsilon,1-\epsilon)$, and using the fact that $d(\epsilon,1-\epsilon)/\log(1/\epsilon)\to 1$ as $\epsilon\downarrow 0$, we arrive at
\begin{equation}
	\liminf\limits_{\epsilon\downarrow 0}\inf\limits_{\pi\in\Pi(\epsilon)}\frac{E_h[\tau(\pi)]}{\log(1/\epsilon)}\geq \frac{1}{R_1^*(P_1,P_2)},\label{eq:lower_bound_equiv}
\end{equation}
where $R_1^*(P_1,P_2)$ is the {\color{black} value} of the following constrained optimisation problem:
\begin{align}
	&R_1^*(P_1,P_2)=\sup\limits_{\kappa}\,\min\limits_{h'\neq h}\,\sum\limits_{a=1}^{K}~\sum\limits_{d=1}^{\infty}~\sum\limits_{i\in\mathcal{S}}\kappa(d,i,a)\, D((P_h^a)^{d}(\cdot|i)\|(P_{h'}^a)^{d}(\cdot|i))\nonumber\\
	&\text{subject to}\nonumber\\
	&\hspace{1cm}\sum\limits_{i\in\mathcal{S}}~\sum\limits_{d=1}^{\infty} d~\kappa(d,i,a)=1\quad \text{ for all }a\in\mathcal{A},\nonumber\\
	&\hspace{1cm} \sum\limits_{d=1}^{\infty}~\sum\limits_{i\in\mathcal{S}}~\sum\limits_{a=1}^K \kappa(d,i,a)=1,\nonumber\\
	&\hspace{1cm} \kappa(d,i,a)\geq 0\quad\text{ for all }d\in\{1,2,\ldots\},~i\in\mathcal{S},~a\in\mathcal{A}.\label{eq:constrained_LPP}
\end{align}
Notice that \eqref{eq:constrained_LPP} constitutes an infinite-dimensional linear programming problem with linear constraints. {\color{black} It is not clear if there exists $\kappa$ that (a) satisfies the constraints in \eqref{eq:constrained_LPP} and (b) attains the supremum in the expression for $R_1^*(P_1, P_2)$}. Also, it is not clear if the constraints in \eqref{eq:constrained_LPP} constitute the tightest set of constraints. From Proposition \ref{prop:lower_bound}, we must of course have $R_1^*(P_1,P_2)\geq R^*(P_1,P_2)$.

We end with a remark that by taking into account the delays and the last observed states of all the arms in deriving the lower bound, as done in Appendix \ref{appndx:proof_of_prop_lower_bound}, the constraint in \eqref{eq:constraint} is automatically captured since any vector $\underline{d}=(d_1,\ldots,d_K)$ of arm delays belongs, by definition, to the subset $\mathbb{S}$ which obeys the constraint in  \eqref{eq:constraint}. Thus, the viewpoint of controlled Markov processes greatly simplifies the analysis of the lower bound. The key insight of this paper is that our `lift' approach of considering the arm delays and the last observed states of all the arms jointly, instead of dealing with the delays and last observed states of each arm separately, makes the problem amenable to analysis.

\section{Restriction to SRS Class Suffices} \label{appndx:an_important_theorem}
An important step in the derivation of the lower bound \eqref{eq:lower_bound} presented in Appendix \ref{appndx:proof_of_prop_lower_bound} is the replacement of the supremum over the set $\Pi(\epsilon)$ appearing in \eqref{eq:lower_bound_5} to the set $\Pi_{\textsf{SRS}}$ of all SRS policies (compare the right hand side of \eqref{eq:lower_bound_5} with that of \eqref{eq:lower_bound_7}). Here, $\Pi(\epsilon)$, which the set of all policies whose probability of error at stoppage is at most $\epsilon$, may potentially include non-SRS policies too. The aforementioned step in the proof of the lower bound is possible thanks to the following theorem which is an analogue of \cite[Theorem 8.8.2]{puterman2014markov} for countable state space controlled Markov processes. We omit the proof of the theorem as it follows straightforwardly from the proof of \cite[Theorem 8.8.2]{puterman2014markov}. 

Recall that for each $\pi^\lambda\in \Pi_{\textsf{SRS}}$, the controlled Markov process $\{(\underline{d}(t), \underline{i}(t)):t\geq K\}$ is, in fact a Markov process. Furthermore, when the trembling hand parameter $\eta>0$, this Markov process is ergodic (Lemma \ref{lem:pi_delta^lambda_is_an_SSRS}).
\begin{thrm} 
\label{thrm:restriction_to_SRS_policies}
	\begin{enumerate}
		\item For each $\pi^\lambda\in \Pi_{\textsf{SRS}}$, $(\underline{d}, \underline{i})\in \mathbb{S}$ and $a\in \mathcal{A}$, let
			\begin{equation}
				\nu^\lambda(\underline{d}, \underline{i}, a) = \mu^\lambda(\underline{d}, \underline{i})~\left(\frac{\eta}{K} + (1-\eta) ~\lambda(a\mid \underline{d}, \underline{i})\right),
			\end{equation}
			where $\eta>0$ is the trembling hand parameter and $\mu^\lambda$ is the unique stationary distribution of the Markov process $\{(\underline{d}(t), \underline{i}(t)):t\geq K\}$ under the SRS policy $\pi^\lambda$. Then, $\nu^\lambda$ is a feasible  solution to \eqref{eq:lower_bound_6_1}-\eqref{eq:lower_bound_6_3}.
			
		\item Let $\nu$ be any feasible solution to \eqref{eq:lower_bound_6_1}-\eqref{eq:lower_bound_6_3}. Then, for each $(\underline{d}, \underline{i})\in \mathbb{S}$, $\sum\limits_{a=1}^{K} ~\nu(\underline{d}, \underline{i}, a) > 0$.
				Let $\lambda^*$ be such that
				$$
				\frac{\eta}{K}+(1-\eta)~\lambda^*(a\mid \underline{d}, \underline{i}) \coloneqq \frac{\nu(\underline{d}, \underline{i}, a)}{\sum\limits_{a=1}^{K} ~\nu(\underline{d}, \underline{i}, a)}, \quad (\underline{d}, \underline{i})\in \mathbb{S}, ~a\in \mathcal{A}.
				$$
				Then, $\nu^{\lambda^*}$ is a feasible solution to \eqref{eq:lower_bound_6_1}-\eqref{eq:lower_bound_6_3} and 
				$$
				\nu^{\lambda^*}(\underline{d}, \underline{i}, a)=\nu(\underline{d}, \underline{i}, a)\quad \text{for all }(\underline{d}, \underline{i})\in \mathbb{S}\text{ and }a\in \mathcal{A}.
				$$
	\end{enumerate}
\end{thrm}

%
\IEEEpeerreviewmaketitle

\bibliographystyle{IEEEtran}
\bibliography{oai_restless_arms}

\begin{thebibliography}{10}
\providecommand{\url}[1]{#1}
\csname url@samestyle\endcsname
\providecommand{\newblock}{\relax}
\providecommand{\bibinfo}[2]{#2}
\providecommand{\BIBentrySTDinterwordspacing}{\spaceskip=0pt\relax}
\providecommand{\BIBentryALTinterwordstretchfactor}{4}
\providecommand{\BIBentryALTinterwordspacing}{\spaceskip=\fontdimen2\font plus
\BIBentryALTinterwordstretchfactor\fontdimen3\font minus
  \fontdimen4\font\relax}
\providecommand{\BIBforeignlanguage}[2]{{%
\expandafter\ifx\csname l@#1\endcsname\relax
\typeout{** WARNING: IEEEtran.bst: No hyphenation pattern has been}%
\typeout{** loaded for the language `#1'. Using the pattern for}%
\typeout{** the default language instead.}%
\else
\language=\csname l@#1\endcsname
\fi
#2}}
\providecommand{\BIBdecl}{\relax}
\BIBdecl

\bibitem{Vaidhiyan2017}
N.~K. Vaidhiyan, S.~Arun, and R.~Sundaresan, ``Neural dissimilarity indices
  that predict oddball detection in behaviour,'' \emph{IEEE Transactions on
  Information Theory}, vol.~63, no.~8, pp. 4778--4796, 2017.

\bibitem{prabhu2017optimal}
G.~R. Prabhu, S.~Bhashyam, A.~Gopalan, and R.~Sundaresan, ``Optimal odd arm
  identification with fixed confidence,'' \emph{arXiv preprint
  arXiv:1712.03682}, 2017.

\bibitem{vaidhiyan2017learning}
N.~K. Vaidhiyan and R.~Sundaresan, ``Learning to detect an oddball target,''
  \emph{IEEE Transactions on Information Theory}, vol.~64, no.~2, pp. 831--852,
  2017.

\bibitem{pnkarthik2019learning}
P.~N. Karthik and R.~Sundaresan, ``Learning to detect an odd markov arm,''
  \emph{IEEE Transactions on Information Theory}, vol.~66, no.~7, pp.
  4324--4348, July 2020.

\bibitem{sripati2010global}
A.~P. Sripati and C.~R. Olson, ``Global image dissimilarity in macaque
  inferotemporal cortex predicts human visual search efficiency,''
  \emph{Journal of Neuroscience}, vol.~30, no.~4, pp. 1258--1269, 2010.

\bibitem{krueger2017evidence}
P.~M. Krueger, M.~K. van Vugt, P.~Simen, L.~Nystrom, P.~Holmes, and J.~D.
  Cohen, ``Evidence accumulation detected in bold signal using slow perceptual
  decision making,'' \emph{Journal of neuroscience methods}, vol. 281, pp.
  21--32, 2017.

\bibitem{zhao2008myopic}
Q.~Zhao, B.~Krishnamachari, and K.~Liu, ``On myopic sensing for multi-channel
  opportunistic access: structure, optimality, and performance,'' \emph{IEEE
  Transactions on Wireless Communications}, vol.~7, no.~12, pp. 5431--5440,
  2008.

\bibitem{whittle1988restless}
P.~Whittle, ``Restless bandits: Activity allocation in a changing world,''
  \emph{Journal of applied probability}, vol.~25, no.~A, pp. 287--298, 1988.

\bibitem{gittins1979bandit}
J.~C. Gittins, ``Bandit processes and dynamic allocation indices,''
  \emph{Journal of the Royal Statistical Society. Series B (Methodological)},
  pp. 148--177, 1979.

\bibitem{liu2012learning}
H.~Liu, K.~Liu, and Q.~Zhao, ``Learning in a changing world: Restless
  multiarmed bandit with unknown dynamics,'' \emph{IEEE Transactions on
  Information Theory}, vol.~59, no.~3, pp. 1902--1916, 2012.

\bibitem{ortner2012regret}
R.~Ortner, D.~Ryabko, P.~Auer, and R.~Munos, ``Regret bounds for restless
  markov bandits,'' in \emph{International Conference on Algorithmic Learning
  Theory}.\hskip 1em plus 0.5em minus 0.4em\relax Springer, 2012, pp. 214--228.

\bibitem{auer2002finite}
P.~Auer, N.~Cesa-Bianchi, and P.~Fischer, ``Finite-time analysis of the
  multiarmed bandit problem,'' \emph{Machine learning}, vol.~47, no. 2-3, pp.
  235--256, 2002.

\bibitem{grunewalder2019approximations}
S.~Gr{\"u}new{\"a}lder and A.~Khaleghi, ``Approximations of the restless bandit
  problem,'' \emph{The Journal of Machine Learning Research}, vol.~20, no.~1,
  pp. 514--550, 2019.

\bibitem{Bubeck2011}
\BIBentryALTinterwordspacing
S.~Bubeck, R.~Munos, and G.~Stoltz, ``Pure {E}xploration in {F}initely-armed
  and {C}ontinuous-armed {B}andits,'' \emph{Theor. Comput. Sci.}, vol. 412,
  no.~19, pp. 1832--1852, Apr. 2011. [Online]. Available:
  \url{http://dx.doi.org/10.1016/j.tcs.2010.12.059}
\BIBentrySTDinterwordspacing

\bibitem{Kaufmann2016}
E.~Kaufmann, O.~Capp{\'e}, and A.~Garivier, ``On the complexity of best-arm
  identification in multi-armed bandit models,'' \emph{The Journal of Machine
  Learning Research}, vol.~17, no.~1, pp. 1--42, 2016.

\bibitem{moulos2019optimal}
V.~Moulos, ``Optimal best markovian arm identification with fixed confidence,''
  in \emph{Advances in Neural Information Processing Systems}, 2019, pp.
  5606--5615.

\bibitem{deshmukh2018controlled}
A.~Deshmukh, S.~Bhashyam, and V.~V. Veeravalli, ``Controlled sensing for
  composite multihypothesis testing with application to anomaly detection,'' in
  \emph{2018 52nd Asilomar Conference on Signals, Systems, and
  Computers}.\hskip 1em plus 0.5em minus 0.4em\relax IEEE, 2018, pp.
  2109--2113.

\bibitem{deshmukh2019sequential}
------, ``Sequential controlled sensing for composite multihypothesis
  testing,'' \emph{arXiv preprint arXiv:1910.12697}, 2019.

\bibitem{prabhu2020sequential}
G.~R. Prabhu, S.~Bhashyam, A.~Gopalan, and R.~Sundaresan, ``Sequential
  multi-hypothesis testing in multi-armed bandit problems: An approach for
  asymptotic optimality,'' \emph{arXiv preprint arXiv:2007.12961}, 2020.

\bibitem{avrachenkov2020whittle}
\BIBentryALTinterwordspacing
K.~Avrachenkov and V.~S. Borkar, ``Whittle index based q-learning for restless
  bandits with average reward,'' 2020. [Online]. Available:
  \url{https://arxiv.org/abs/2004.14427}
\BIBentrySTDinterwordspacing

\bibitem{milgrom2002envelope}
P.~Milgrom and I.~Segal, ``Envelope theorems for arbitrary choice sets,''
  \emph{Econometrica}, vol.~70, no.~2, pp. 583--601, 2002.

\bibitem{borkar1988control}
V.~S. Borkar, ``Control of markov chains with long-run average cost
  criterion,'' in \emph{Stochastic Differential Systems, Stochastic Control
  Theory and Applications}.\hskip 1em plus 0.5em minus 0.4em\relax Springer,
  1988, pp. 57--77.

\bibitem{puterman2014markov}
M.~L. Puterman, \emph{Markov decision processes: discrete stochastic dynamic
  programming}.\hskip 1em plus 0.5em minus 0.4em\relax John Wiley \& Sons,
  2014.

\bibitem{levin2017markov}
D.~A. Levin and Y.~Peres, \emph{Markov chains and mixing times}.\hskip 1em plus
  0.5em minus 0.4em\relax American Mathematical Soc., 2017, vol. 107.

\bibitem{victor1999general}
H.~Victor \emph{et~al.}, ``A general class of exponential inequalities for
  martingales and ratios,'' \emph{The Annals of Probability}, vol.~27, no.~1,
  pp. 537--564, 1999.

\bibitem{Chernoff1959}
H.~Chernoff, ``Sequential design of experiments,'' \emph{The Annals of
  Mathematical Statistics}, vol.~30, no.~3, pp. 755--770, 1959.

\bibitem{kontoyiannis2005relative}
I.~Kontoyiannis, L.~A. Lastras-Monta{\~n}o, and S.~P. Meyn, ``Relative entropy
  and exponential deviation bounds for general markov chains,'' in
  \emph{Proceedings. International Symposium on Information Theory, 2005. ISIT
  2005.}\hskip 1em plus 0.5em minus 0.4em\relax IEEE, 2005, pp. 1563--1567.

\end{thebibliography}

%








\end{document}